\documentclass[article,fleqn,usenatbib]{mnras}

\usepackage[T1]{fontenc}
\usepackage{ae,aecompl}

\usepackage{graphicx}	
\usepackage{amsmath}	
\usepackage{amssymb}	
\usepackage{latexsym}
\usepackage{amsfonts}
\usepackage{epstopdf}
\usepackage{rotating}
\usepackage{lscape}
\usepackage{changepage}
\usepackage{url}
\usepackage{color}
\usepackage{footnote}
\usepackage{cleveref}
\crefname{section}{§}{§§}
\Crefname{section}{§}{§§}
\usepackage{float}
\usepackage{subfig}
\usepackage{caption}
\usepackage{booktabs}
\usepackage{fixltx2e}
\usepackage{threeparttable}
\usepackage{floatpag}
\makesavenoteenv{itemize}
\usepackage{booktabs}
\usepackage{hyperref}
\def\aj{AJ}                   
            
\def\apj{ApJ}                
\def\apjl{ApJ}                
\def\apjs{ApJS}              
\def\ao{Appl.~Opt.}          
             
\def\aap{A\&A}

\def\mnras{MNRAS}

\def\pasa{PASA}
\def\pasp{PASP}
\def\pasj{PASJ}

\def\ssr{Space~Sci.~Rev.}

\def\iaucirc{IAU~Circ.}

\def\memsai{Mem.~Soc.~Astron.~Italiana}

\newcommand{\ignore}[1]{}

\title[Effective Temperatures of CV white dwarfs]{Effective Temperatures of Cataclysmic Variable White Dwarfs as a Probe of their Evolution}

\author[A. F. Pala et al.]{A.~F.~Pala$^{1}$\thanks{E-mail:
A.F.Pala@warwick.ac.uk},
B.~T.~G{\"a}nsicke$^{1}$, 
D.~Townsley$^{2}$, 
D.~Boyd$^{3}$,
M.~J.~Cook$^{4}$, \newauthor
D.~De~Martino$^{5}$, 
P.~Godon$^{6,7}$, 
J.~B.~Haislip$^{8}$, 
A.~A.~Henden$^{4}$, 
I.~Hubeny$^{9}$, \newauthor
K.~M.~Ivarsen$^{8}$, 
S.~Kafka$^{4}$, 
C.~Knigge$^{10}$, 
A.~P.~LaCluyze$^{8}$,  
K.~S.~Long$^{11,12}$, \newauthor
T.~R.~Marsh$^{1}$, 
B.~Monard$^{13}$, 
J.~P.~Moore$^{8}$, 
G.~Myers$^{4}$, 
P.~Nelson$^{4}$,  
D.~Nogami$^{14}$, \newauthor
A.~Oksanen$^{15}$,
R.~Pickard$^{3}$, 
G.~Poyner$^{3}$, 
D.~E.~Reichart$^{8}$, 
D.~Rodriguez~Perez$^{4}$, \newauthor
M.~R.~Schreiber$^{16}$, 
J.~Shears$^{3}$, 
E.~M.~Sion$^{6}$, 
R.~Stubbings$^{4}$, 
P.~Szkody$^{17}$, 
M.~Zorotovic$^{16}$ \\
$^{1}$Department of Physics, University of Warwick, Coventry, CV4 7AL, UK\\
$^{2}$Department of Physics and Astronomy, University of Alabama, Tuscaloosa, AL 35405, USA\\
$^{3}$British Astronomical Association, Variable Star Section, Burlington House, Piccadilly, London, W1J ODU, UK\\
$^{4}$American Association of Variable Star Observers, Cambridge, MA 02138, USA\\
$^{5}$INAF -- Osservatorio Astronomico di Capodimonte, Napoli, I--80131, Italy \\
$^{6}$Astrophysics \& Planetary Science, Villanova University, Villanova, PA 19085, USA\\
$^{7}$Visiting at the Johns Hopkins University, Henry A. Rowland Department of Physics and Astronomy, Baltimore, MD 21218, USA \\
$^{8}$Skynet Robotic Telescope Network, University of North Carolina, Chapel Hill, NC 27599, USA\\
$^{9}$Steward Observatory, The University of Arizona, Tucson, AZ 85721, USA\\
$^{10}$School of Physics and Astronomy, University of Southampton, Southampton, SO17 1BJ, UK\\
$^{11}$Space Telescope Science Institute, 3700 San Martin Drive, Baltimore, MD 21218, USA\\
$^{12}$Eureka Scientific, Inc. 2452 Delmer Street, Suite 100, Oakland, CA 94602--3017, USA\\
$^{13}$CBA Kleinkaroo, Calitzdorp, South Africa\\
$^{14}$Department of Astronomy, Graduate School of Science, Kyoto University, Kitashirakawa--Oiwake--cho, Sakyo--ku, Kyoto, 606--8502, Japan\\
$^{15}$Caisey Harlingten Observatory, Caracoles 166, San Pedro de Atacama, Chile\\
$^{16}$Instituto de F{\'i}sica y Astronom{\'i}a, Universidad de Valpara{\'i}so, 2360102 Valparaiso, Chile\\
$^{17}$Department of Astronomy, University of Washington, Seattle, WA 98195--1580, USA\\}

\date{Accepted 2016 December 14. Received 2016 December 10; in original form 2016 November 12}

\pubyear{2016}

\begin{document}
\label{firstpage}
\pagerange{\pageref{firstpage}--\pageref{lastpage}}
\maketitle

\begin{abstract}
We present \textit{HST} spectroscopy for 45 cataclysmic variables (CVs), observed with \textit{HST}/COS and \textit{HST}/STIS. 
For 36 CVs, the white dwarf is recognisable through its broad Ly$\alpha$ absorption profile and we measure the white dwarf effective temperatures ($T_{\mathrm{eff}}$) by fitting the \textit{HST} data assuming $\log\,g=8.35$, which corresponds to the average mass for CV white dwarfs ($\simeq\,0.8\,\mathrm{M}_\odot$). Our results nearly double the number of CV white dwarfs with an accurate temperature measurement.\\
We find that CVs above the period gap have, on average, higher temperatures ($\langle  T_{\mathrm{eff}} \rangle \simeq 23\,000\,$K) and exhibit much more scatter compared to those below the gap ($\langle  T_{\mathrm{eff}} \rangle \simeq 15\,000\,$K). While this behaviour broadly agrees with theoretical predictions, some discrepancies are present: (i) all our new measurements above the gap are characterised by lower temperatures ($T_{\mathrm{eff}} \simeq 16\,000 - 26\,000\,$K) than predicted by the present day CV population models ($T_{\mathrm{eff}} \simeq 38\,000 - 43\,000\,$K); (ii) our results below the gap are not clustered in the predicted narrow track and exhibit in particular a relatively large spread near the period minimum, which may point to some shortcomings in the CV evolutionary models.\\ 
Finally, in the standard model of CV evolution, reaching the minimum period, CVs are expected to evolve back towards longer periods with mean accretion rates $\dot{M}\lesssim 2 \times 10^{-11}\,\mathrm{M}_\odot\,\mathrm{yr}^{-1}$, corresponding to $T_\mathrm{eff}\lesssim 11\,500$\,K. We do not unambiguously identify any such system in our survey, suggesting that this major component of the predicted CV population still remains elusive to observations.

\end{abstract}




\section{Introduction}
Cataclysmic variables (CVs) are close interacting binaries containing a white dwarf accreting from a Roche--lobe filling low--mass companion star \citep{Warner}. In the absence of a strong magnetic field ($B \lesssim 10\,\mathrm{MG}$), the flow of material lost by the secondary gives rise to an accretion disc around the white dwarf primary.
  
In the standard model of CV evolution, the stability of the accretion process on long time scales (of the order of $10^9\,\mathrm{Gyr}$, \citeauthor{Kolb_age} \citeyear{Kolb_age}) requires a mass ratio $q = M_2/M_1 \lesssim 5/6$ (\citeauthor{libro} \citeyear{libro}, $M_1$ and $M_2$ denote the white dwarf and secondary mass) and a mechanism of angular momentum loss (AML) which continuously shrinks the system and keeps the secondary in touch with its Roche lobe. In the frequently referenced Interrupted Magnetic Braking Scenario, at least two different AML mechanisms drive CV evolution (\citeauthor{Rappaport1983} \citeyear{Rappaport1983}, \citeauthor{Paczynski} \citeyear{Paczynski}, \citeauthor{SpruitRitter} \citeyear{SpruitRitter}). For orbital periods $P_\mathrm{orb}\,\gtrsim\,3\,\mathrm{h}$, the angular momentum is predominantly removed by magnetic braking (MB) which arises from a stellar wind associated with the magnetic activity of the secondary, with typical mass transfer rates of $\dot{M}\,\sim\,10^{-9} - 10^{-8} \,\mathrm{M}_\odot \, \mathrm{yr}^{-1}$ \citep{SpruitRitter}. For $P_\mathrm{orb} \lesssim 3\,\mathrm{h}$, the dominant AML mechanism is gravitational wave radiation (GR), with typical $\dot{M} \sim 5 \times 10^{-11} \,\mathrm{M}_\odot \, \mathrm{yr}^{-1}$ \citep{Patterson1984}.

Another fundamental ingredient for CV evolution is the internal structure of the secondary and its response to mass loss (\citeauthor{Knigge_period_gap} \citeyear{Knigge_period_gap}, \citeauthor{Knigge2011} \citeyear{Knigge2011}). At $P_\mathrm{orb} \simeq 3\,\mathrm{h}$, the donor star has become fully convective. Magnetic braking via a stellar wind is thought to be greatly reduced and the system evolves as a detached binary through the period range $2\,\mathrm{h} \lesssim P_\mathrm{orb} \lesssim 3\,\mathrm{h}$ (the period gap). Following the diminution of MB \citep{Rappaport1983}, the system evolves through GR only, which reduces the orbital separation bringing the secondary again into contact with its Roche--lobe at $P_\mathrm{orb} \simeq 2\,\mathrm{h}$.

In the final phase of CV evolution ($P_\mathrm{orb} \simeq 80\,\mathrm{min}$), the donor starts behaving like a degenerate object, ceasing to contract in response to continue mass loss, and the system evolves back towards longer periods (the so--called `period bouncers'), with mean accretion rates $\dot{M} \lesssim 2 \times 10^{-11}\,\mathrm{M}_\odot\,\mathrm{yr}^{-1}$.

Since different period ranges are characterised by different mass transfer rates, $P_\mathrm{orb}$ (one of the most easily measured system parameter in CVs) provides a first rough estimate of the accretion rate and the evolutionary stage of the system. A more precise measurement of the mean accretion rate can be derived from the white dwarf effective temperature ($T_\mathrm{eff}$): its value is set by compressional heating of the accreted material \citep{Sion1995}, providing a constraint on the secular mean of the mass--transfer rate $\langle \dot{M} \rangle$, averaged over the thermal time--scale of the white dwarf envelope, ($10^3$ -- $10^5$ yr) and is one of the best available tests for the present models of CV evolution \citep{TowBild}.

While there are now over 1100 CVs with an orbital period determination \citep{RitterKolb}, comparatively little is known about their accreting white dwarfs: reliable\footnote{A temperature measurement is defined reliable when the white dwarf can be unambiguously detected either in the ultraviolet spectrum (from a broad Ly$\alpha$ absorption profile and, possibly, sharp absorption metal lines) or its ingress and egress can be seen in eclipse light curves.} $T_\mathrm{eff}$ measurements are available for only 43 CV white dwarfs \citep{Tow_and_Boris2009}, only 32 have an accurate mass determination \citep{Zorotovic} and only 11 have both. To improve our knowledge of CV evolution, it is essential to increase the number of objects with an accurate $T_\mathrm{eff}$ and mass measurement.

Since CV white dwarfs are relatively hot objects ($T_\mathrm{eff} \gtrsim 10\,000\,\mathrm{K}$) their spectral energy distribution peaks in the ultraviolet. At these wavelengths, the contamination from the accretion flow and the secondary star is often small or negligible compared to their contribution at optical wavelengths, and therefore space--based ultraviolet observations are necessary for a white dwarf $T_\mathrm{eff}$ determination (\citeauthor{Szkody2002} \citeyear{Szkody2002}, \citeauthor{Boris2006} \citeyear{Boris2006}, \citeauthor{BDPav} \citeyear{BDPav}, and many others). For this purpose, we carried out a large \textit{Hubble Space Telescope} (\textit{HST}) program in which we obtained high--resolution ultraviolet spectroscopy of 40 CVs with the \textit{Cosmic Origin Spectrograph} (COS). We complemented our sample with eight systems (three of which are in common with the COS sample) observed during two programs carried out with the \textit{Space Telescope Imaging Spectrograph} (STIS).

\begin{figure}
 \includegraphics[angle=-90,width=0.48\textwidth]{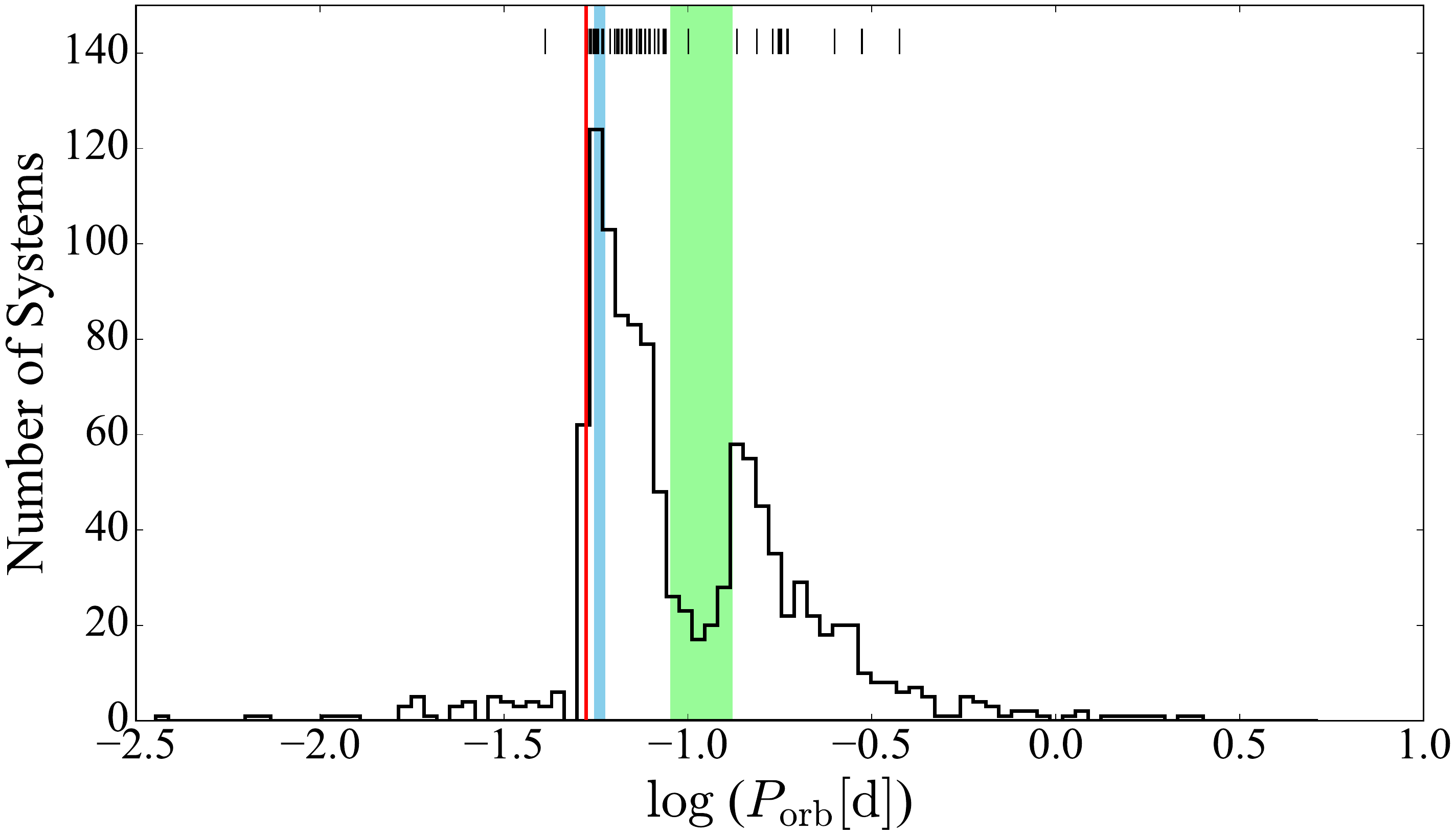}
 \caption{The orbital period distribution of 1144 semi--detached binaries containing a white dwarf and a Roche--lobe filling low--mass secondary (\citeauthor{RitterKolb} \citeyear{RitterKolb}, 7th Edition, Release 7.21, March 2014). The systems visible at short orbital periods ($P_\mathrm{orb} \lesssim 75\,$ min) are the AM CVn stars and CVs hosting an evolved donor. The green band highlights the period gap ($2.15\,\mathrm{h} \lesssim P_\mathrm{orb} \lesssim 3.18\,\mathrm{h}$, \citeauthor{Knigge_period_gap} \citeyear{Knigge_period_gap}), the blue box indicates the period spike ($80\,\mathrm{min} \lesssim P_\mathrm{orb} \lesssim 86\,\mathrm{min}$, \citeauthor{period_minimum} \citeyear{period_minimum}), the red line shows the period minimum ($P_\mathrm{min} = 76.2\,\mathrm{min}$, \citeauthor{Knigge_period_gap} \citeyear{Knigge_period_gap}), while the vertical lines along the top mark the orbital periods of the objects in our survey.}\label{orbitalperiod}
\end{figure}

The targets were selected to sample the entire orbital period range of the galactic CV population (Figure\,\ref{orbitalperiod}); four objects with previous $T_{\mathrm{eff}}$ measurements were included for comparison with our results (CU\,Vel, DV\,UMa, GW\,Lib and SDSS\,J103533.02+055158). In particular, we selected two eclipsing systems (DV\,UMa and SDSS\,J103533.02+055158) to compare our spectroscopic analysis with results obtained from modelling the eclipse light curve (e.g. \citeauthor{Feline} \citeyear{Feline}, \citeauthor{1035} \citeyear{1035}, \citeyear{Littlefair2008}), which is the most commonly used method for determining white dwarf $T_{\mathrm{eff}}$ from ground based observations. 

Here we present the \textit{HST} observations and the results of our spectral analysis, which almost doubles the number of objects with reliable $T_{\mathrm{eff}}$ measurements, providing $\langle \dot{M} \rangle$ and testing the present day CV population evolution models.

\begin{table*}
  \caption{Log of the \textit{HST}/COS observations. The systems are ordered by orbital period. The \textit{V}--band magnitudes listed here are the quiescent values reported from the literature and do not represent the brightness during the COS observations (see Figures~\ref{lightcurves_quiescence} and \ref{lightcurves}).}\label{table_Log_COS_obs}
  \begin{tabular}{@{}lcccccccc@{}}
  \toprule
  System                  	&	 $P_\mathrm{orb}$ 	&	 Type     	&	 \textit{V}    	&	 $E(B-V)$ 	&	 Observation  	&	    Number     	&	 Total             	&	 State  	\\
                          	&	 (min)            	&	          	&	(\textrm{mag}) 	&	 (\textrm{mag})   	&	     date     	&	      of       	&	 exposure          	&	         	\\
                          	&		&	          	&	               	&	                  	&	              	&	    orbits     	&	 time (\textrm{s}) 	&	        	\\
\midrule                          	
V485\,Cen                  	&	 59.03  	&	 SU UMa 	&	  17.7         	&	 $0.071^a$        	&	  2013 Mar 16 	&	 4             	&	  9907             	&	 \textit{q}  	\\
GW\,Lib                    	&	 76.78  	&	 WZ Sge 	&	  16.5         	&	 $0.03^b$         	&	  2013 May 30 	&	 3             	&	  7417             	&	 \textit{q}  	\\
SDSS\,J143544.02+233638.7  	&	 78.00  	&	 WZ Sge?	&	  18.2         	&	 $0.029^a$        	&	  2013 Mar 09 	&	 3             	&	  7123             	&	 \textit{q}  	\\
OT\,J213806.6+261957       	&	 78.10  	&	 WZ Sge 	&	  16.1         	&	 $0.063^a$        	&	  2013 Jul 25 	&	 2             	&	  4760             	&	 \textit{q} 	\\
SDSS\,J013701.06--091234.8 	&	 79.71  	&	 SU UMa 	&	  18.7         	&	 $0.024^c$        	&	  2013 Sep 13 	&	 4             	&	 10505             	&	 \textit{q}    	\\
SDSS\,J123813.73--033932.9 	&	 80.52  	&	 WZ Sge 	&	  17.8         	&	 $0.006^c$        	&	  2013 Mar 01 	&	 3             	&	  7183             	&	 \textit{q} 	\\
PU\,CMa                    	&	 81.63  	&	 SU UMa 	&	  16.2         	&	 $0.09^a$         	&	  2013 Mar 22 	&	 2             	&	  4757             	&	 \textit{eq}	\\
V1108\,Her                 	&	 81.87  	&	 WZ Sge 	&	  17.7         	&	 $0.025^c$        	&	  2013 Jun 06 	&	 3             	&	  7327             	&	 \textit{q}  	\\
ASAS\,J002511+1217.2       	&	 82.00  	&	 WZ Sge 	&	  17.5         	&	 $0.025^c$        	&	  2012 Nov 15 	&	 3             	&	  7183             	&	 \textit{q}  	\\
SDSS\,J103533.02+055158.4  	&	 82.22  	&	 WZ Sge?	&	  18.8         	&	 $0.009^c$        	&	  2013 Mar 08 	&	 5             	&	 12282             	&	 \textit{q} 	\\
CC\,Scl                    	&	 84.10  	&	 DN/IP  	&	  17.0         	&	 $0.013^a$        	&	  2013 Jun 29 	&	 2             	&	  4668             	&	 \textit{q} 	\\
SDSS\,J075507.70+143547.6  	&	 84.76  	&	 WZ Sge?	&	  18.2         	&	 $0.028^a$        	&	  2012 Dec 14 	&	 3             	&	  7183             	&	 \textit{q}  	\\
1RXS\,J105010.8--140431    	&	 88.56  	&	 WZ Sge 	&	  17.0         	&	 $0.018^c$        	&	  2013 May 10 	&	 3             	&	  7363             	&	 \textit{q} 	\\
MR\,UMa                    	&	 91.17  	&	 SU UMa 	&	  16.7         	&	 $0.02^a$         	&	  2013 Apr 04 	&	 2             	&	  4401             	&	 \textit{q}  	\\
QZ\,Lib                    	&	 92.36  	&	 WZ Sge 	&	  17.5         	&	 $0.054^c$        	&	  2013 Apr 26 	&	 3             	&	  7512             	&	 \textit{q}   	\\
SDSS\,J153817.35+512338.0  	&	 93.11  	&	 CV     	&	  18.6         	&	 $0.012^a$        	&	  2013 May 16 	&	 2             	&	  4704             	&	 \textit{q} 	\\
1RXS\,J023238.8--371812    	&	 95.04  	&	 WZ Sge 	&	  18.8         	&	 $0.027^a$        	&	  2012 Nov 01 	&	 5             	&	 12556             	&	 \textit{q}           	\\
SDSS\,J093249.57+472523.0  	&	 95.48  	&	 DN?    	&	  17.9         	&	 $0.014^a$        	&	  2013 Jan 11 	&	 2             	&	  4326             	&	 \textit{eq}	\\
BB\,Ari                    	&	 101.20 	&	 SU UMa 	&	  17.9         	&	 $0.105^a$        	&	  2013 Sep 27 	&	 2             	&	  4817             	&	 \textit{q}  	\\
DT\,Oct                    	&	 104.54 	&	 SU UMa 	&	  16.5         	&	 $0.145^a$        	&	  2013 May 20 	&	 2             	&	  4875             	&	 \textit{eq}	\\
IY\,UMa                    	&	 106.43 	&	 SU UMa 	&	  18.4         	&	 $0.012^c$        	&	  2013 Mar 30 	&	 2             	&	  4195             	&	 \textit{q}  	\\
SDSS\,J100515.38+191107.9  	&	 107.60 	&	 SU UMa 	&	  18.2         	&	 $0.025^a$        	&	  2013 Jan 31 	&	 3             	&	  7093             	&	 \textit{q}    	\\
RZ\,Leo                    	&	 110.17 	&	 WZ Sge 	&	  19.2        	&	 $0.029^c$        	&	  2013 Apr 11 	&	 4             	&	 10505             	&	 \textit{q} 	\\
CU\,Vel                    	&	 113.04 	&	 SU UMa 	&	  17.0         	&	 $<0.02^d$        	&	  2013 Jan 18 	&	 2             	&	  4634             	&	 \textit{q} 	\\
AX\,For                    	&	 113.04 	&	 SU UMa 	&	  18.5         	&	 $0.027^c$        	&	  2013 Jul 11 	&	 3             	&	  7483             	&	 \textit{q} 	\\
SDSS\,J164248.52+134751.4  	&	 113.60 	&	  DN    	&	  18.6         	&	 $0.063^a$        	&	  2012 Oct 12 	&	 3             	&	  7240             	&	 \textit{eq}	\\
QZ\,Ser                    	&	 119.75 	&	 SU UMa 	&	  17.9         	&	 $0.038^c$        	&	  2013 Jun 21 	&	 4             	&	 10505             	&	 \textit{q} 	\\
IR\,Com                    	&	 125.34 	&	 DN     	&	  18.0         	&	 $0.016^c$        	&	  2013 Feb 08 	&	 3             	&	  6866             	&	 \textit{q}     	\\
SDSS\,J001153.08--064739.2 	&	 144.40 	&	 U Gem  	&	  17.8         	&	 $0.029^a$        	&	  2012 Nov 09 	&	 5             	&	 12601             	&	 \textit{q}  	\\
OR\,And                    	&	 195.70 	&	 VY Scl 	&	  18.2         	&	 $0.158^a$        	&	  2013 Jul 10 	&	 3             	&	  7361             	&	 \textit{hs}	\\
BB\,Dor                    	&	 221.90 	&	 VY Scl 	&	  18.0         	&	 $0.03^a$         	&	  2013 Feb 13 	&	 3             	&	  7272             	&	 \textit{is}	\\
SDSS\,J040714.78--064425.1 	&	 245.04 	&	 U Gem  	&	  17.8         	&	 $0.08^a$         	&	  2013 Jan 24 	&	 2             	&	  4315             	&	 \textit{q} 	\\
CW\,Mon                    	&	 254.30 	&	 U Gem  	&	  16.8         	&	 $0.044^c$        	&	  2012 Nov 30 	&	 2             	&	  4774             	&	 \textit{eq} 	\\
V405\,Peg                  	&	 255.81 	&	  CV    	&	  18.2         	&	 $0.064^c$        	&	  2012 Dec 07 	&	 3             	&	  7303             	&	 \textit{eq}         	\\
HS\,2214+2845              	&	 258.02 	&	 U Gem  	&	  16.8         	&	 $0.078^a$        	&	  2013 Jun 18 	&	 2             	&	  4680             	&	 \textit{q} 	\\
BD\,Pav                    	&	 258.19 	&	 U Gem  	&	  15.4         	&	 $0.0^b$          	&	  2013 Jun 14 	&	 3             	&	  7375             	&	 \textit{q?}	\\
SDSS\,J100658.41+233724.4  	&	 267.71 	&	 U Gem  	&	  18.3         	&	 $0.018^c$        	&	  2014 Mar 20 	&	 5             	&	 12494             	&	 \textit{q} 	\\
HM\,Leo                    	&	 268.99 	&	 U Gem  	&	  17.5         	&	 $0.021^c$        	&	  2013 Feb 22 	&	 4             	&	  9979             	&	 \textit{eq}	\\
SDSS\,J154453.60+255348.8	&	 361.84	&	 U Gem? 	&	  16.9         	&	 $0.032^c$        	&	  2013 Apr 22 	&	    --             	&	    --             	&	 \textit{?}      	\\
HS\,0218+3229              	&	 428.02 	&	 U Gem  	&	  15.5         	&	 $0.05^c$         	&	  2012 Dec 22 	&	 3             	&	  7159             	&	 \textit{q}  	\\
HS\,1055+0939              	&	 541.88 	&	 U Gem  	&	  15.5         	&	 $0.025^a$        	&	  2013 Apr 15 	&	 2             	&	  4688             	&	 \textit{eq}	\\

\bottomrule
\end{tabular}
\begin{tablenotes}
\item \textbf{Notes.} The $E(B-V)$ data have been retrieved from: (\textit{a}) the NASA/IPAC Extragalactic Database (NED); (\textit{b}) GW\,Lib: \citet{EBV_GWLib}, BD\,Pav: \citet{BDPav}; (\textit{c}) the three--dimensional map of interstellar dust reddening based on Pan--STARRS\,1 and 2MASS photometry \citep{panstar}; (\textit{d}) this work. The NED data are the galactic colour excess and represent an upper limit for the actual reddening while the Pan--STARRS data are the actual reddening for the systems with a known distance. The last column reports the state of the system during the \textit{HST} observations: \textit{eq}, early quiescence, i.e. observed close to an outburst; \textit{q}, quiescence; \textit{hs}, high state; \textit{is}, intermediate state.
\end{tablenotes}
\end{table*}

\begin{figure*}
 \subfloat{%
  \includegraphics[angle=-90,width=0.48\textwidth]{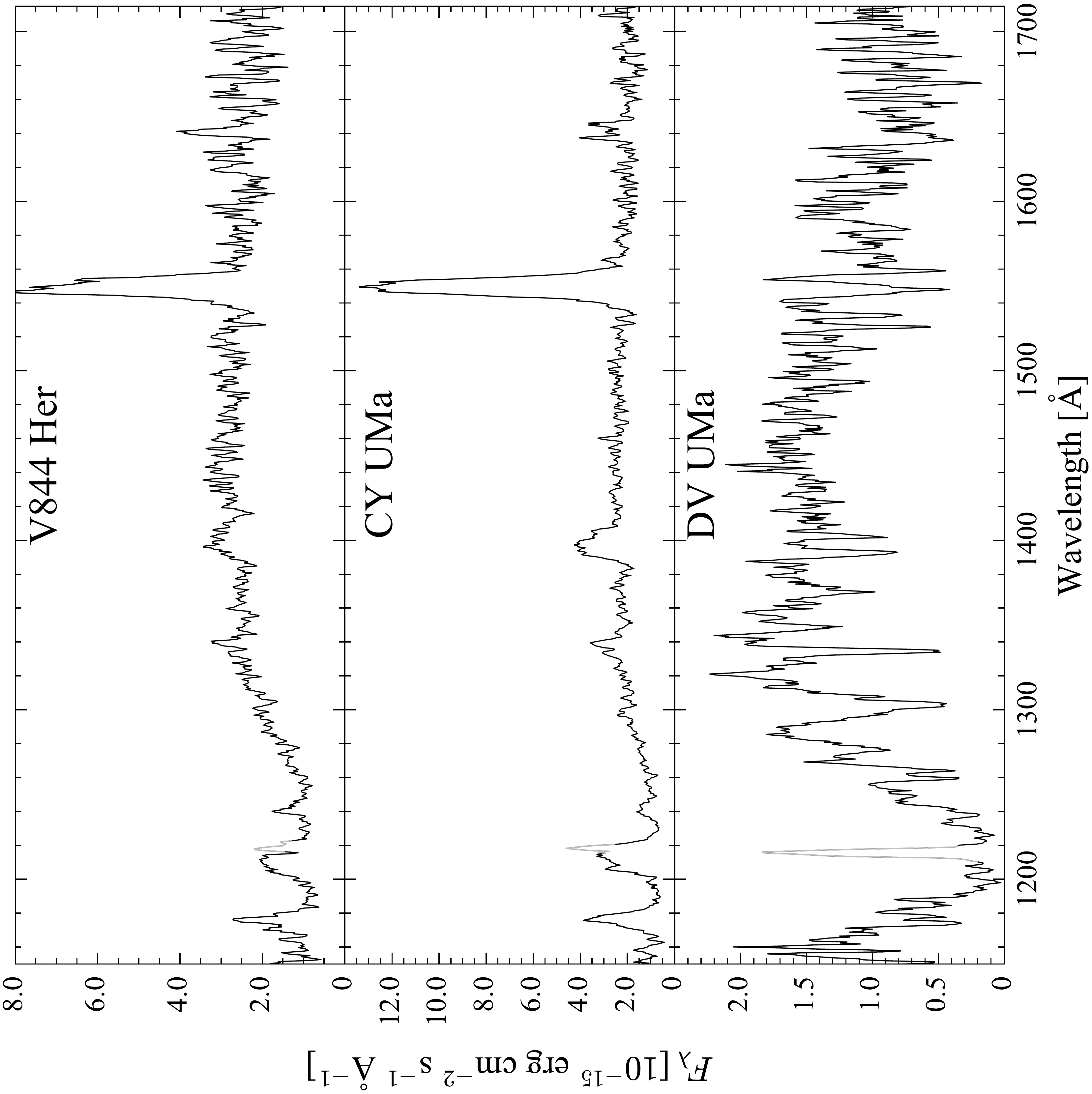} \qquad %
  }
 \subfloat{%
  \includegraphics[angle=-90,width=0.48\textwidth]{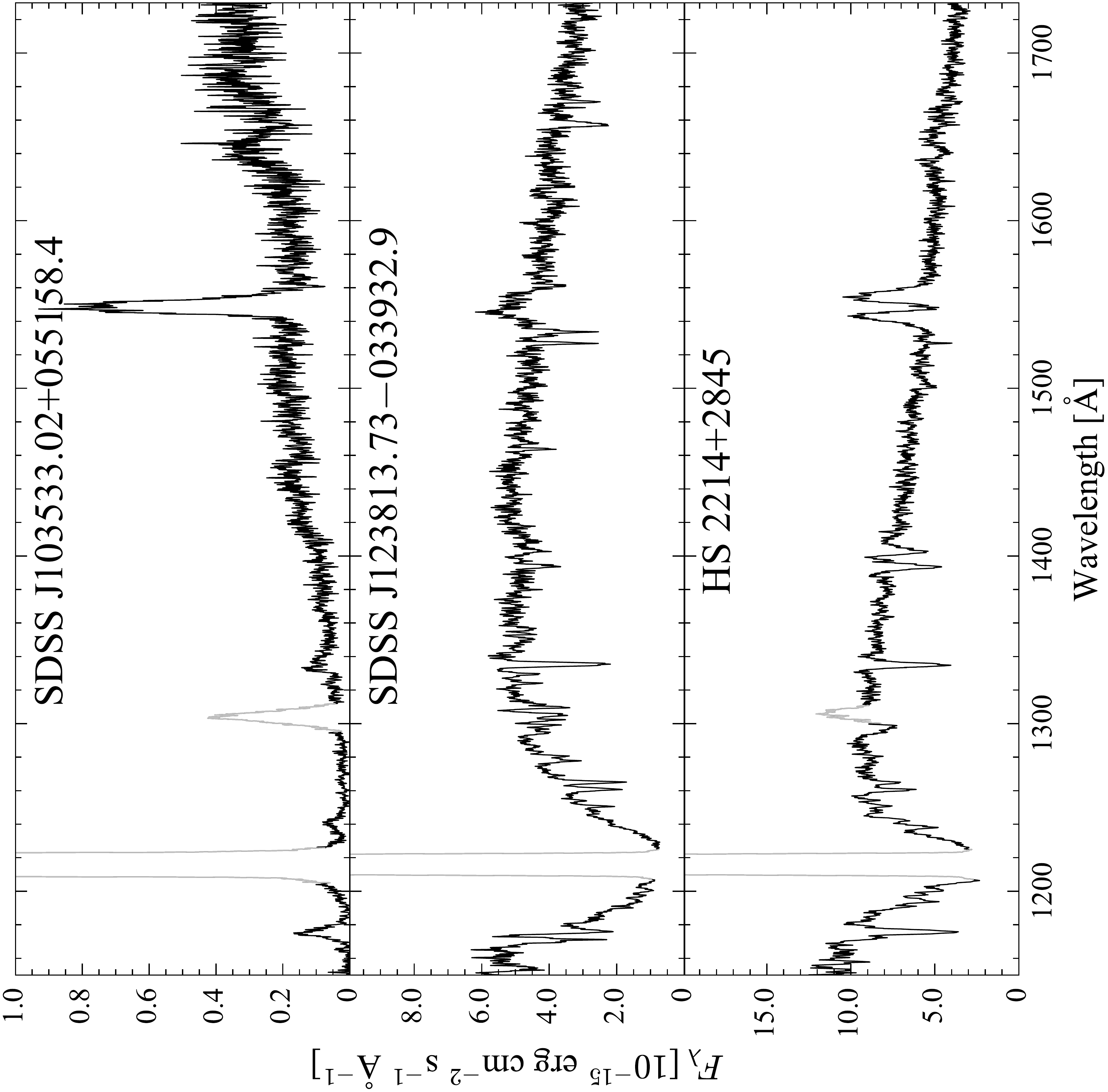}
  }%
 \caption{\textit{HST}/STIS spectra (left) and \textit{HST}/COS spectra (right) for six objects in which the white dwarf dominates the far--ultraviolet. The geocoronal emission lines of Ly$\alpha$ ($1216\,$\AA) and \hbox{O\,{\sc i}} ($1302\,$\AA) are plotted in grey.}\label{figure_spectra}  
 \end{figure*}

\begin{figure*}
 \subfloat{%
  \includegraphics[angle=-90,width=0.48\textwidth]{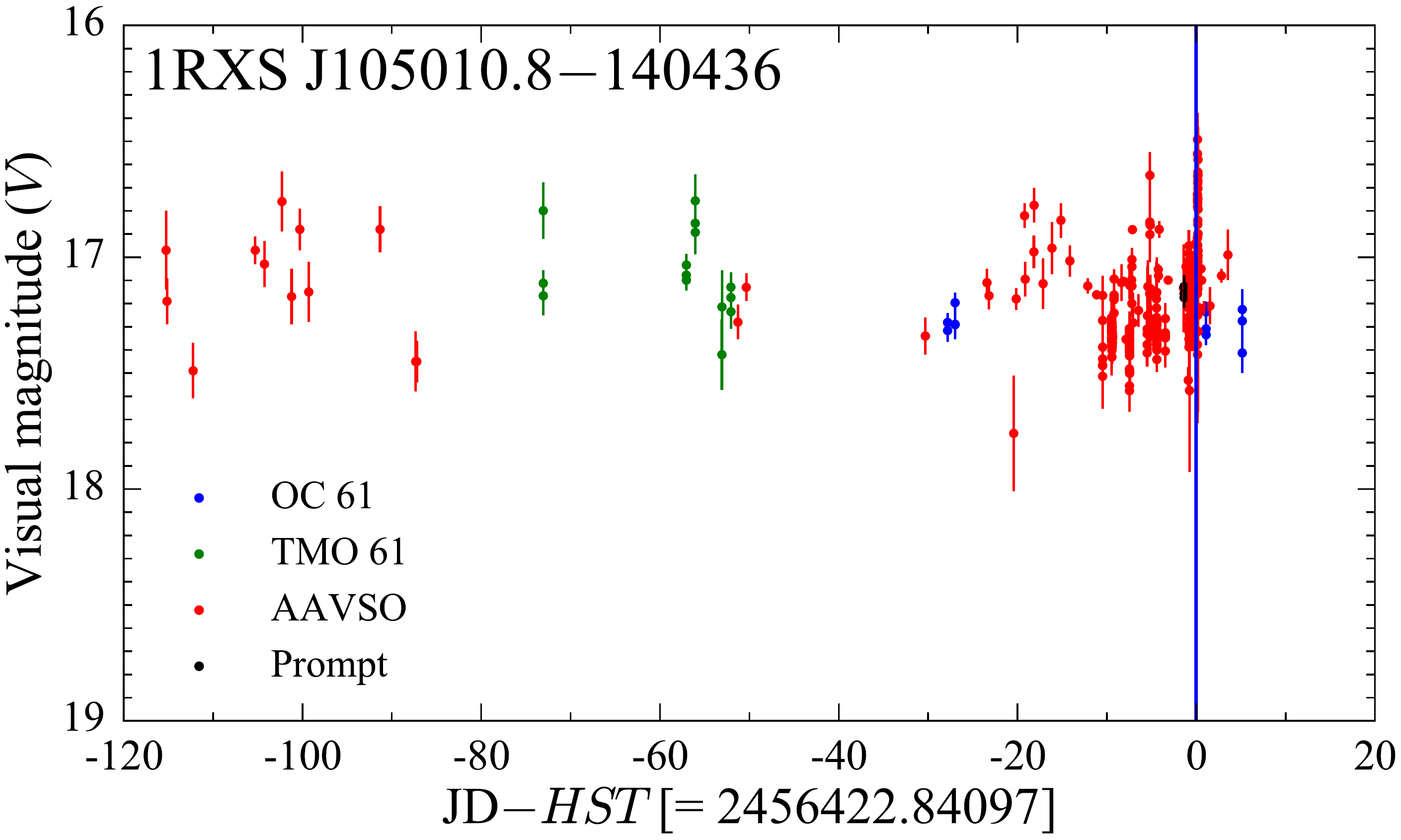} \qquad %
  }
 \subfloat{%
  \includegraphics[angle=-90,width=0.48\textwidth]{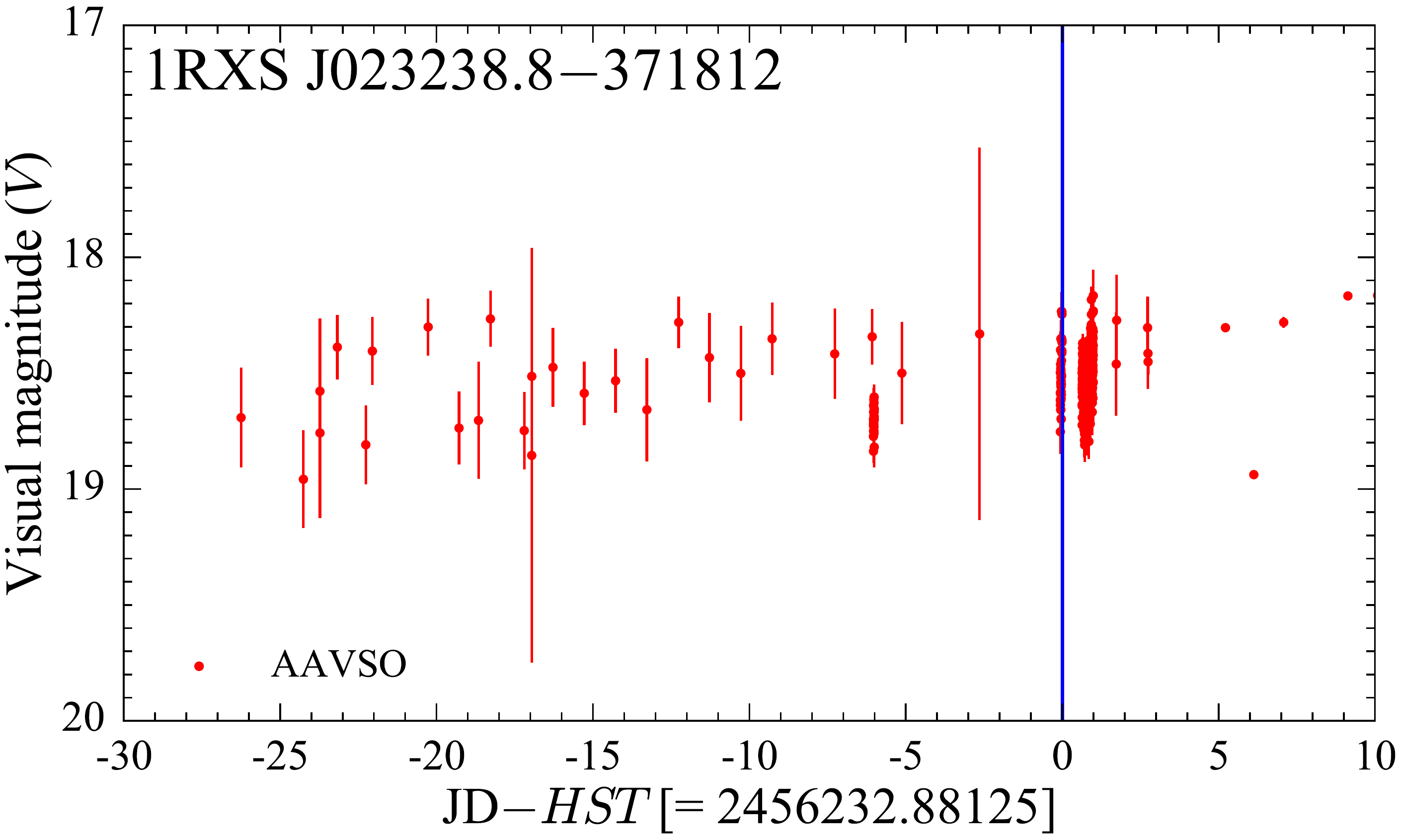}
  }\\%
 \subfloat{%
  \includegraphics[angle=-90,width=0.48\textwidth]{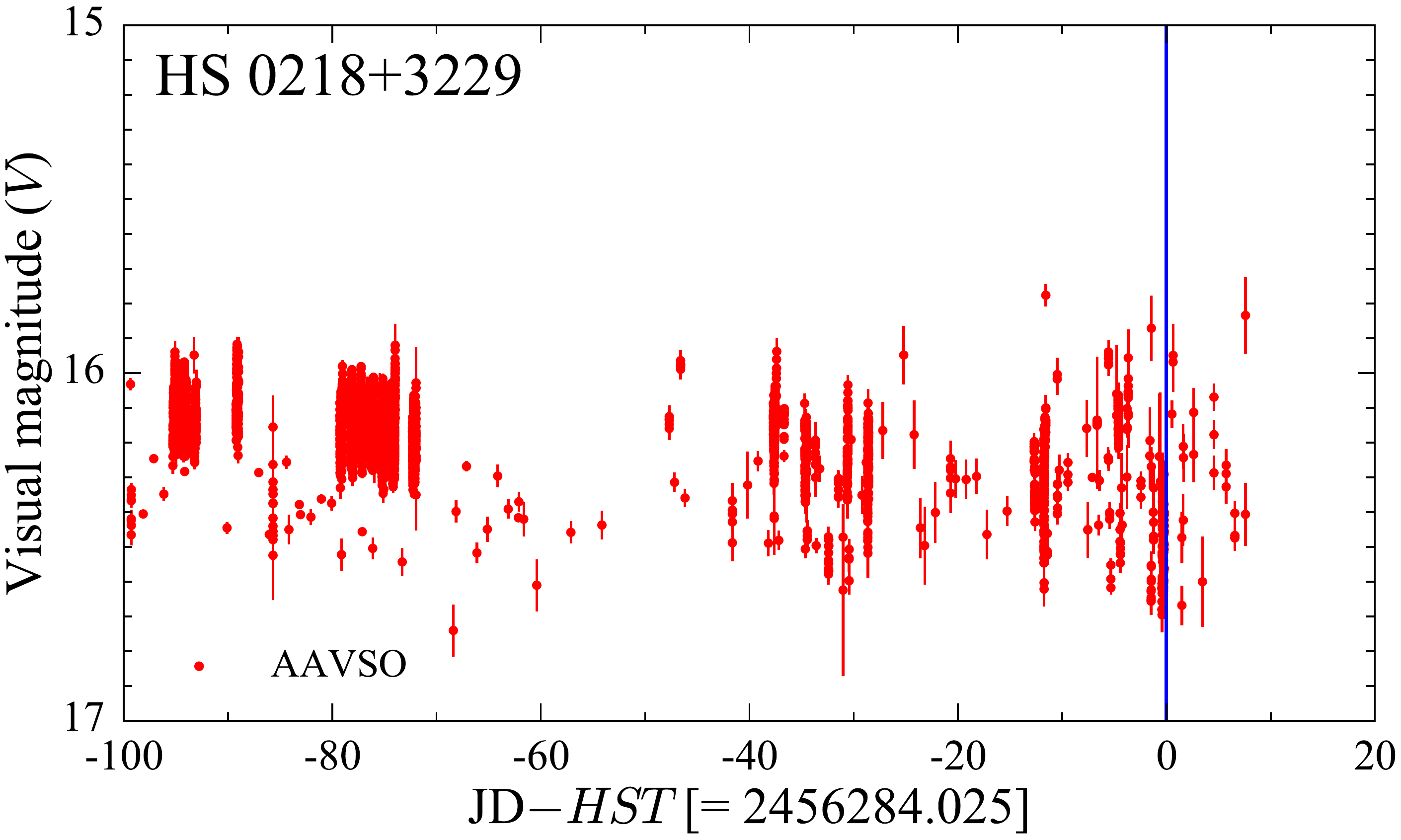} \qquad %
  }
 \subfloat{%
  \includegraphics[angle=-90,width=0.48\textwidth]{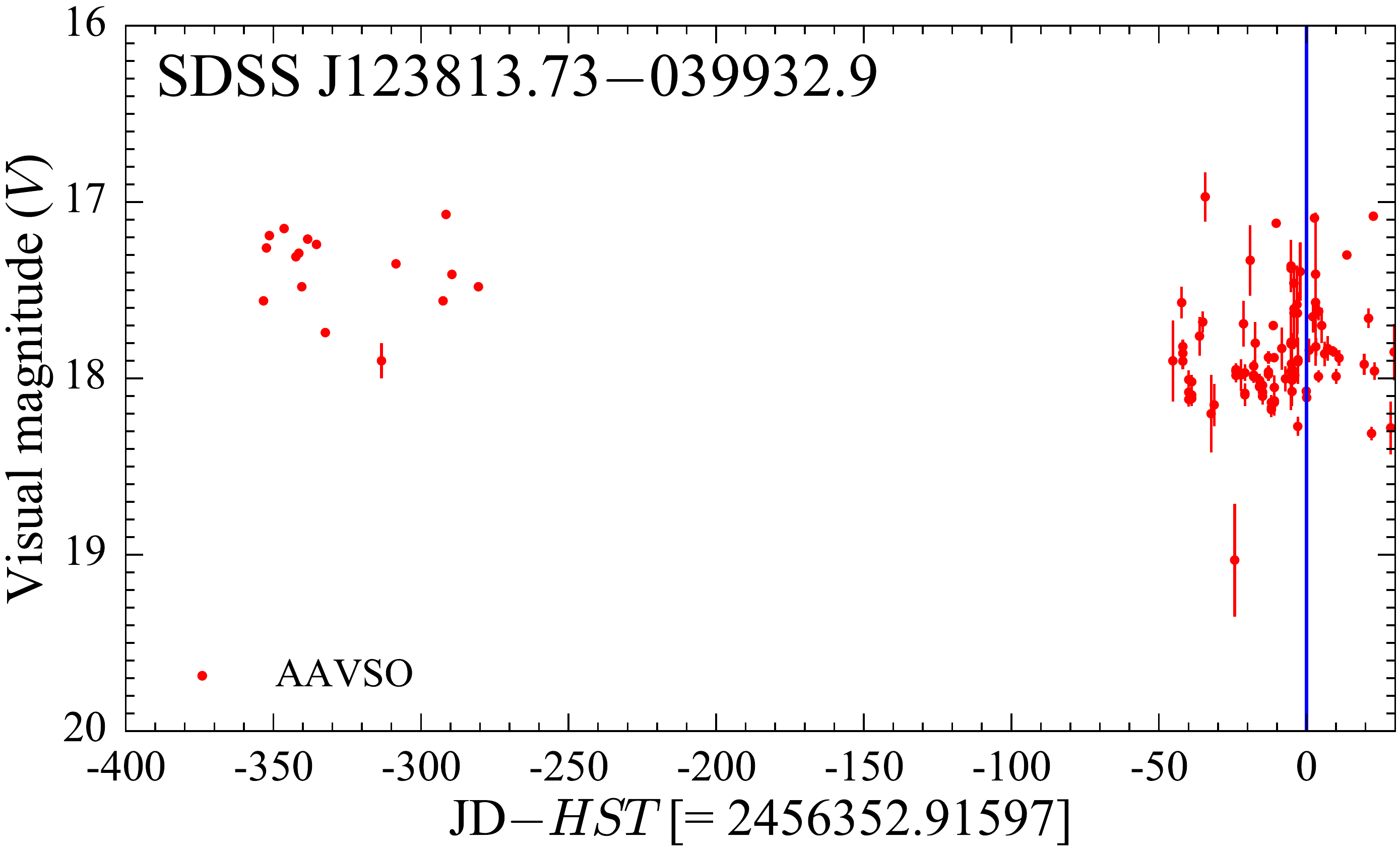}
  }
\caption{Sample light curves for four targets in our sample which have been observed in quiescence. The blue line represents the date of the \textit{HST} observation, which has been subtracted from the Julian date on the x--axis. Note the different time range on the x--axis. }%
\label{lightcurves_quiescence}
\end{figure*}

\begin{figure*}
 \subfloat{%
  \includegraphics[angle=-90,width=0.48\textwidth]{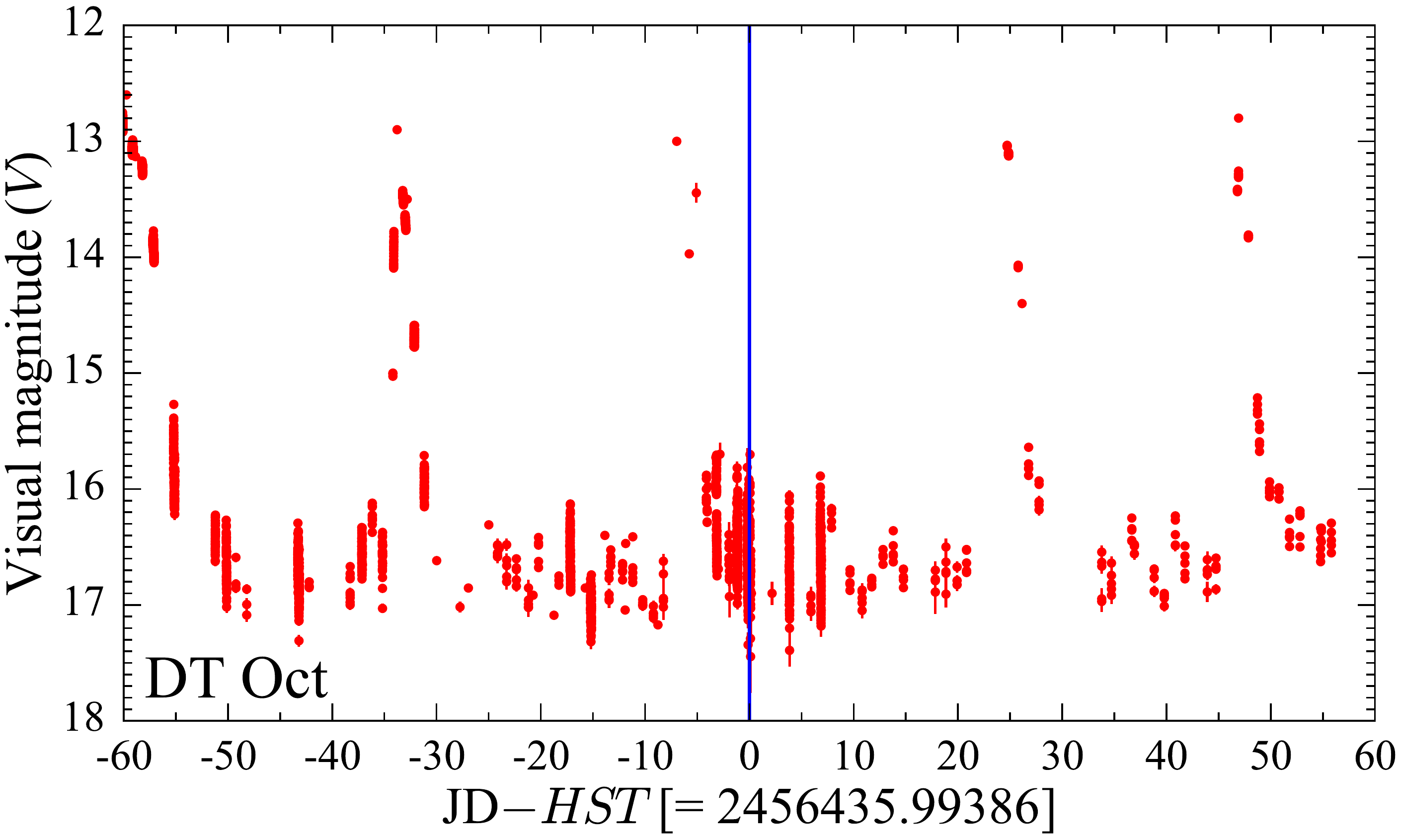} \qquad %
  }
 \subfloat{%
  \includegraphics[angle=-90,width=0.48\textwidth]{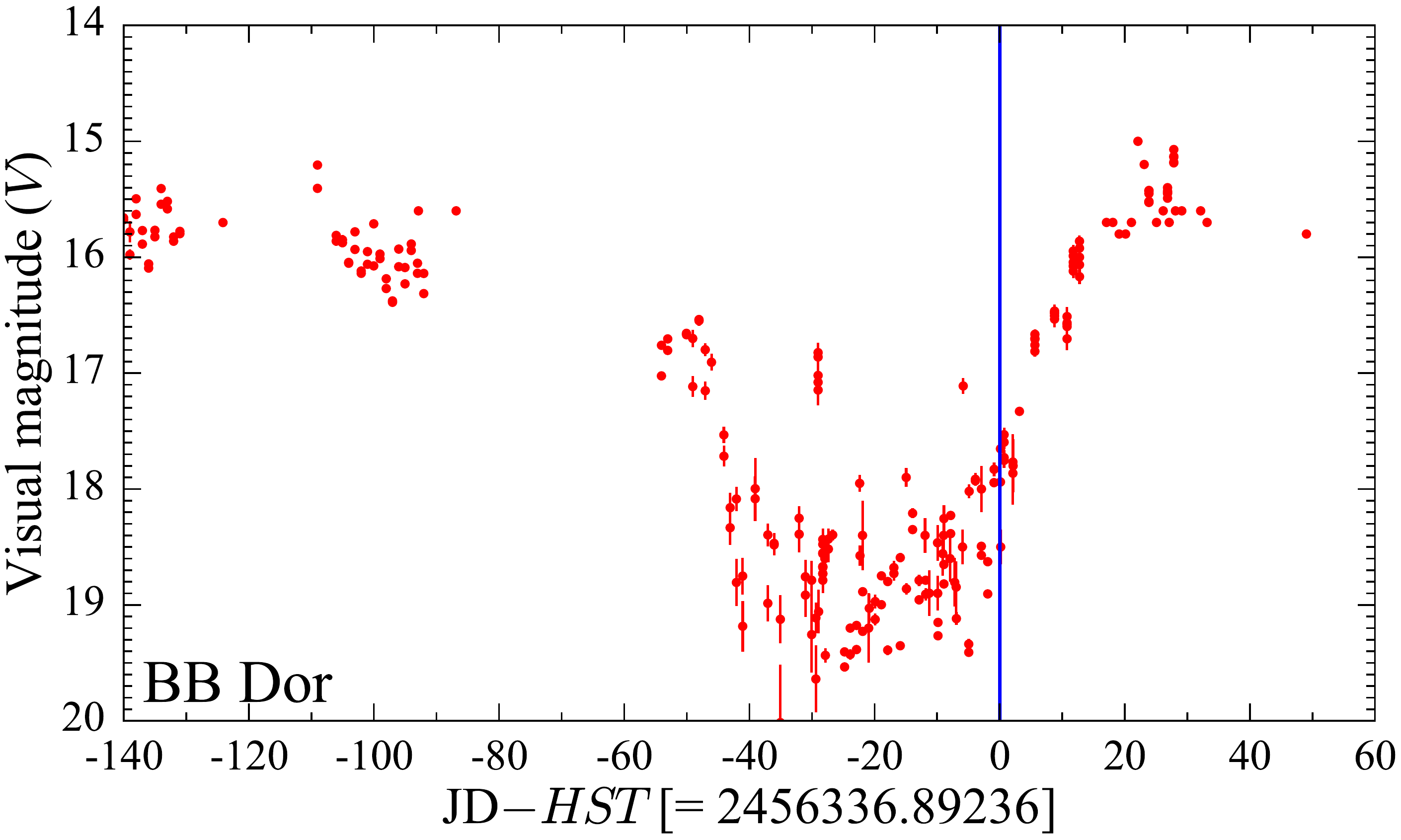}
  }\\%
 \subfloat{%
  \includegraphics[angle=-90,width=0.48\textwidth]{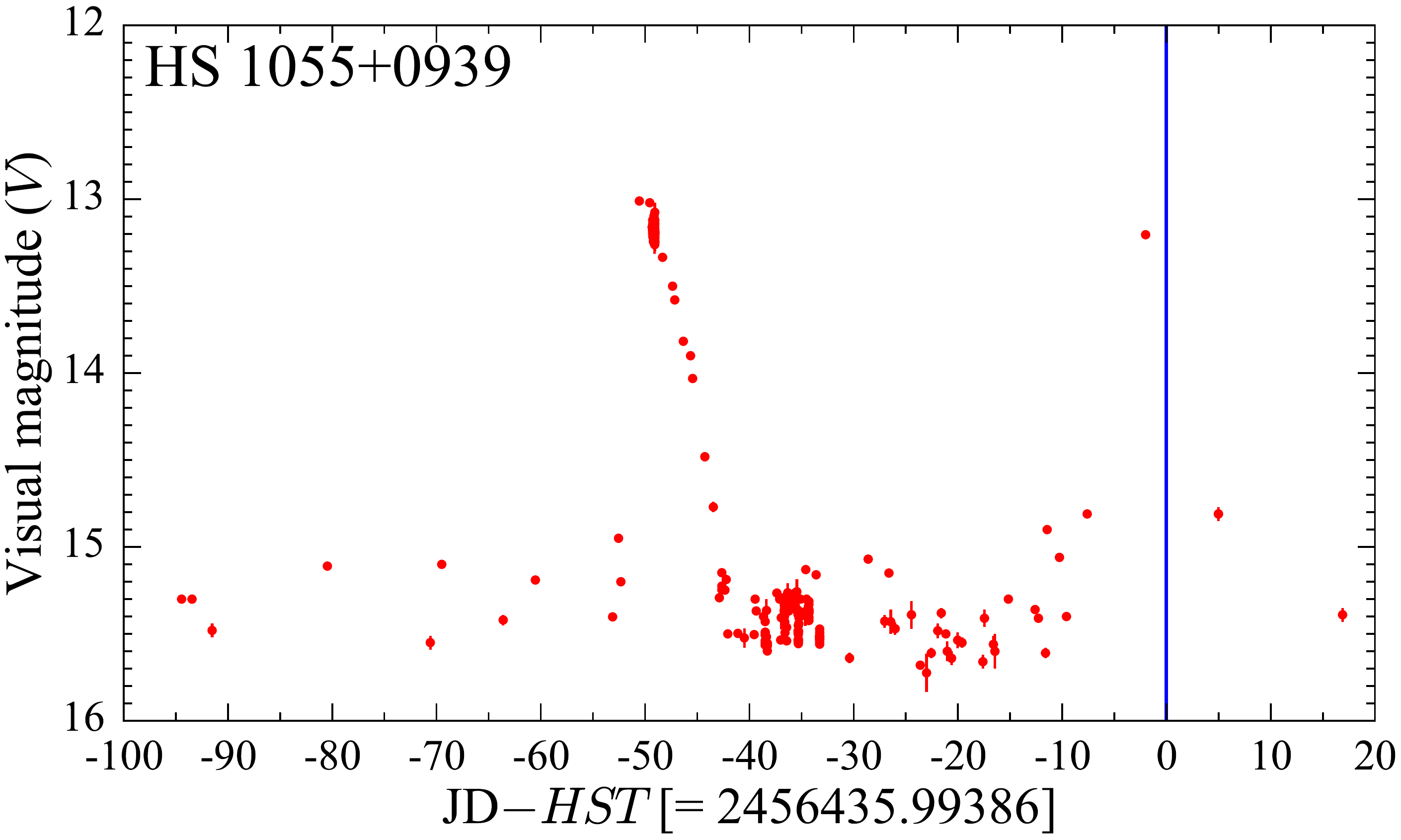} \qquad %
  }
 \subfloat{%
  \includegraphics[angle=-90,width=0.48\textwidth]{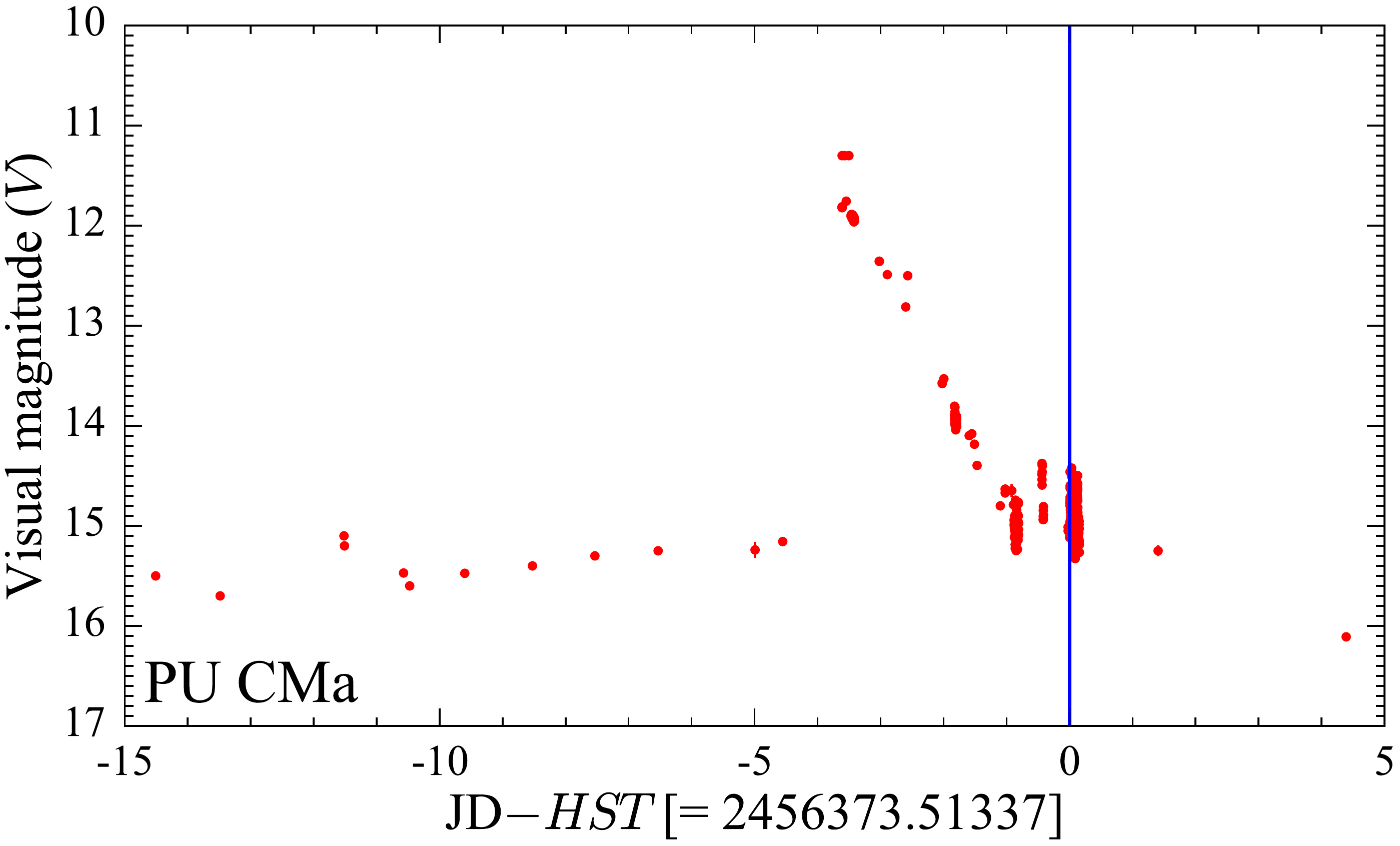}
  }
\caption{Sample light curves for four targets in our sample which have been observed close to an outburst or in an intermediate state. The blue line represents the date of the \textit{HST} observation, which has been subtracted from the Julian date on the x--axis. Note the different time range on the x--axis. The data have been retrieved from the AAVSO database.}%
\label{lightcurves}
\end{figure*}

\section{Observations}\label{sec:Obs}
\subsection{COS observations}\label{subsec:COS_Obs}
The \textit{Cosmic Origin Spectrograph} (COS) data were collected in 122 \textit{HST} orbits from October 2012 to March 2014 (program ID 12870, Table~\ref{table_Log_COS_obs}). Each CV was observed using the Primary Science Aperture (PSA) for two to five consecutive spacecraft orbits. 
The data from all the orbits were summed to produce an average ultraviolet spectrum for each object. We used the G140L grating, which has a nominal resolution of $R\,\simeq 3000$, at the central wavelength of $1105\,$\AA$\,$ and the far--ultraviolet (FUV) channel, covering the wavelength range $1105-2253\,$\AA. The detector sensitivity quickly decreases in the red portion of the spectrum, reducing the useful range to $\simeq\,1150-1730\,$\AA.

The COS FUV detector consists of a photon--counting micro--channel plate which converts the incoming photons into electronic pulses. An excessive photon flux could result in permanent damages, and even the loss, of the detector. This is very important in the framework of CV observations since these objects are characterised by periods of quiescence, in which the accretion onto the white dwarf is greatly reduced, interrupted by bright outbursts (thermal instabilities in the disc which cause a variation in the mass transfer rate through the disc, \citeauthor{Osaki1974} \citeyear{Osaki1974}, \citeauthor{outburst_theory} \citeyear{outburst_theory}, \citeauthor{Meyer_Meyer} \citeyear{Meyer_Meyer}). During an outburst, CVs typically brighten by 2--5 magnitudes, occasionally up to 9 magnitudes (\citeauthor{Warner} \citeyear{Warner}, \citeauthor{GWLib_9a} \citeyear{GWLib_9a}, \citeauthor{GWLib_9b} \citeyear{GWLib_9b}). This increase in luminosity occurs rapidly, on the time--scale of about a day, first at optical wavelengths and hours to a day later in the ultraviolet (\citeauthor{UV_delay1} \citeyear{UV_delay1}, \citeauthor{UV_delay2} \citeyear{UV_delay2}, \citeauthor{UV_delay3} \citeyear{UV_delay3}). Therefore CVs can easily reach, and exceed, the COS detector safety threshold. The outburst recurrence time ranges from weeks to decades and these events are unpredictable. To avoid damage to the detectors an intensive monitoring of each target was required before the observations. This monitoring program was carried out by the global amateur community and some additional small telescopes (AAVSO, Prompt and many others), and only their outstanding support has made this \textit{HST} survey possible.

Among our 41 COS targets, only SDSS\,J154453.60+255348.8 could not be observed since it showed strong variations in its optical brightness in the days before the \textit{HST} observation.

After an outburst, the white dwarf does not cool immediately to its quiescent temperature since it has been heated by the increased infall of material. The time required to cool back to the quiescent temperature has been modelled and is related to the outburst amplitude and duration (\citeauthor{Sion1995} \citeyear{Sion1995}, \citeauthor{Dean_Lars} \citeyear{Dean_Lars}, \citeauthor{cooling} \citeyear{cooling}). It has been observed that it can vary from days or weeks up to several years (\citeauthor{Knox_cooling} \citeyear{Knox_cooling}, \citeauthor{Boris1996} \citeyear{Boris1996}, \citeauthor{Slevinsky1999} \citeyear{Slevinsky1999}, \citeauthor{cooling} \citeyear{cooling}, see \citeauthor{Paula_GWLib2016} \citeyear{Paula_GWLib2016} for an extreme case). Therefore the effective temperature measured in a system in which an outburst has recently occurred provides only an upper limit for its quiescent effective temperature and, consequently, for $\langle \dot{M} \rangle$.

\begin{figure}
 \includegraphics[angle=-90,width=0.47\textwidth]{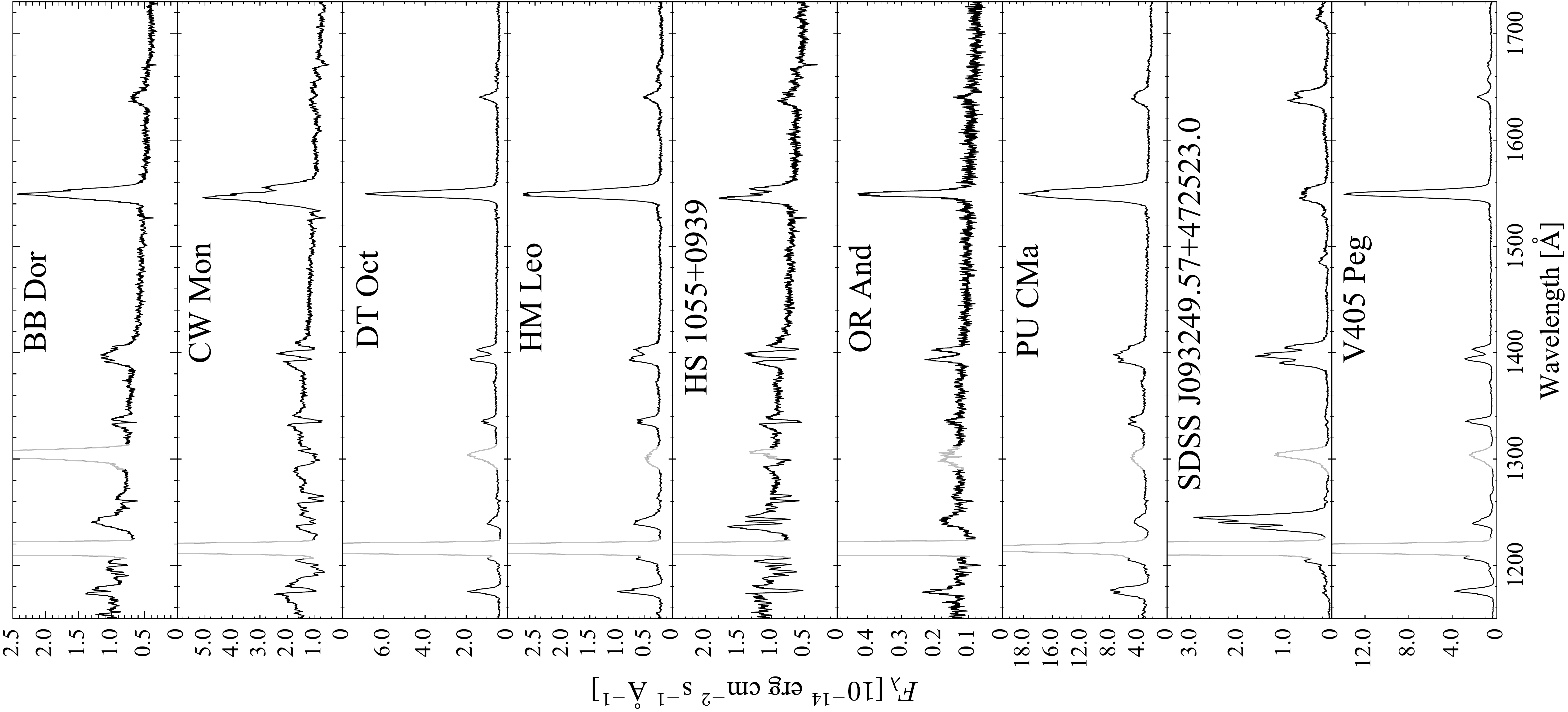}
 \caption{\textit{HST}/COS spectra of the nine CVs dominated by emission from the accretion flow, outshining the white dwarf. Strong emission lines of \ion{C}{iii} ($1175\,$\AA), \ion{N}{v} ($1242\,$\AA), \ion{C}{ii} ($1335\,$\AA), \ion{Si}{iv} ($1400\,$\AA), \ion{C}{iv} ($1550\,$\AA) and \ion{He}{ii} ($1640\,$\AA) are clearly visible. The wavelength ranges affected by the geocoronal emission lines of Ly$\alpha$ ($1216\,$\AA) and \ion{O}{i} ($1302\,$\AA) are plotted in grey.}\label{figure_disc} 
\end{figure}

The right panels of Figure~\ref{figure_spectra} show three sample COS ultraviolet spectra of quiescent CVs in which the white dwarf dominates the emission, as seen from the broad Ly$\alpha$ absorption centred on $1216\,$\AA. The shape of the Ly$\alpha$ line changes with $T_{\mathrm{eff}}$, becoming more defined and narrower in the hotter white dwarfs, while the continuum slope of the spectrum becomes steeper. 

While the majority of the CVs in our sample have been observed in quiescence (Figure~\ref{lightcurves_quiescence}), obvious signatures of the white dwarf were missing in the spectra of nine systems: seven of them experienced an outburst shortly before the \textit{HST} observation while the remaining two systems are VY\,Scl stars (Figure~\ref{lightcurves}). This class of CVs is characterised by high mean accretion rates which keep the disc usually in a stable hot state, equivalent to a dwarf nova in permanent outburst. In this high state, the disc dominates the spectral appearance even in the far ultraviolet. However, occasionally the accretion rate drops (low state) and unveils the white dwarf (e.g. \citeauthor{TTArietis} \citeyear{TTArietis}, \citeauthor{DWUMa} \citeyear{DWUMa}, \citeauthor{MVLyrae} \citeyear{MVLyrae}). The VY\,Scl stars in our sample (OR\,And and BB\,Dor) were caught in the high state and intermediate state respectively, preventing a detection of the white dwarf. For the purpose of completeness, the spectra of these nine objects are shown in Figure \ref{figure_disc}, but a detailed analysis of these systems is outside the scope of this paper and will be presented elsewhere (e.g. \citeauthor{patrick_bbdor} \citeyear{patrick_bbdor}). In Section~\ref{sec:DataAnalysis}, we analyse the 31 systems in which the white dwarf signature is easily recognisable from a broad Ly$\alpha$ absorption profile.

\begin{table*}
  \caption{Log of the \textit{HST}/STIS observations. The systems are ordered by orbital period. The fourth column reports the instrumental magnitudes measured from the acquisition images, which were obtained through the $F28\,\times\,50LP$ filter.}\label{table_Log_STIS_obs}
  \begin{tabular}{@{}lcccccccc@{}}
  \toprule
    System         	&	 $P_\mathrm{orb}$ 	&	 Type      	&	 \textit{$F28\,\times\,50LP$} 	&	 $E(B-V)$ 	&	 Observation 	&	 Number  	&	 Total             	&	 State 	\\
                   	&	 (min)            	&	           	&	(\textrm{mag})                	&	 (\textrm{mag})   	&	     date    	&	  of     	&	 exposure          	&	     	\\
                   	&		&	           	&	                              	&	                  	&	             	&	 orbits  	&	 time (\textrm{s}) 	&	 	\\
\midrule                   	
V485\,Cen          	&	 59.03  	&	 SU UMa         	&	  17.9                   	&	 $0.071^a$        	&	 2004 Apr 15 	&	  1      	&	 900               	&	 \textit{q} 	\\
V844\,Her          	&	 78.69  	&	 SU UMa         	&	  17.4                   	&	 $0.013^c$        	&	 2003 Feb 23 	&	  1      	&	 900               	&	 \textit{q} 	\\
UV\,Per            	&	 93.44  	&	 SU UMa         	&	  17.9                   	&	 $0.0^b$          	&	 2002 Oct 11 	&	  1      	&	 900               	&	 \textit{q}	\\
RZ\,Sge            	&	 98.32  	&	 SU UMa         	&	  18.0                   	&	 $0.302^a$        	&	 2003 Jun 13 	&	  1      	&	 900               	&	 \textit{q}	\\
CY\,UMa            	&	 100.18 	&	 SU UMa         	&	  17.3                   	&	 $0.012^a$        	&	 2002 Dec 27 	&	  1      	&	 830               	&	 \textit{q}	\\
QZ\,Ser            	&	 119.75 	&	 SU UMa         	&	  16.9                   	&	 $0.038^c$        	&	 2003 Oct 07 	&	  1      	&	 900               	&	 \textit{q}	\\
DV\,UMa            	&	 123.62 	&	 SU UMa         	&	  18.6                   	&	 $0.014^c$        	&	 2004 Feb 08 	&	  1      	&	 900               	&	 \textit{q}	\\
BD\,Pav            	&	 258.19 	&	  U Gem         	&	  14.6                   	&	 $0.0^b$          	&	 2003 Feb 09 	&	  1      	&	 600               	&	 \textit{q}	\\

\bottomrule
\end{tabular}
\begin{tablenotes}
\item \textbf{Notes.} The $E(B-V)$ data have been retrieved from: (\textit{a}) the NASA/IPAC Extragalactic Database (NED); (\textit{b}) UV\,Per: \citet{Paula_UVPer}, BD\,Pav: \citet{BDPav}; (\textit{c}) the three--dimensional map of interstellar dust reddening based on Pan--STARRS\,1 and 2MASS photometry \citep{panstar}. The NED data are the galactic colour excess and represent an upper limit for the actual reddening while the Pan--STARRS data are the actual reddening for the systems with a known distance. The last column reports the state of the system during the  \textit{HST} observations: \textit{q}, quiescence.
\end{tablenotes}
\end{table*}

\subsection{STIS observations}\label{subsec:STIS_Obs}
The \textit{Space Telescope Imaging Spectrograph} (STIS) data were collected as part of two snapshot programs (programs ID 9357 and 9724) in Cycle 11 and 12 (Table~\ref{table_Log_STIS_obs}). Snapshot programs are designed to fill short gaps in the weekly \textit{HST} observing schedule, therefore each CV was observed with exposure times of only 600 to 900 seconds, through the $52"\,\times\,0.2"$ aperture. We used the G140L grating at the central wavelength $1425\,$\AA\, and the FUV--MAMA detector to obtain spectra covering the wavelength range $1150-1700\,$\AA\, at a nominal resolution of $R\,\simeq\,1000$. 

Similar to the COS--FUV detector, the MAMA detector is also subject to damage by excessive illumination but, contrary to the COS program, the snapshot programs did not have a long term schedule which could allow detailed monitoring before the STIS observations. We analysed the STIS acquisition images to determine the source brightness immediately before the spectrum acquisition. These images were acquired with the $F28\,\times\,50LP$ filter (central wavelength of $7228.5\,$\AA\, and full width at half--maximum of $2721.6\,$\AA). We determined the corresponding magnitude following the method described in \citet{STIS_filter}. By comparison with published magnitudes and the system brightness in the Sloan Digital Sky Survey (SDSS) and AAVSO Photometric All--Sky Survey (APASS), we verified that all the targets were observed in quiescence.

These snapshot programs produced ultraviolet spectra of 69 CVs (e.g. \citeauthor{BorisSTIS} \citeyear{BorisSTIS}, \citeauthor{HS2231} \citeyear{HS2231}, \citeauthor{PabloSTIS} \citeyear{PabloSTIS}). Here we analyse the remaining eight objects in which the white dwarf dominates the ultraviolet emission. We show three sample STIS spectra of quiescent CVs in the left panel of Figure~\ref{figure_spectra} where the broad Ly$\alpha$ absorption centred on $1216\,$\AA\, reveals the white dwarf photosphere.

Three of the eight systems observed by STIS (BD\,Pav, QZ\,Ser and V485\,Cen) are in common with the COS sample and we present a detailed comparison of the two datasets in Section~\ref{with_COS_STIS1}.

Adding the five additional CVs observed with STIS to the 31 from the COS survey results in our total sample of 36 CVs in which the white dwarf dominates the overall spectral emission.

\section{Data analysis}\label{sec:DataAnalysis} 
Ultraviolet spectra are particularly suitable for the study of CV white dwarfs because, at these wavelengths, the spectral energy distribution is often dominated by the white dwarf while the donor star does not contribute at all. However, some contamination from the disc, the bright spot (the region where the ballistic stream from the companion intersects the disc) or the boundary layer (the interface between the accretion disc and the white dwarf) can dilute the emission from the white dwarf photosphere. As described in Section~\ref{subsec:COS_Obs}, the overall shape of a white dwarf ultraviolet spectrum is determined by its effective temperature: hotter white dwarfs have narrow Ly$\alpha$ profiles and steep blue continua, while cooler white dwarfs are characterised by broad Ly$\alpha$ profiles and flat continua. Therefore the white dwarf effective temperature can be accurately measured by fitting the \textit{HST} data with synthetic white dwarf atmosphere models.

\begin{figure*}
 \includegraphics[angle=180,width=0.8\textwidth]{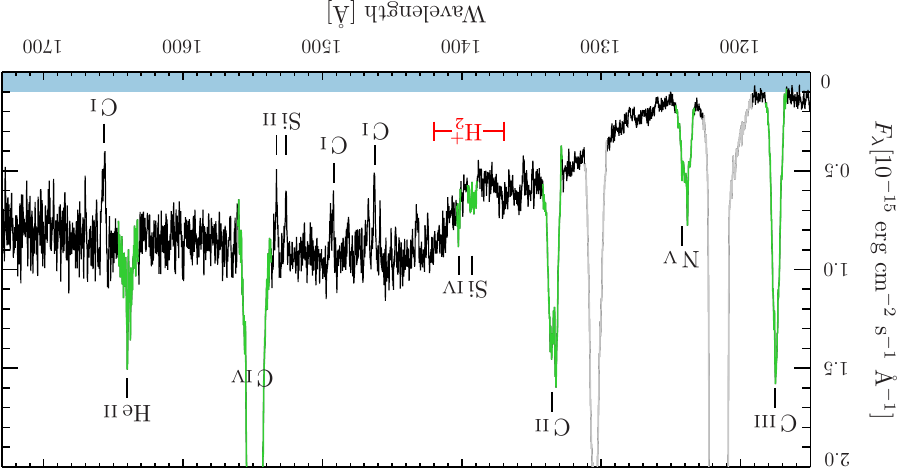}
 \caption{\textit{HST}/COS spectrum of 1RXS J023238.8--371812. Plotted in green are the emission lines of \ion{C}{iii} ($1175\,$\AA), \ion{N}{v} ($1242\,$\AA), \ion{C}{ii} ($1335\,$\AA), \ion{Si}{iv} ($1400\,$\AA), \ion{C}{iv} ($1550\,$\AA) and \ion{He}{ii} ($1640\,$\AA). The red band underlines the position of the H$_2^{+}$ quasi--molecular absorption bands. The blue band highlights the presence of a second, flat continuum component which, in this case, contributes $\simeq 10$ per cent of the observed flux. The geocoronal emission lines of Ly$\alpha$ ($1216\,$\AA) and \ion{O}{i} ($1302\,$\AA) are plotted in grey.}\label{esempio} 
\end{figure*}

In Figure~\ref{esempio}, we show the COS data of 1RXS\,J023238.8--371812 as an example of a typical CV spectrum from our sample. The broad Ly$\alpha$ absorption at $\simeq 1216\,$\AA, the quasi--molecular absorption bands of H$_2^{+}$ at $\simeq 1400\,$\AA\, and several sharp absorption lines (e.g. \ion{C}{i} $1460$/$1490$\,\AA, \ion{Si}{ii} $1526$/$1533$\,\AA, \ion{C}{i} $1657\,$\AA) reveal the white dwarf photosphere. The white dwarf dominates the ultraviolet emission but the non--zero flux detected in the core of the Ly$\alpha$ suggests the presence of a secondary continuum component whose origin is still not clear. It has been argued that this second component could arise from the bright spot, from a hot boundary layer, or from the disc (\citeauthor{Long1993} \citeyear{Long1993}, \citeauthor{Godon2004} \citeyear{Godon2004}, \citeauthor{Boris2005} \citeyear{Boris2005}). In addition we detect emission lines, which are broadened by the Keplerian velocity distribution of the disc. A model to fit the data has to account for these three contributions (white dwarf, second component and emission lines) detected in a CV spectrum.

We used \textsc{tlusty} and \textsc{synspec} (\citeauthor{Hubeny1988} \citeyear{Hubeny1988}; \citeauthor{Hubeny} \citeyear{Hubeny}) to compute a grid of synthetic spectra of white dwarf atmospheres, covering $T_{\mathrm{eff}} = 10\,000 - 70\,000\,\mathrm{K}$ in steps of $100\,\mathrm{K}$, and metal abundances $Z$ of $0.01$, $0.10$, $0.20$, $0.50$ and $1.00$ times their solar values\footnote{Using a single metallicity is sufficient to account for the presence of the metal lines in the fit, which are relatively weak. Possible deviations of single element abundances from the overall scaling with respect to the solar values do not affect the results of the $\chi^2$ minimisation routine. In fact, varying the abundances of individual elements has no effect on the measured $T_\mathrm{eff}$.}.

Ideally, a spectral fit to the ultraviolet data would provide both $T_{\mathrm{eff}}$ and the surface gravity ($\log g$). However these two quantities correlate: an increase in the temperature translates into a larger fraction of ionised hydrogen and narrower Lyman and Balmer lines; this effect can be counterbalanced by a higher gravity which increases pressure broadening. It is not possible to break this degeneracy from the sole analysis of the \textit{HST} data since they only provide the Ly$\alpha$ absorption profile which, in the case of cool CV white dwarfs, is limited to only the red wing of the line. 
Therefore we needed to make an assumption on the surface gravity. Most previously published work analysing CV white dwarf ultraviolet spectra assumed $\log g = 8.00$, corresponding to $0.6\,\mathrm{M}_\odot$, the average mass of isolated white dwarfs (\citeauthor{Koester1979} \citeyear{Koester1979}, \citeauthor{Liebert2005} \citeyear{Liebert2005}, \citeauthor{Kepler2007} \citeyear{Kepler2007}), unless an independent white dwarf mass determination was available. However, \citeauthor{Zorotovic} (\citeyear{Zorotovic}) demonstrated that the average mass of white dwarfs in CVs is actually higher than that of isolated white dwarfs, $\simeq 0.8\,\mathrm{M}_\odot$ ($\log g \simeq 8.35$). Since the canonical assumption $\log g = 8.00$ does not reflect the observed average mass of CV white dwarfs, we generated our grid of models assuming $\log g = 8.35$.

\begin{table}
 \centering
 \caption{Results for 1RXS\,J023238.8--371812 for the two different fitting methods for the disc emission lines, in which we either mask the emission lines (\textit{Mask}) or include them as Gaussian profiles (\textit{Gaussian fit}). We fixed $\log g = 8.35$. The last two rows (average) report the mean and the standard deviations for each method.}\label{table1rxsj}
  \begin{tabular}{@{}lcccc@{}}
  \toprule 
   Emission &   Second  & $T_\mathrm{eff}$ (K) & dof     & $\chi^2$ \\
    lines   & component &                      &         &          \\ 
\midrule
Mask                &    blackbody      &    $13\,543 \pm 36$    &  $1856$ &  $1794$  \\
Mask                &    power law      &    $13\,574 \pm 36$    &  $1856$ &  $1808$  \\
Mask                &    constant       &    $13\,462 \pm 28$    &  $1857$ &  $1861$  \\
\midrule
Gaussian fit        &    blackbody      &    $13\,540 \pm 34$    &  $2191$ &  $2170$  \\
Gaussian fit        &    power law      &    $13\,573 \pm 34$    &  $2191$ &  $2186$  \\
Gaussian fit        &    constant       &    $13\,466 \pm 26$    &  $2192$ &  $2245$  \\
\midrule
Mask                &    average        &    $13\,526 \pm 57$    &         &          \\
Gaussian fit        &    average        &    $13\,526 \pm 55$    &         &          \\
\bottomrule
\end{tabular}
\end{table}

The COS ultraviolet spectra are contaminated by geocoronal emission of Ly$\alpha$ and \ion{O}{i} $1302\,$\AA, and we masked these wavelength regions for our spectral analysis. Furthermore, we noticed the presence of an additional continuum component in all our systems, which contributes $\simeq 10 - 30$ per cent of the observed flux. In order to account for this, we included in the fit a blackbody, a power law or a constant flux\footnote{The constant is a special case of a power law and accounts for the easiest case with no slope, as it requires only one free parameter (a scaling factor), while power law and blackbody require two (respectively, a temperature and an exponent, plus a scaling factor). For this reason, we included the constant flux as an additional mode for the second component.} (in $F_\mathrm{\lambda}$), which can reproduce different slopes in the second component and are representative of different physical processes (thermalised emission in the case of the blackbody, optically thin emission in the case of the power law). Using a $\chi^2$ minimisation routine, we fitted the grid of model spectra to the \textit{HST} data and measured the effective temperatures of the 36 CV white dwarfs. 

To investigate the influence of the disc emission lines on our fitting procedure, we carried out our spectral analysis using two different methods: (i) we masked all the emission lines (\textit{Mask}); (ii) we included the emission lines as Gaussian profiles, allowing three free parameters: amplitude, wavelength and width (\textit{Gaussian fit}). 

\begin{figure*}
\thisfloatpagestyle{empty}
 \subfloat{%
  \label{1rxsj_a}%
  \includegraphics[width=0.40\textwidth,angle=-90]{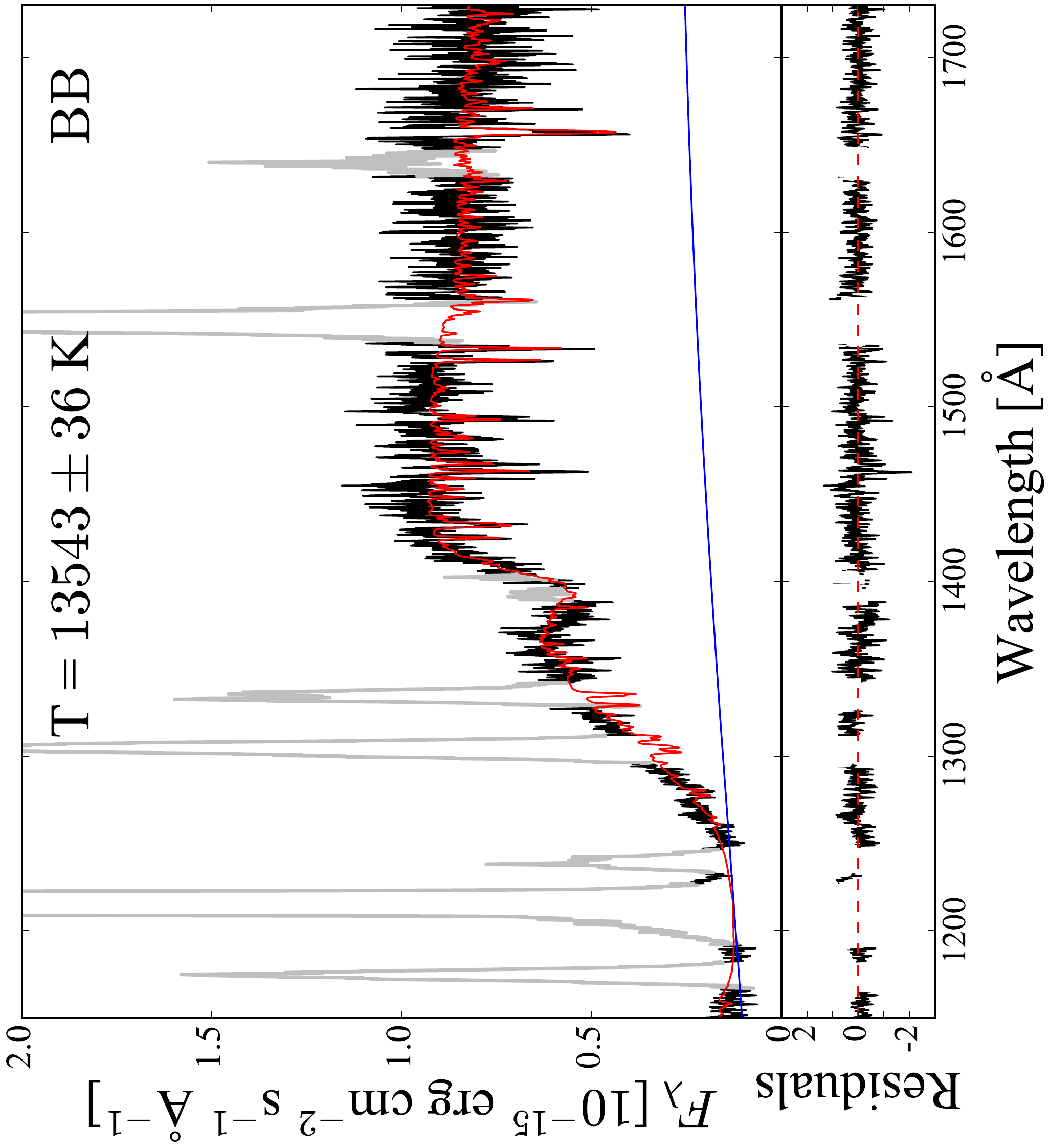} \qquad\qquad %
  \includegraphics[width=0.40\textwidth,angle=-90]{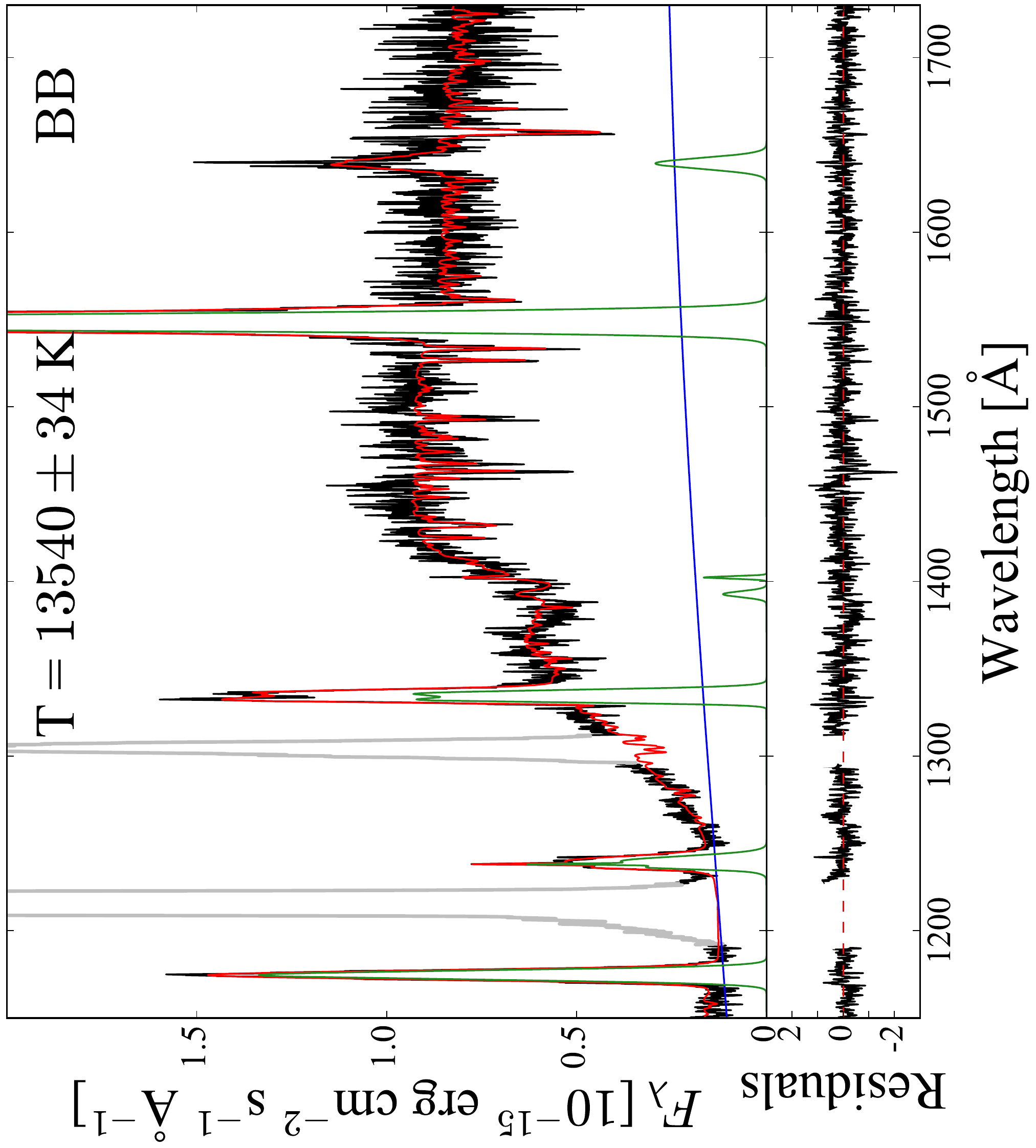}
  }\\%
 
 \subfloat{%
  \label{1rxsj_b}%
  \includegraphics[width=0.40\textwidth,angle=-90]{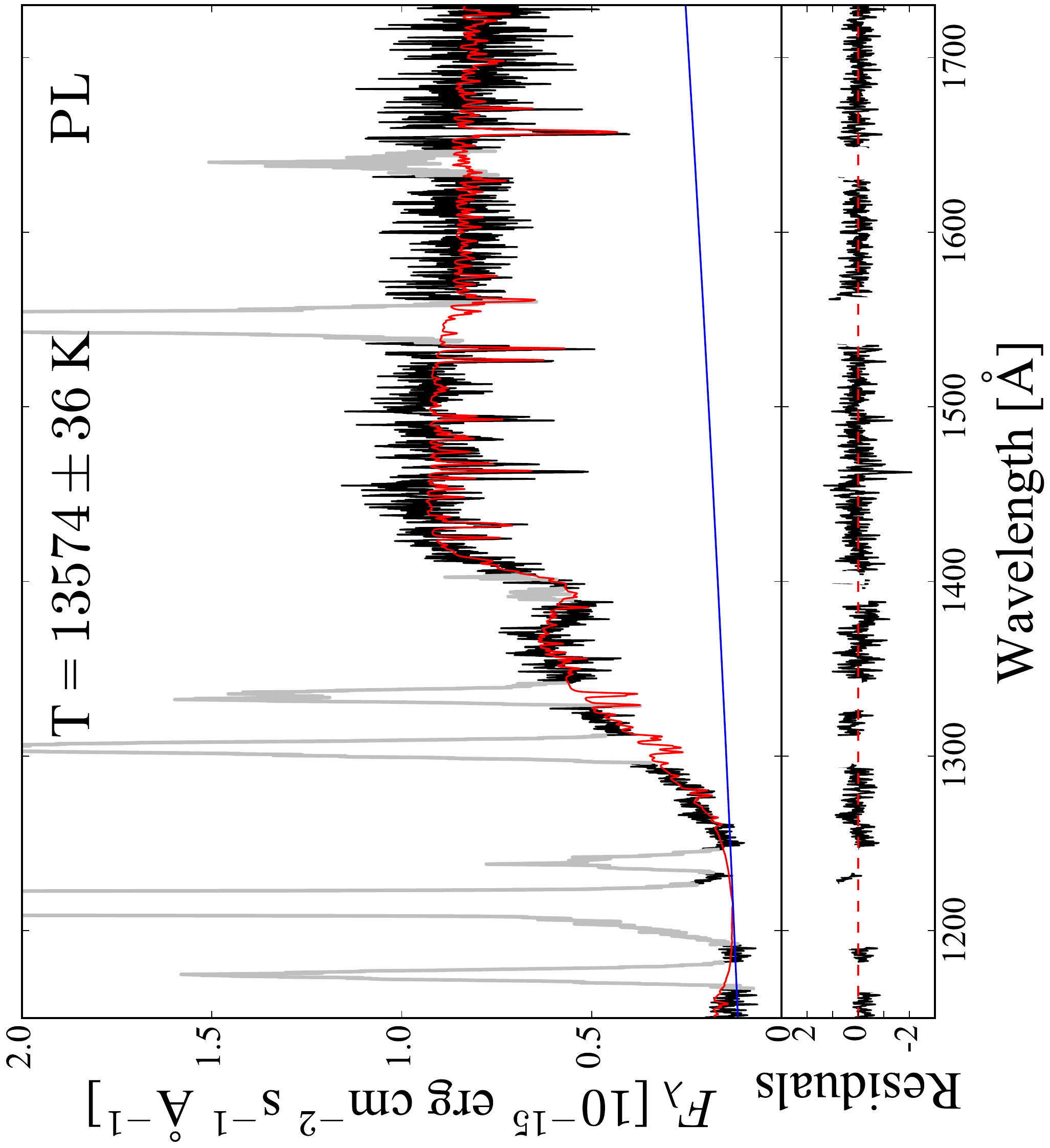} \qquad\qquad %
  \includegraphics[width=0.40\textwidth,angle=-90]{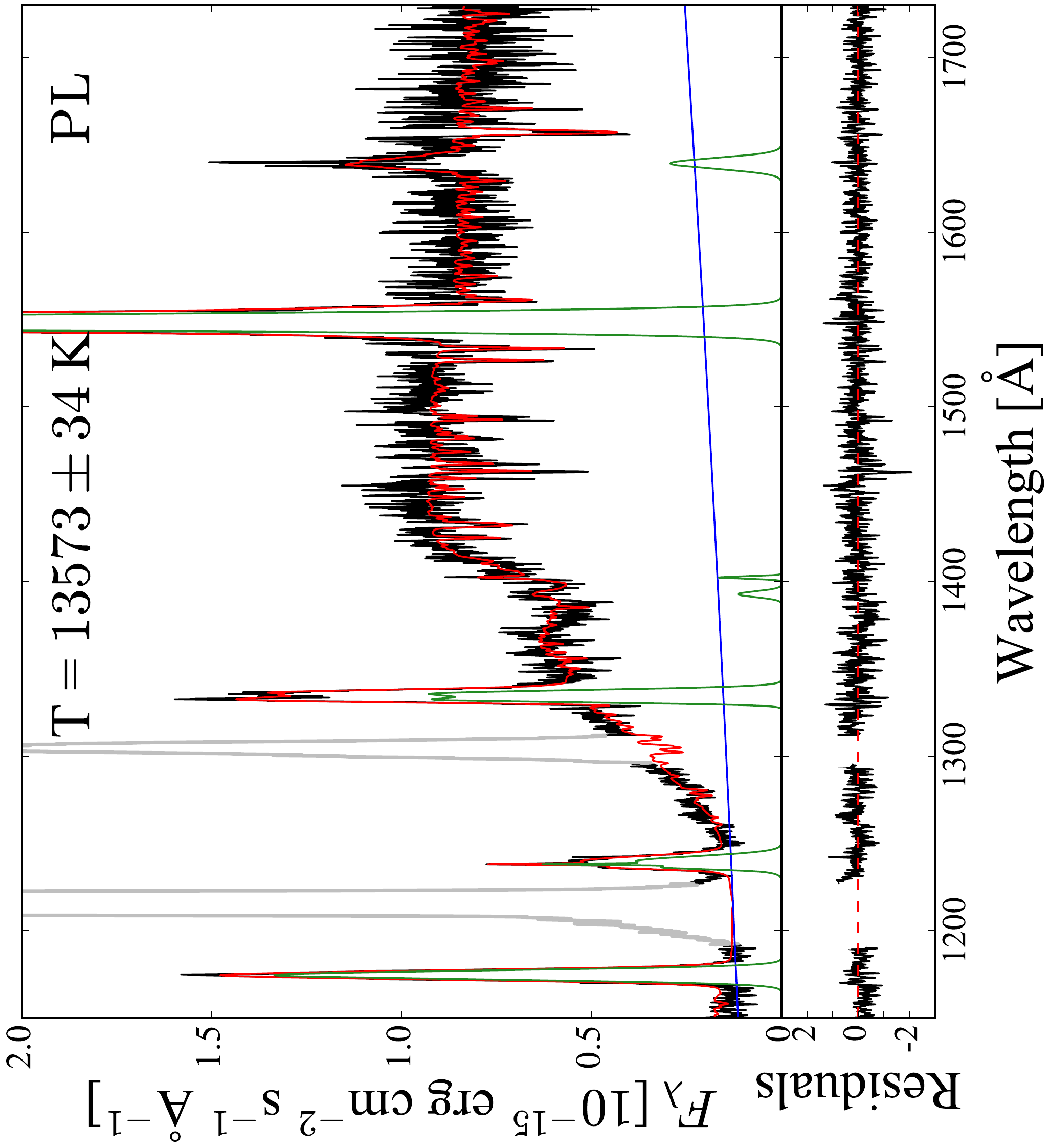}
  }\\%
  
 \subfloat{%
 \label{1rxsj_c}%
 \includegraphics[width=0.40\textwidth,angle=-90]{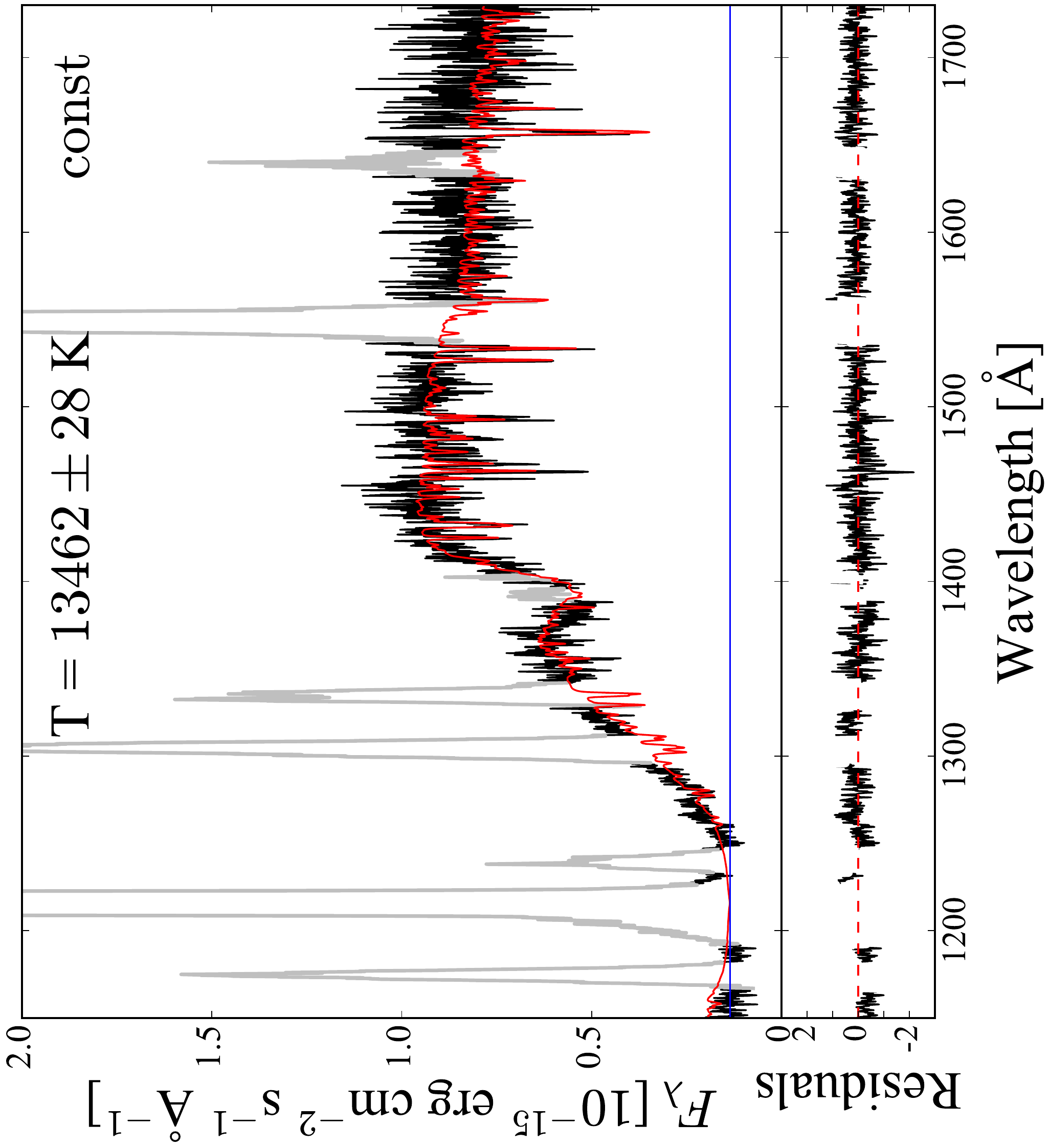} \qquad\qquad %
 \includegraphics[width=0.40\textwidth,angle=-90]{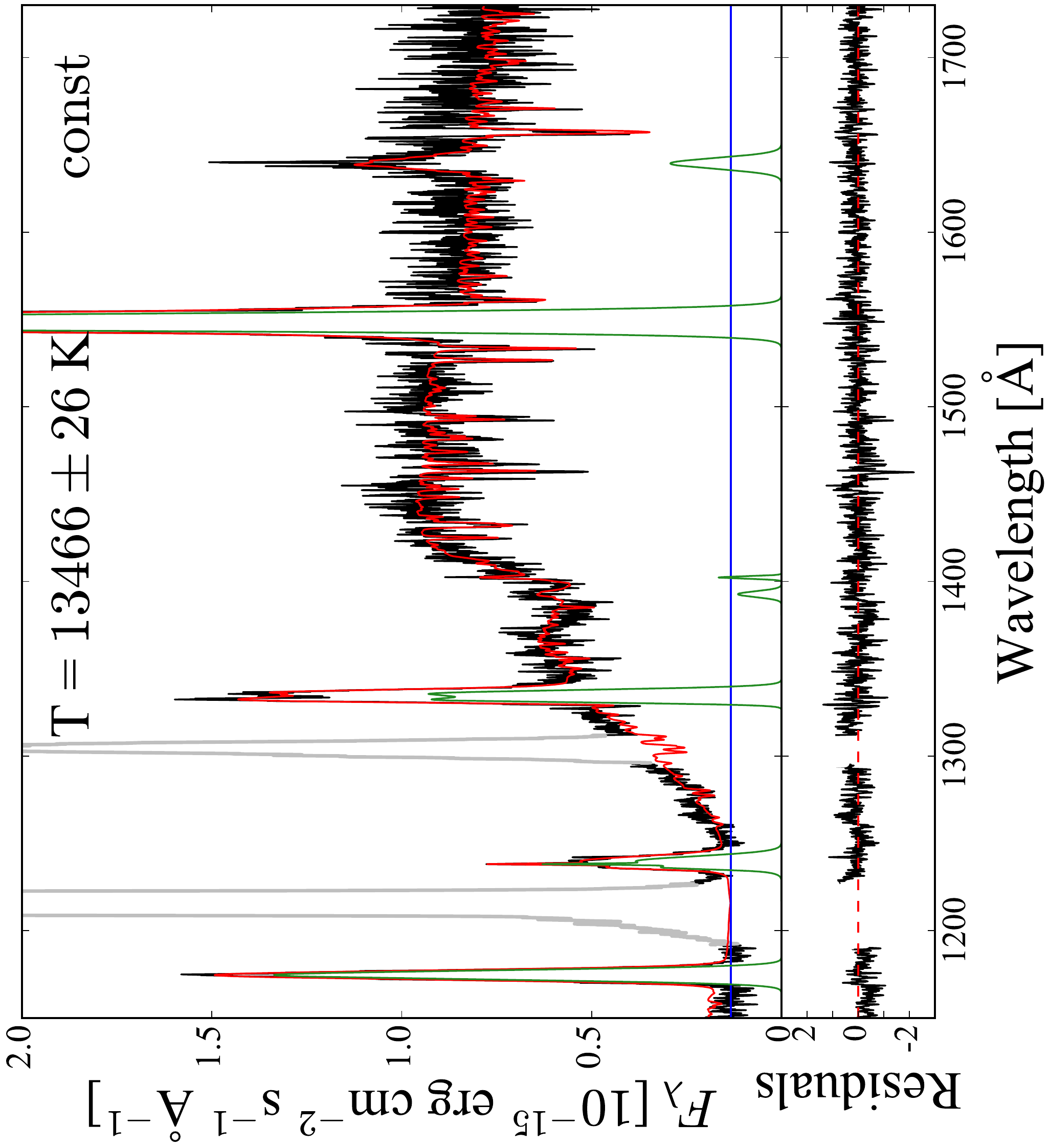}
 }%
\caption{\textit{HST}/COS spectrum of 1RXS\,J023238.8--371812 (black -- the disc emission lines of \ion{C}{iii} ($1175\,$\AA), \ion{N}{v} ($1242\,$\AA), \ion{C}{ii} ($1335\,$\AA), \ion{C}{iv} ($1550\,$\AA) and \ion{He}{ii} ($1640\,$\AA) are clearly visible; weak emission from \ion{Si}{iv} can be identified at $\simeq1400\,$\AA), along with the best--fit model (red) for the two different fitting methods: masking the emission lines (left panels) and including them as Gaussian profiles (right panels). The flux in the core of Ly$\alpha$ does not go to zero, implying the presence of a second continuum component (blue) which has been included in the fit in the form of a blackbody (BB), a power law (PL), or a constant (const). Masking the emission lines from the disc (grey in the left panels) or including them as a Gaussian profile (green in the right panels) returns very similar effective temperatures, within the statistical errors (Table \ref{table1rxsj}). Note that the geocoronal emission lines, Ly$\alpha$ ($1216\,$\AA) and \ion{O}{i} ($1302\,$\AA) have always been masked.}%
\label{1rxsj}
\end{figure*}

We illustrate the differences between the two methods using 1RXS\,J023238.8--371812. Figure~\ref{1rxsj} shows the COS spectrum along with the best--fit models obtained masking the emission lines (left panels) and including the lines in the fit (right panels), for all three different second components (from top to bottom: blackbody, power law and constant). The temperatures measured with the two methods typically only differ by $\simeq 3\,\mathrm{K}$ (see Table~\ref{table1rxsj}), demonstrating that including or masking the disc lines has no influence on the fit result (see also \citeauthor{Paula_second_component} \citeyear{Paula_second_component}). 
Therefore, to use as much of the data as possible, we decided to include the lines in the fit (\textit{Gaussian fit}).\par
The uncertainties of the individual fits listed in Table~\ref{table1rxsj} are purely statistical, as derived from the fitting procedure. They are unrealistically small and do not reflect the real uncertainties, which are instead dominated by several systematic effects, analysed in the following sections.\par

\subsection{The unknown nature of the second component}\label{subsec:sec_comp}
The first systematic uncertainty in our fitting procedure arises from the unknown nature of the additional emission component. The blackbody, the power law or the constant are not constrained outside the ultraviolet wavelength range, and only serve to account for different slopes in the detected additional continuum component. They thus represent a very simplified model of the additional continuum contribution and it is likely that none of them provides a realistic physical description of this emission component.

From a statistical point of view, and owing to the limited wavelength coverage of our data, we cannot discriminate among the three of them, and they all result in fits of similar quality and in very similar temperatures for the white dwarf (see Table~\ref{table1rxsj}). These differences in $T_\mathrm{eff}$ reflect a systematic effect related to presence of this additional flux. For this reason, we decided to adopt as final $T_\mathrm{eff}$ measurement the mean of the results obtained with the three different additional components. To evaluate the magnitude of the related systematic uncertainty, we calculated both the standard deviation of the $T_\mathrm{eff}$ values obtained with the three different additional components and the sum in quadrature of the statistical errors of the individual fits. We assumed as final and more realistic estimate for the systematic uncertainty the larger of these two. Additional systematic effects are analysed in the following sections.

\begin{table}
 \centering
 \caption{$T_\mathrm{eff}$ for 1RXS\,J023238.8--371812 allowing different $\log g$, corresponding to different white dwarf masses. These results have been obtained fitting the data including the emission lines as Gaussian profiles, and using three different models for the second component.}\label{table1rxsj_lg}
  \begin{tabular}{@{}lccc@{}}
  \toprule
                   & $T_\mathrm{eff}$ (K) & $T_\mathrm{eff}$ (K) & $T_\mathrm{eff}$ (K) \\
  Second           &  $\log g = 8.00$     &   $\log g = 8.35$    & $\log g = 8.60$     \\ 
 component         &  $M_{\mathrm{WD}} = 0.6\, \mathrm{M}_\odot$ & $M_{\mathrm{WD}} = 0.8\, \mathrm{M}_\odot$ & $M_{\mathrm{WD}} = 1.0\, \mathrm{M}_\odot$ \\ 
\midrule
blackbody          &  $13\,016 \pm 28$      &  $13\,540 \pm 34$      &  $13\,938 \pm 37$  \\
power law          &  $13\,045 \pm 29$      &  $13\,573 \pm 34$      &  $13\,972 \pm 37$  \\
constant           &  $12\,949 \pm 23$      &  $13\,466 \pm 26$      &  $13\,848 \pm 28$  \\
\midrule
average            &  $13\,003 \pm 49$      &  $13\,526 \pm 55$      &  $13\,919 \pm 64$  \\
\bottomrule
\end{tabular}
\end{table}

\subsection{The unknown white dwarf mass}\label{subsec:unknown_mass}
\begin{figure*}
 \centering
 \includegraphics[width=0.5\textwidth,angle=-90]{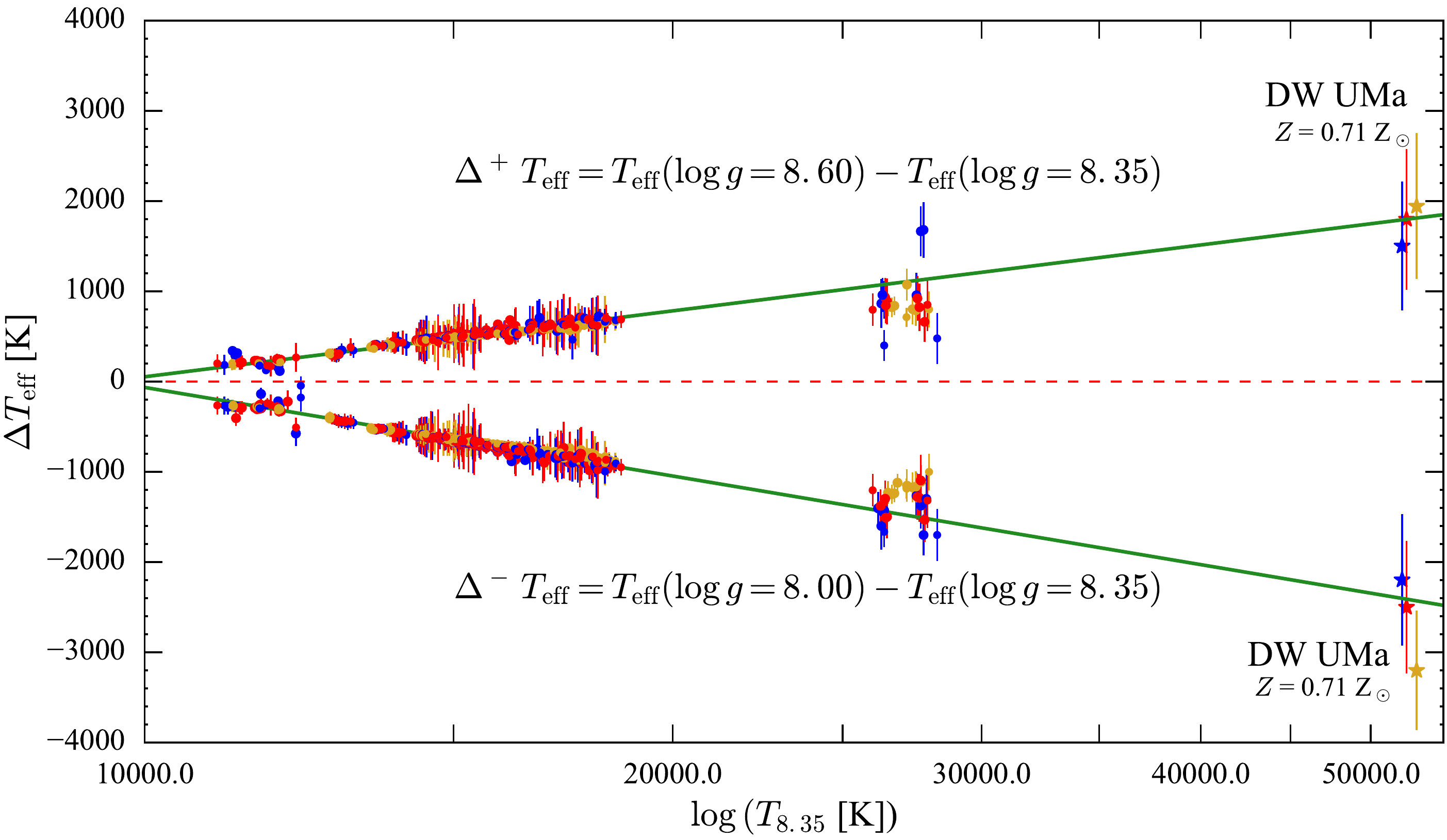}\\
  \caption{Best--fit to $\Delta^+\,T_\mathrm{eff} = T_\mathrm{eff}(\log g = 8.60) - T_\mathrm{eff}(\log g = 8.35)$ and $\Delta^-\,T_\mathrm{eff} = T_\mathrm{eff}(\log g = 8.00) - T_\mathrm{eff}(\log g = 8.35)$, as a function of $T_\mathrm{eff}(\log g = 8.35)$, for the five metallicities and the three additional continuum second components (red for blackbody, blue for power law and gold for constant). The fit excludes the effective temperatures of (i) the eclipsing systems IY\,UMa, SDSS\,J040714.78--064425.1 and SDSS\,J100658.41+233724.4 for which disc absorption along the line of sight makes the identification of the continuum flux level difficult (Section~\ref{eclipsing}); (ii) CC\,Scl and SDSS\,J164248.52+134751.4 for which a reliable effective temperature determination was not possible (Section~\ref{subsec:confronto}); (iii) SDSS\,J153817.35+512338.0 because the core of its narrow Ly$\alpha$ line is strongly contaminated by geocoronal airglow emission which makes the data less sensitive to a change in the model surface gravity. DW\,UMa was not included in the fit, but serves as independent test of the fit.}\label{systematic_uncertainty}
\end{figure*}

As pointed out above, the temperature and mass of a white dwarf correlate. Table~\ref{table1rxsj_lg} reports the effective temperatures of 1RXS\,J023238.8--371812 allowing different $\log g$ (= different white dwarf masses). These results show that an assumption on $\log g$ translates into an accurate measurement of the temperature \textit{for a given mass} ($\simeq 50\,\mathrm{K}$), while the systematic uncertainty introduced considering different surface gravities is typically an order of magnitude larger ($\simeq 500\,\mathrm{K}$). Therefore the dominant source of error is the (unknown) white dwarf mass and we used the following approach to evaluate how this systematic uncertainty depends on $T_{\mathrm{eff}}$. 

As shown by \citet{Zorotovic}, the average mass of CV white dwarfs is $0.83 \pm 0.23 \mathrm{M_\odot}$ and we  therefore investigated the correlation between effective temperature and $\log\,g$ in the mass range $0.6\,\mathrm{M_\odot} (\log\,g\,=\,8.00)$ to $1.0\,\mathrm{M_\odot} (\log\,g\,=\,8.60)$. We fitted our \textit{HST} data assuming $\log\,g\,=\,8.00$, $\log\,g\,=\,8.35$ and $\log\,g\,=8.60$ for each metallicity and for each type of second component, resulting in 15 effective temperatures (five metallicities $\times$ three second components) for each $\log\,g$, for each object. We define\footnote{We found that the effective temperature derived for different metallicities were consistent within the uncertainties. To simplify our analysis, we do not distinguish between different $Z$ in this discussion.} as $T_{8.00} = T_\mathrm{eff}\,(\log\,g=8.00)$, $T_{8.35} = T_\mathrm{eff}\,(\log\,g=8.35)$ and $T_{8.60} = T_\mathrm{eff}\,(\log\,g=8.60)$, and calculated $\Delta^+ \,T_\mathrm{eff} = T_{8.00} -T_{8.35}$, and $\Delta^- \,T_\mathrm{eff} = T_{8.60} - T_{8.35}$.

We excluded from this analysis the eclipsing systems IY\,UMa, SDSS\,J040714.78--064425.1 and SDSS\,J100658.41+233724.4 for which disc absorption along the line of sight makes the identification of the continuum flux level difficult (Section~\ref{eclipsing}).
We also did not include CC\,Scl and SDSS\,J164248.52+134751.4 for which a reliable effective temperature determination was not possible (Section~\ref{subsec:confronto}), and SDSS\,J153817.35+512338.0 because the core of its narrow Ly$\alpha$ line is strongly contaminated by geocoronal airglow emission which makes the data less sensitive to a change in the model surface gravity.

The remaining objects have typically $T_\mathrm{eff} \lesssim 21\,000\,\mathrm{K}$, with the exception of HS2214+2845 ($T_\mathrm{eff} \simeq 26\,000\,\mathrm{K}$). To better constrain $\Delta^+\,T_\mathrm{eff}$ and $\Delta^-\,T_\mathrm{eff}$ and to verify the validity of this relationship at high temperatures, we included in our analysis two additional hotter objects: SS\,Aur ($T_\mathrm{eff} = 34\,000 \pm 2\,000\,\mathrm{K}$ for $\log\,g=8.8$, \citeauthor{BDPav} \citeyear{BDPav}) and DW\,UMa ($T_\mathrm{eff} = 50\,000 \pm 1\,000\,\mathrm{K}$ for $\log\,g = 8.00$ and $Z = 0.71\,\mathrm{Z}_\odot$, \citeauthor{DWUMa1} \citeyear{DWUMa1}). Both these objects have been observed with STIS, which allows the removal of the contamination from geocoronal emission and therefore, differently from SDSS1538 in our sample, they can be used to constrain the relationship at high temperatures. 

We retrieved the STIS spectrum of SS Aur and the out--of--eclipse STIS low--state spectrum of DW UMa \citep{DWUMa} from the \textit{HST} data archive. Following the same method as for the CVs in our sample, we fitted these data assuming $\log g = 8.00$, $\log g = 8.35$ and $\log g = 8.60$. For SS\,Aur we found: 
$T_{8.00} = 26\,269 \pm 275\,\mathrm{K}$, 
$T_{8.35} = 27\,507 \pm 282\,\mathrm{K}$ and 
$T_{8.60} = 28\,402 \pm 270\,\mathrm{K}$ 
for $Z = 0.1\,\mathrm{Z}_\odot$.
In the case of DW UMa, we assumed $Z = 0.71\,\mathrm{Z}_\odot$, and obtained
$T_{8.00} = 49\,900 \pm 844\,\mathrm{K}$, in agreement with \citet{DWUMa1}, $T_{8.35} = 52\,532 \pm 893\,\mathrm{K}$ and $T_{8.60} = 54\,282 \pm 986\,\mathrm{K}$, respectively.

Figure~\ref{systematic_uncertainty} show the trend of $\Delta^+\,T_\mathrm{eff}$ and $\Delta^-\,T_\mathrm{eff}$ as a function of $T_\mathrm{eff}\,(\log\,g=8.35)$. These correlations are well fit with the following relation:

\begin{equation}\label{equation1}
\Delta\,T_\mathrm{eff} = a \log (T_\mathrm{eff} \, b)\\
\end{equation}

\noindent{where $a = 1054(2)\,\mathrm{K}$, $b = 0.0001052(7)\,\mathrm{K}^{-1}$ for $\Delta^+\,T_\mathrm{eff}$, and $a = -1417(2)\,\mathrm{K}$, $b = 0.0001046(5)\,\mathrm{K}^{-1}$ for $\Delta^-\,T_\mathrm{eff}$.}\par

We included the SS\,Aur temperatures in the fit of $\Delta^+\,T_\mathrm{eff}$ and $\Delta^-\,T_\mathrm{eff}$ to better constrain the relationship at high temperatures. In contrast, we did not include the DW\,UMa temperatures and we only overplot them in Figure~\ref{systematic_uncertainty} to illustrate that our best--fit is in good agreement with this independent measurement.\par
The two best--fit curves represent the systematic uncertainty $\sigma_{T_{\mathrm{eff}}}$ due to the unknown mass of the white dwarf. The two curves are not symmetric (since the relationship between $\log\,g$ and the white dwarf mass is not linear), resulting in asymmetric error bars
with $\Delta^-\,T_\mathrm{eff} > \Delta^+\,T_\mathrm{eff}$. However, to simplify our discussion, we adopted as final uncertainty the larger  value derived from $\Delta^-\,T_\mathrm{eff}$:

\begin{equation}\label{equation2}
\sigma_{T_{\mathrm{eff}}} = 1417 \log (0.0001046 \, T_\mathrm{eff} )\\
\end{equation}

In summary, we found that the systematic uncertainty due to the unknown white dwarf mass lies in the range $300 - 1800\,\mathrm{K}$. Once the masses for each system are accurately determined, the degeneracy between $T_\mathrm{eff}$ and $\log\,g$ will be broken, reducing the uncertainties in $T_\mathrm{eff}$ to those related to the unknown nature of the second additional component (see Section~\ref{subsec:sec_comp}).

\begin{figure}
 \centering
 \includegraphics[angle=-90,width=0.48\textwidth]{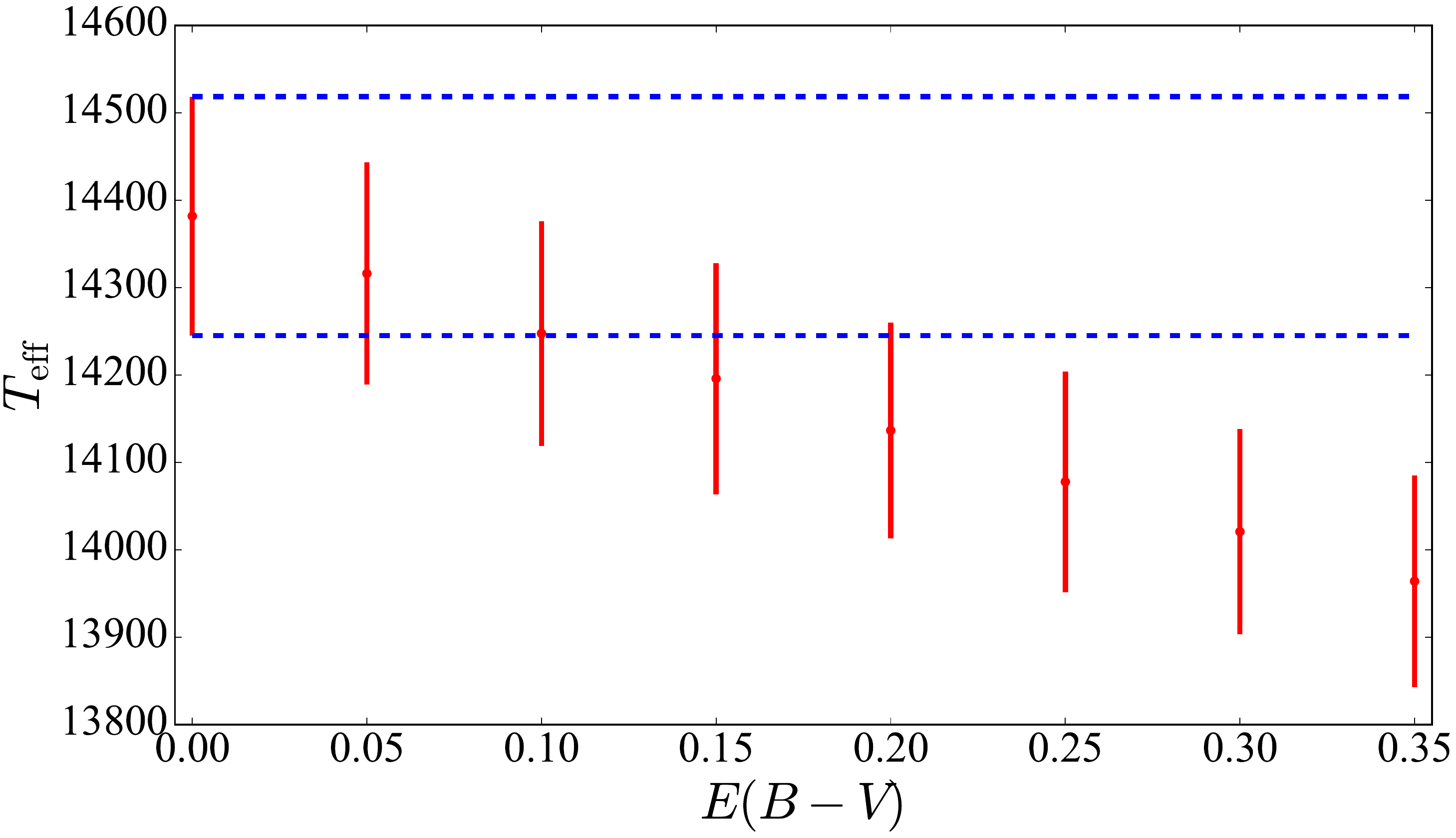}\\
   \caption{Best--fit effective temperatures for UV\,Per, obtained by reddening the \textit{HST}/STIS data for increasing values of $E(B-V)$. The blue lines show the systematic uncertainties on the effective temperature derived from the fit to the original data ($E(B-V) = 0$), following the method described in Section~\ref{subsec:sec_comp}. They establish the threshold above which the effect of interstellar absorption on our $T_\mathrm{eff}$ is greater than the systematic uncertainties, i.e. $E(B-V) \simeq 0.1$.}\label{t_ebv}
  \end{figure}

\subsection{Reddening}\label{subsec:reddening}
Reddening due to interstellar dust along the line of sight can introduce an additional systematic uncertainty in the effective temperature determination as it affects the overall slope of the ultraviolet spectra. However, CVs are intrinsically faint and thus we are observationally biased towards nearby systems, for which extinction is usually negligible. Moreover, owing to their colour similarity with quasars, CVs are often discovered by extragalactic surveys (such as SDSS), which cover high Galactic latitudes, and are therefore not heavily affected by reddening. To verify that reddening is a minor contribution to the total error budget, for all the systems in our sample, we compiled the colour excess of our CVs (Table~\ref{table_Log_COS_obs} and \ref{table_Log_STIS_obs}) using the three--dimensional map of interstellar dust reddening based on Pan--STARRS\,1 and 2MASS photometry \citep{panstar} wherever possible, i.e. when the distance is known and the object is inside the field covered by this map. For the remaining objects, we report either the value from the literature where available or the galactic $E(B-V)$ from the NASA/IPAC Extragalactic Database (NED), which only represents an upper limit for the actual reddening. Finally, for CU\,Vel, we determined its colour excess below (Section~\ref{CUVel_section}). 

To establish how the interstellar absorption affects our analysis, we considered one of the systems with zero colour excess, UV\,Per, and reddened its spectrum using the relationship given by \citet{cardelli} for a range of values in $E(B-V)$. Since we previously determined the effective temperature for this CV white dwarf, $T_\mathrm{eff} [E(B-V) = 0] = 14\,389 \pm 578\,$K, we can use this ``artificially reddened'' dataset to study the variation of $T_\mathrm{eff}$ as a function of reddening. We fitted the reddened spectra following the prescription from Sections~\ref{sec:DataAnalysis} and \ref{subsec:sec_comp} and show the derived temperatures in Figure~\ref{t_ebv}. The interstellar absorption introduces a variation in $T_\mathrm{eff}$ greater than the systematic uncertainties defined in Section~\ref{subsec:sec_comp} (blue dashed lines) for $E(B-V) \gtrsim 0.1$, which we assumed as the threshold above which the effect of the reddening cannot be neglected. 
Only one system in our sample has a colour excess significantly higher than this value: RZ\,Sge, $E(B-V)= 0.302$, as returned from the NED database. However Sagittarius lies on the Galactic plane and this $E(B-V)$ represents an estimate over the entire Galactic column. Given the typical distances of CVs, the actual reddening of RZ\,Sge is likely to be lower than that.
We assumed the mass--radius relationship from \citet{m-r_relationship} and, using the scaling factors from the fit to the STIS data for different surface gravities (Section~\ref{subsec:unknown_mass}), we estimated the distance to RZ\,Sge to lie in the range $262\,\mathrm{pc}\,\lesssim\,d\,\lesssim\,335\,\mathrm{pc}$, for $0.6\,\mathrm{M}_\odot\,\leq\,M_\mathrm{WD}\,\leq\,0.8\,\mathrm{M}_\odot$. From the three--dimensional map of interstellar dust reddening \citep{panstar} we found $0.044\,\lesssim\,E(B-V)\,\lesssim 0.070$, well below the threshold we established above. We therefore considered negligible the reddening for this system.

Finally, interstellar gas along the line of sight can also contaminate the observed spectrum with additional Ly$\alpha$ absorption. Using the relation between interstellar neutral Hydrogen column density and reddening from \citet{lya_interstellar}, we determined that this contribution is always of the order of few \r{A}ngstrom ($\simeq 3\,$\AA\, for $E(B-V)= 0.05$ up to $\simeq 7\,$\AA\, for $E(B-V)= 0.3$) and therefore much narrower than the white dwarf Ly$\alpha$ absorption line. In fact, the interstellar Ly$\alpha$ absorption is located in the spectral region that we always masked owing to geocoronal airglow emission, which typically has a width of $\simeq 18\,$\AA, and therefore we concluded that interstellar Ly$\alpha$ absorption has no effect on our results.

\begin{figure}
\thisfloatpagestyle{empty}
 \centering
 {%
  \subfloat{
  \includegraphics[width=0.4\textwidth,angle=-90]{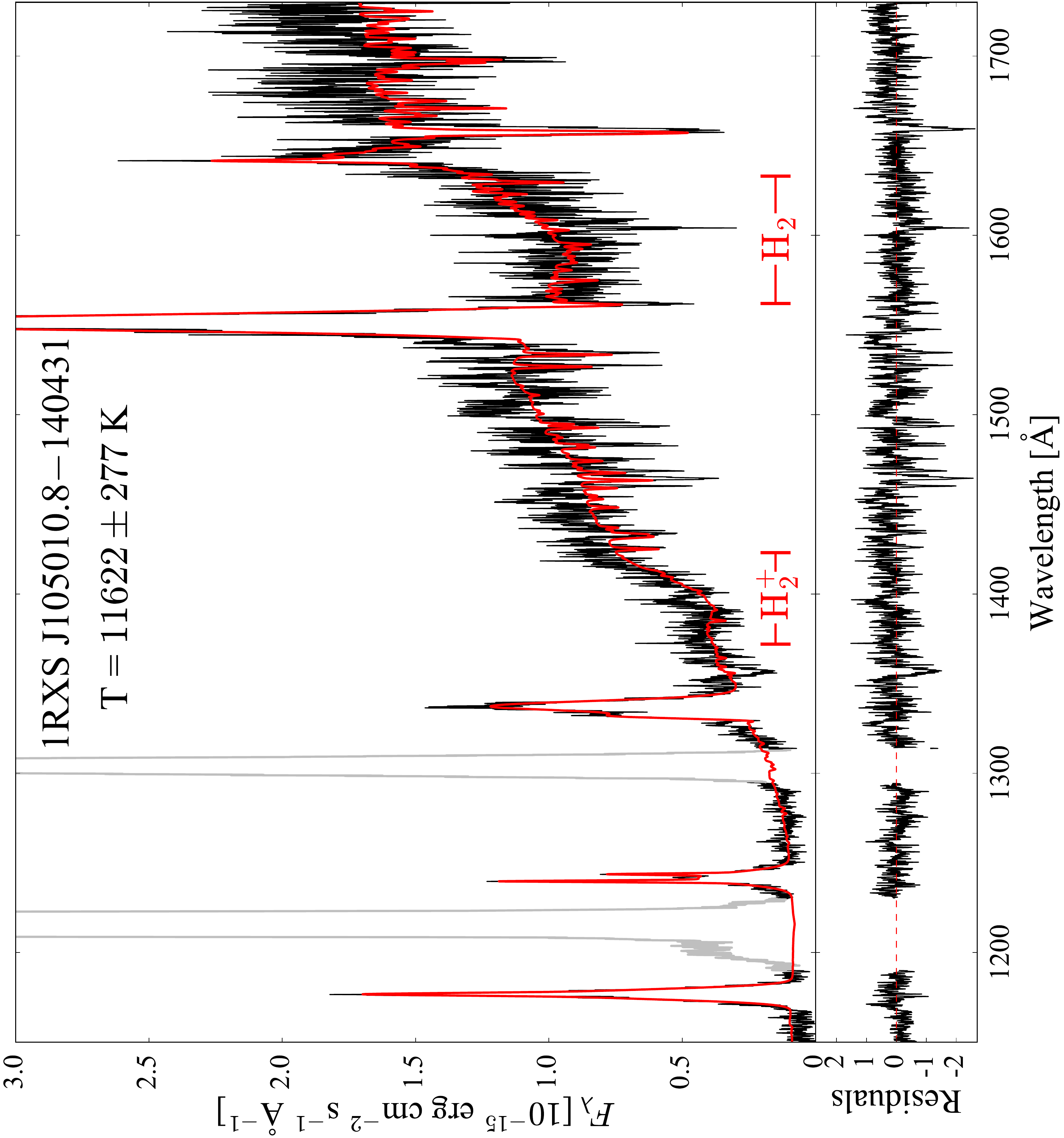} %
  }\\%
 
 \subfloat{
  \includegraphics[width=0.4\textwidth,angle=-90]{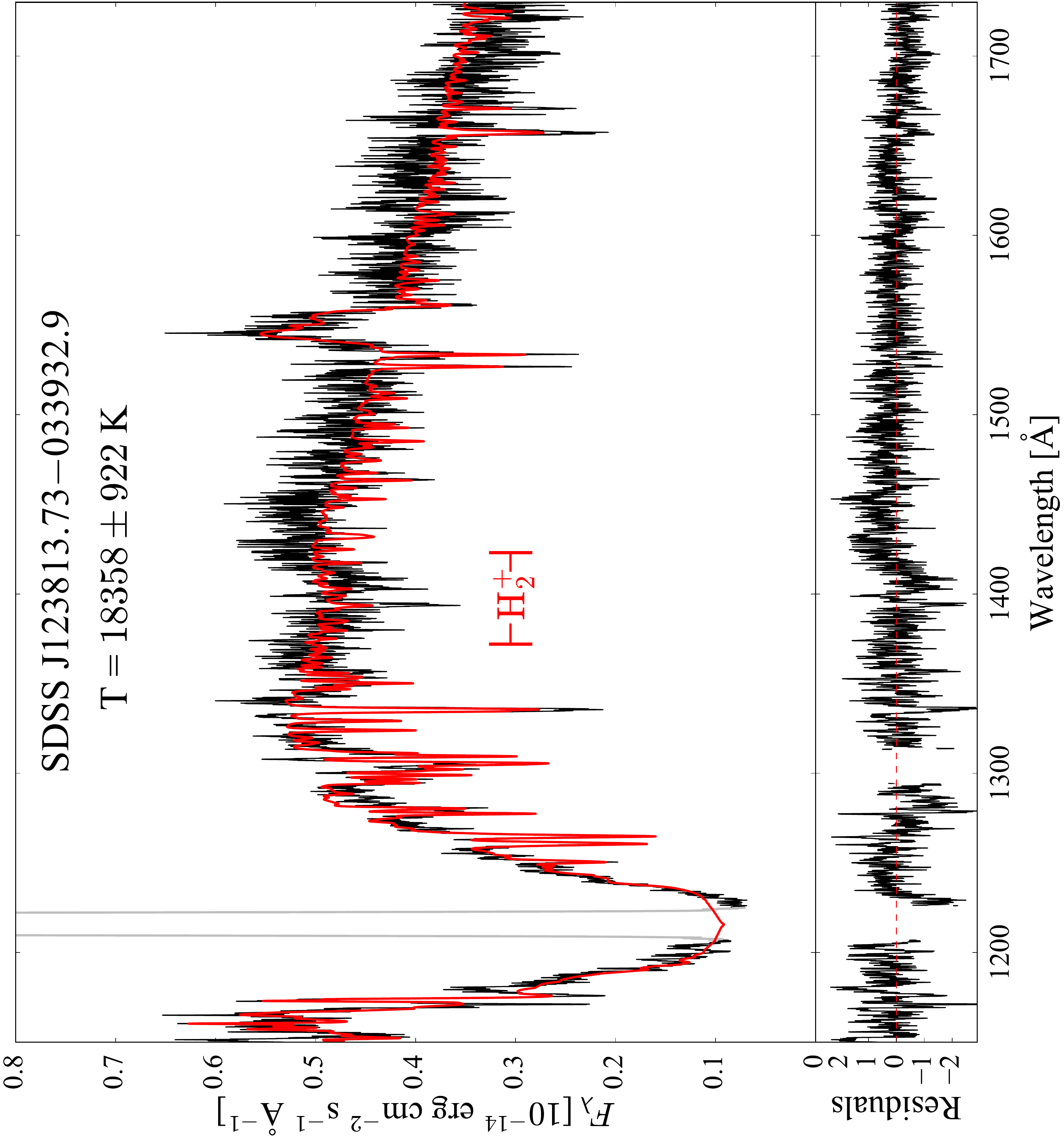} %
  }\\%
  
 \subfloat{
 \includegraphics[width=0.4\textwidth,angle=-90]{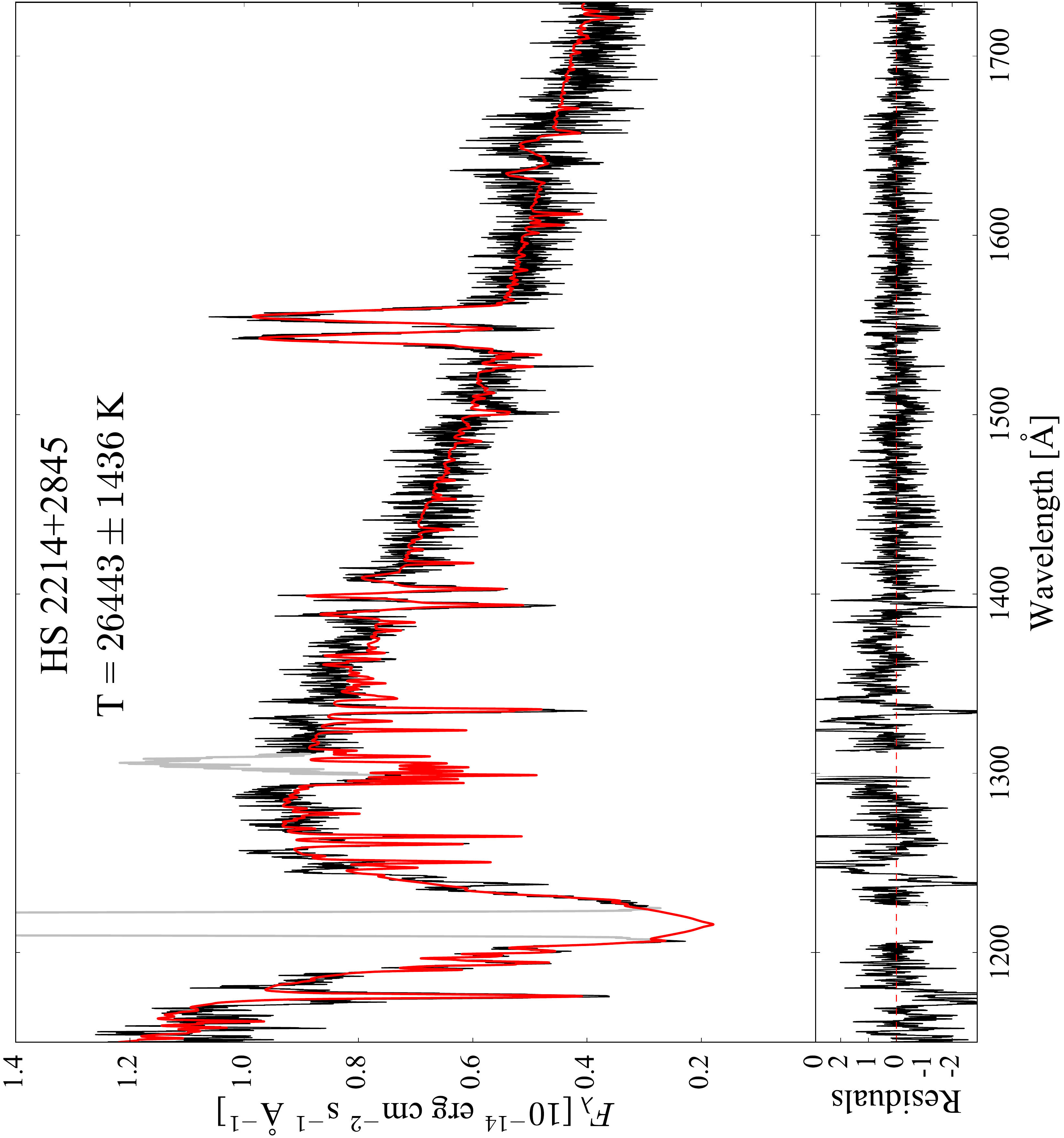} %
 }%
\caption{Ultraviolet spectra (black) of cool, warm and hot CV white dwarfs in our sample along with the best--fit model (red) assuming a blackbody second component. 
The quasi--molecular absorption bands of H$_2^{+}$ and of H$_2$ are visible at $\simeq 1400\,$\AA\, and $\simeq 1600\,$\AA\, for $T_\mathrm{eff}\,\lesssim 19\,000\,\mathrm{K}$ and $T_\mathrm{eff}\,\lesssim 13\,500\,\mathrm{K}$, respectively (\citeauthor{quasi-molecular} \citeyear{quasi-molecular}, \citeauthor{Koester1985} \citeyear{Koester1985}). The geocoronal emission lines of Ly$\alpha$ ($1216\,$\AA) and \ion{O}{i} ($1302\,$\AA) are plotted in grey.}\label{plot-fit}}
\end{figure}

\subsection{Other possible systematic effects}\label{subsec:additional_systematics}
To review all the systematic uncertainties in a way as comprehensive as possible, we need to discuss the possibility of Ly$\alpha$ absorption from the additional non--white dwarf component and inaccuracies in the instrument calibration.

Although not much is known about the physical origin of the second component, it is likely to arise from a hot optically thick (e.g. the bright spot) or a hot optically thin medium surrounding a cooler optically thick layer (e.g. the disc or the boundary layer). It is therefore possible that the second component can contribute, to some extent, to the observed Ly$\alpha$ absorption. The Ly$\alpha$ profile is, along with the spectral slope, the main tracer of the white dwarf temperature, and such a hypothetical contamination could systematically affect our results. \citet{absorp_sec_comp} investigated an additional Ly$\beta$ absorption in VW\,Hyi finding that it most likely originates from a hot spot region whose emission can be approximated by a stellar model with $\log\,g\simeq 4$. The Lyman absorption lines arising from such an environment are consequently significantly narrower than the white dwarf Ly$\alpha$ absorption itself, which is broadened by the higher pressure on the white dwarf surface at $\log g \simeq 8.35$. 
Thus, although the model we used to describe the additional component does not account for the possibility of absorption in the Ly$\alpha$ region, such contamination (if present) would not appreciably affect our effective temperature determination.

Finally, limitations in the instrument calibration may affect our results. \citet{cos_calib} report that the systematic uncertainties in the COS G140L flux calibration are less than two per cent for $\lambda < 1200\,$\AA\, and one per cent for $1200\,$\AA\,$< \lambda < 1900\,$\AA, with an increase up to six per cent at $\lambda = 2150\,$\AA. Owing to a decrease in the detector sensitivity in the red portion of the spectrum, we only considered the wavelength range $1150\,$\AA\,$ < \lambda < 1730\,$\AA, for which systematic uncertainties are less than two per cent. The maximum effect on our results can be evaluated by multiplying the COS spectra with a linear function with a two per cent slope. To assess also for possible dependencies with the temperature of the white dwarf, we choose three systems representative of cool, warm and hot white dwarfs: 1RXS\,J105010.8--140431, SDSS\,J123813.73--033932.9 and HS\,2214+2845. We fitted their spectra after applying the two per cent slope in flux calibration and find that the resulting $T_\mathrm{eff}$ values are in agreement, within the uncertainties, with the one we derived from the original COS data. We therefore conclude that our analysis is not affected by the very small uncertainty in the COS instrument calibration.

Three objects in our sample have been observed both with STIS and COS. In Section~\ref{with_COS_STIS1} we compare the two datasets and, for this comparison to be reliable, we needed to verify that the STIS and COS calibrations agree. To do so, we retrieved from the \textit{HST} archive the STIS and COS data of the flux standard WD\,0308--565. We overplot the one available STIS spectrum and the COS data acquired at different epochs, finding that the STIS spectrum matches the flux level of all the COS data. With a linear fit to the ratio between the two datasets, we determined that they differ, on average, by $\simeq 3$ per cent. This comparison proves that uncertainties in the STIS calibration are comparable to the systematic uncertainties of COS and therefore they are negligible in our analysis. Finally, we can also conclude that the differences in the STIS and COS effective temperatures discussed in Sections~\ref{with_COS_STIS1} are not related to calibration issues of the two instruments.\\

We followed the procedure outlined in the previous Sections to fit the \textit{HST} data of the 36 CVs and summarize the results in Table~\ref{table1}. Figure~\ref{plot-fit} shows three examples of best--fit models obtained with this procedure: 1RXS\,J105010.8--140431 (top panel), SDSS\,J123813.73--033932.9 (middle panel) and HS\,2214+2845 (bottom panel), which are representative of different temperature regimes (cool, intermediate and hot, respectively). All the spectra,  along with their best--fit models, are available in the online material.

\begin{table*}
\thisfloatpagestyle{empty}
 \caption{Characteristics of the 45 CV white dwarfs observed with COS and STIS.}\label{table1}
 \setlength{\tabcolsep}{0.13cm}
  \begin{tabular}{@{}lcccr|ccccc@{}}
  \toprule
   System         & $P_\mathrm{orb}$ &   d               &  $M_{\mathrm{WD}}$    & References &    $Z$    & $T_\mathrm{eff}$ & $\pm$ &    WD       & Instrument\\
                  & (min)            &  (pc)             &($\mathrm{M}_{\odot}$) &            &($Z_\odot$)&   (K)            &   (K) & contribution&  \\
\midrule
V485\,Cen                  & 59.03  &                    &                   & 1            & 0.5  & 15\,200 & 655  & 56\% & STIS \\ 
GW\,Lib                    & 76.78  & $140^{+30}_{-20}$  &                   & 2, 3         & 0.2  & 16\,995$\downarrow$ & 812 & 85\% & COS \\ 
SDSS\,J143544.02+233638.7  & 78.00  &                    &                   & 4            & 0.01 & 11\,940 & 315  & 84\% & COS  \\ 
OT\,J213806.6+261957       & 78.10  &                    &                   & 5            & 0.2  & 16\,292 & 753  & 69\% & COS  \\
V844\,Her                  & 78.69  & $290 \pm 30$       &                   & 3, 6         & 0.1  & 14\,850 & 622  & 61\% & STIS \\ 
SDSS\,J013701.06--091234.8 & 79.71  & $300 \pm 80$       &                   & 7, 8         & 0.1  & 14\,547 & 594  & 77\% & COS  \\   
SDSS\,J123813.73--033932.9 & 80.52  & $110$              &                   & 9            & 0.5  & 18\,358 & 922  & 75\% & COS  \\
PU\,CMa                    & 81.63  &                    &                   & 10           &      &         &      &      & COS  \\
V1108\,Her                 & 81.87  & $130 \pm 30$       &                   & 11, 12       & 1.0  & 13\,643 & 503  & 81\% & COS  \\ 
ASAS\,J002511+1217.2       & 82.00  & $130 \pm 30$       &                   & 12, 13       & 0.1  & 12\,830 & 416  & 77\% & COS  \\
SDSS\,J103533.02+055158.4  & 82.22  & $170 \pm 12$       & $0.835 \pm 0.009$ & 14, 15       & 0.01 & 11\,620 & 44*  & 85\% & COS  \\
CC\,Scl                    & 84.10  &                    &                   & 16           & 0.2  & 16\,855:& 801  & 35\% & COS  \\
SDSS\,J075507.70+143547.6  & 84.76  &                    &                   & 17           & 0.5  & 15\,862 & 716  & 90\% & COS  \\
1RXS\,J105010.8--140431    & 88.56  & $100 \pm 50$       &                   & 18, 19       & 0.1  & 11\,622 & 277  & 89\% & COS  \\
MR\,UMa                    & 91.17  &                    &                   & 20           & 0.2  & 15\,182$\downarrow$ & 654 & 69\% & COS  \\ 
QZ\,Lib                    & 92.36  & $120 \pm 50$       &                   & 19           & 0.01 & 11\,303 & 238  & 64\% & COS  \\
SDSS\,J153817.35+512338.0  & 93.11  &                    &                   & 17           & 0.01 & 33\,855 &1785  & 94\% & COS  \\
UV\,Per                    & 93.44  &                    &                   & 21           & 0.2  & 14\,389 & 578  & 75\% & STIS \\ 
1RXS\,J023238.8--371812    & 95.04  & $160$              &                   & 22           & 0.2  & 13\,527 & 491  & 78\% & COS  \\  
SDSS\,J093249.57+472523.0  & 95.48  &                    &                   & 17           &      &         &      &      & COS  \\
RZ\,Sge           & 98.32  &                    &                   & 20           & 0.5  & 15\,287 & 663  & 56\% & STIS \\    
CY\,UMa                    & 100.18 &                    &                   & 23           & 0.1  & 15\,232 & 658  & 61\% & STIS \\        
BB\,Ari                    & 101.20 &                    &                   & 24, 25       & 0.2  & 14\,948$\downarrow$ & 632 & 83\% & COS \\ 
DT\,Oct                    & 104.54 &                    &                   & 24, 25       &      &         &      &        & COS \\
IY\,UMa          & 106.43 & $190 \pm 60$       & $0.79 \pm 0.04$ & 26, 27  & 1.0  & 17\,750$\downarrow$ & 1000* & 77\% &  COS \\
SDSS\,J100515.38+191107.9  & 107.60 &                    &                   & 17           & 0.2  & 15\,944 & 723  & 74\% & COS  \\   
RZ\,Leo                    & 110.17 & $340 \pm 110$      &                   & 28, 29       & 0.5  & 15\,014 & 638  & 83\% & COS  \\
CU\,Vel                    & 113.04 & $150 \pm 50$       &                   & 29, 30       & 0.1  & 15\,336 & 668  & 89\% & COS  \\
AX\,For                    & 113.04 & $370^{+20}_{-60}$  &                   & 25, 31       & 1.0  & 16\,571$\downarrow$ & 777 & 68\% & COS  \\
SDSS\,J164248.52+134751.4  & 113.60 &                    &                   & 32           & 1.0  & 17\,710:& 871  & 48\% & COS  \\
QZ\,Ser                    & 119.75 & $460^{+150}_{-110}$&                   & 33           & 0.2  & 14\,481 & 587  & 68\% & STIS \\
DV\,UMa                    & 123.62 & $504 \pm 30$       & $1.098 \pm 0.024$ & 15, 34       & 1.0  & 18\,874 & 182* & 92\% & STIS \\   
IR\,Com                    & 125.34 & $300$              &                   & 22, 35       & 1.0  & 16\,618 & 781  & 86\% & COS  \\    
SDSS\,J001153.08--064739.2 & 144.40 &                    &                   & 36           & 0.1  & 13\,854 & 525  & 63\% & COS  \\ 
OR\,And                    & 195.70 &                    &                   & 25           &      &         &      &      & COS  \\
BB\,Dor                    & 221.90 & $1500 \pm 500$     &                   & 37          &      &         &      &      & COS  \\
SDSS\,J040714.78--064425.1 & 245.04 &                    &                   & 38           & 1.0  & 20\,885 &1104  & 67\% & COS  \\
CW\,Mon                    & 254.30 & $297$              &                   & 25, 39           &      &         &      &      & COS  \\
V405\,Peg                  & 255.81 & $149^{+26}_{-20}$  &                   & 40           &      &         &      &      & COS  \\
HS\,2214+2845              & 258.02 &                    &                   & 41           & 0.5  & 26\,443$\downarrow$ &1436 & 84\% &  COS \\
BD\,Pav                    & 258.19 & $500$              &                   & 42, 43       & 0.5  & 17\,775 & 876  & 92\% & STIS \\
SDSS\,J100658.41+233724.4  & 267.71 & $676 \pm 40$       & $0.78 \pm 0.12$   & 44           & 1.0  & 16\,000 & 1000*& 96\% & COS  \\
HM\,Leo                    & 268.99 & $350$              &                   & 45           &      &         &      &      & COS  \\  
HS\,0218+3229              & 428.02 & $1000$             &                   & 46           & 0.2  & 17\,990 & 893  & 70\% & COS \\ 
HS\,1055+0939              & 541.88 &                    &                   & 25           &      &         &      &       & COS  \\
\bottomrule 
\end{tabular}
\begin{tablenotes}
\item \textbf{Notes.} For each object, its orbital period, distance and white dwarf mass are compiled from the literature. The last columns report the results from this work: metallicity, effective temperatures, the systematic uncertainties arising from the unknown white dwarf mass, the percentage of the white dwarf contribution to the total flux and the instrument used. The white dwarf contribution has been calculated assuming a constant flux as second component. For the four systems highlighted with a star, a precise mass measurement is available. While the uncertainties on the effective temperature of IY\,UMa and SDSS1006 are dominated by the presence of the curtain of veiling gas (Section~\ref{eclipsing}), the uncertainties reported for SDSS1035 and DV\,UMa are those related the unknown nature of the second additional component (Section~\ref{subsec:sec_comp}). In particular, for DV\,UMa, we report the effective temperature obtained assuming log g = 8.78 (Section~\ref{subsec:degeneracy}). The values flanked by a downwards arrow represent upper limits for the temperature of the systems. The values flanked by a colon represent unreliable effective temperature and are not considered in the discussion.\\
\item \textbf{References.} 
(1) \citet{V485Cen}, 
(2) \citet{GWLib_dist}, (3) \citet{GWLib_period} 
(4) \citet{sdss1435}, 
(5) \citet{otj2138}, 
(6) \citet{V844Her}, 
(7) \citet{sdss0137_1}, (8) \citet{0137}, 
(9) \citet{1238}, 
(10) \citet{pucma}, 
(11) \citet{V1108Her}, (12) \citet{ASAS0025}, 
(13) \citet{Templeton}, 
(14) \citet{1035}, (15) \citet{SDSS1035_2011} 
(16) \citet{ccscl1}, 
(17) \citet{period_minimum}, 
(18) \citet{1rxsj10510}, (19) \citet{1rxsj10510_distance}, 
(20) \citet{mruma}, 
(21) \citet{UVPer}, 
(22) \citet{Patterson}, 
(23) \citet{cyuma}, 
(24) \citet{bbari}, (25) \citet{RitterKolb}, 
(26) \citet{iyuma}, (27) \citet{iyuma_dist}, 
(28) \citet{rzleo}, (29) \citet{Mennickent2},  
(30) \citet{cuvel_period}, 
(31) \citet{AXFor}, 
(32) \citet{sdss1642}, 
(33) \citet{QZSer}, 
(34) \citet{Feline}, 
(35) \citet{ircom}, 
(36) \citet{sdss0011}, 
(37) \citet{bbdor}, 
(38) \citet{0407}, 
(39) \citet{CWMon_dist}, 
(40) \citet{V405Peg}, 
(41) \citet{HS2214}, 
(42) \citet{bdpav_period}, (43) \citet{BDPav}, 
(44) \citet{SDSS1006}, 
(45) \cite{hmleo}, 
(46) \citet{hs0218} 
\end{tablenotes}
\end{table*}

\section{Discussion}\label{sec:Discussion}
\subsection{$\log g$ correction of published $T_\mathrm{eff}$ values}\label{subsec:correzione}
The two most commonly used techniques to measure effective temperatures of CV white dwarfs are spectroscopy, that is by fitting ultraviolet or optical spectra with white dwarf atmosphere models, and photometry, i.e. from the analysis of light curves to study the white dwarf ingress and egress in eclipsing systems. \citeauthor{Tow_and_Boris2009} (\citeyear{Tow_and_Boris2009}, hereafter TG09) present an analysis of CV white dwarf effective temperatures and selected only those systems for which a reliable $T_\mathrm{eff}$ determination is available. They consider a measurement unreliable when obtained from (i) spectra in which the  white dwarf could not be unambiguously detected or, (ii) in the case of eclipse light curve analyses, oversimplified models or data of poor quality were used. Following these criteria, TG09 compiled 43 systems with a reliable temperature measurement (see their table\,1). 

Among the 43 measurements from TG09, 15 were obtained from light curve analyses (which also delivers the white dwarf mass) or have a white dwarf mass measurements independent from the spectral fit. For the remaining 28 objects an independent mass determination was not available and the white dwarf $T_\mathrm{eff}$ was evaluated via spectroscopic analyses, following methods similar to the one described here, but assuming $\log g = 8.00$. As discussed in Section~\ref{sec:DataAnalysis}, this assumption does not reflect the average mass in CV white dwarfs, $M \simeq 0.8\,\mathrm{M}_\odot$, corresponding to $\log g  = 8.35$. To combine our results with those of TG09, we need therefore to evaluate the corresponding $T_\mathrm{eff}$ for $\log g = 8.35$ for those 28 objects. 

The systematic correction that we need to apply is $\Delta T_\mathrm{eff} = T_\mathrm{eff}\,(\log\,g=8.35) - T_\mathrm{eff}\,(\log\,g=8.00)$, which is the opposite of the quantity $\Delta^- T_\mathrm{eff}$ we calculated in Section~\ref{subsec:unknown_mass}. Using the relationship $- \Delta^- T_\mathrm{eff} = 1417 \times \log (T_\mathrm{eff} \times 0.0001046)$, we corrected the $T_\mathrm{eff}$ values for those 28 systems from TG09 which did not have an independent mass measurement to the average CV white dwarf mass (i.e. $\log\,g=8.35)$, therefore enabling a consistent combinations of these values with the 36 temperatures that we derived for the COS+STIS data.

\begin{figure*}
 \centering
 \subfloat{%
  \label{AXFor}
  \includegraphics[angle=-90,width=0.48\textwidth]{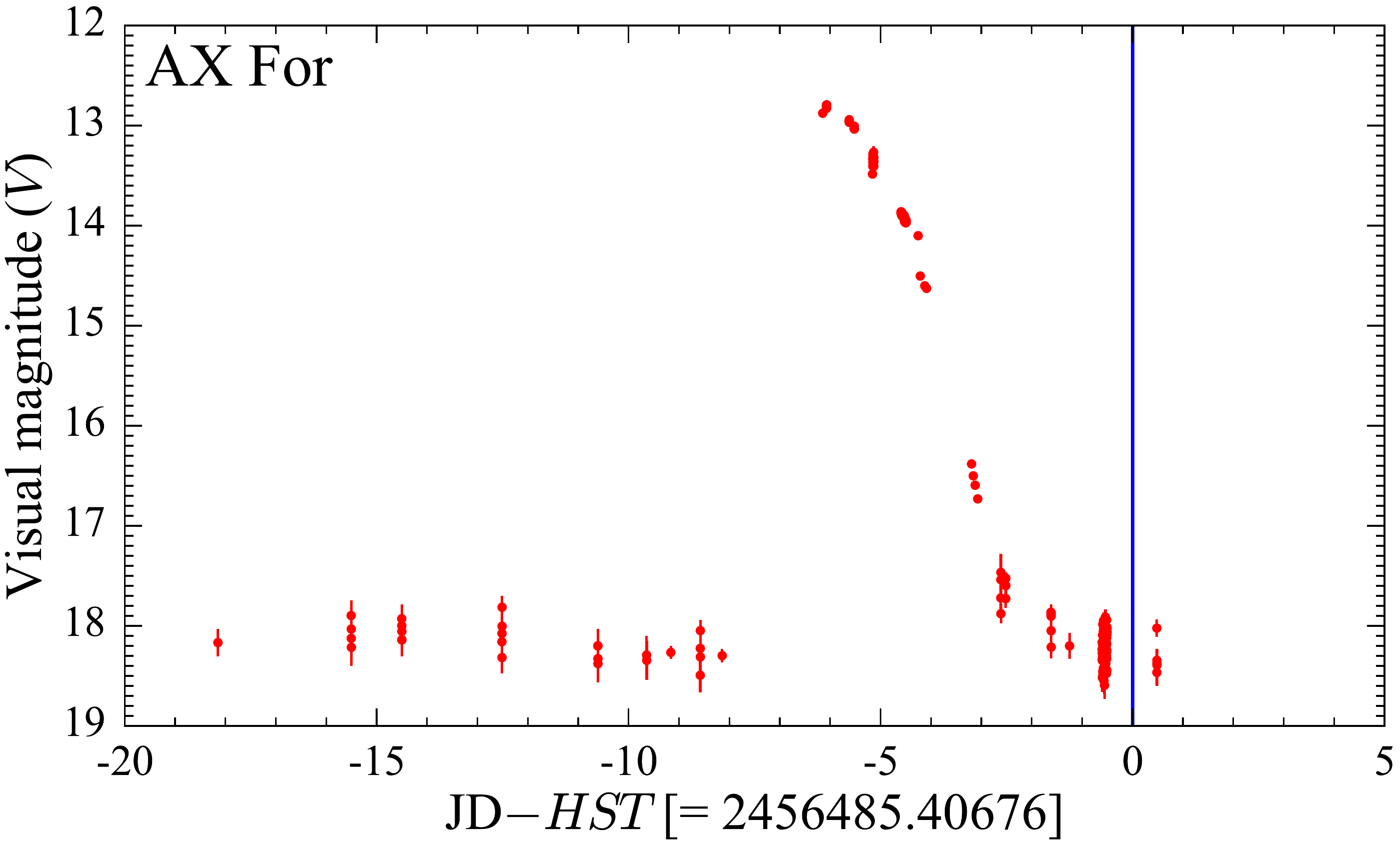} \qquad %
  }
 \subfloat{%
  \label{BBAri}
  \includegraphics[angle=-90,width=0.48\textwidth]{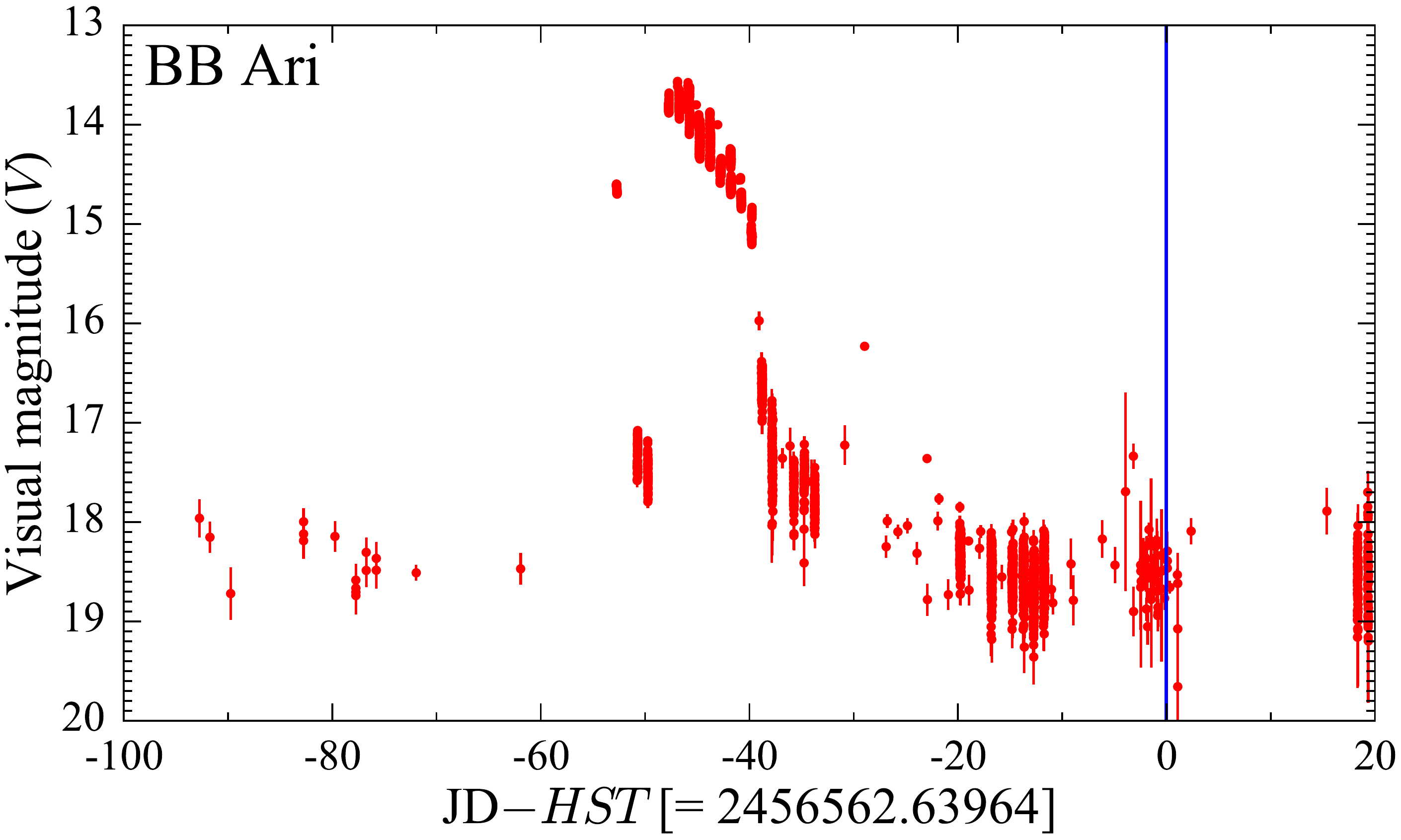}
  }\\%
 \subfloat{%
  \label{HS2214}
  \includegraphics[angle=-90,width=0.48\textwidth]{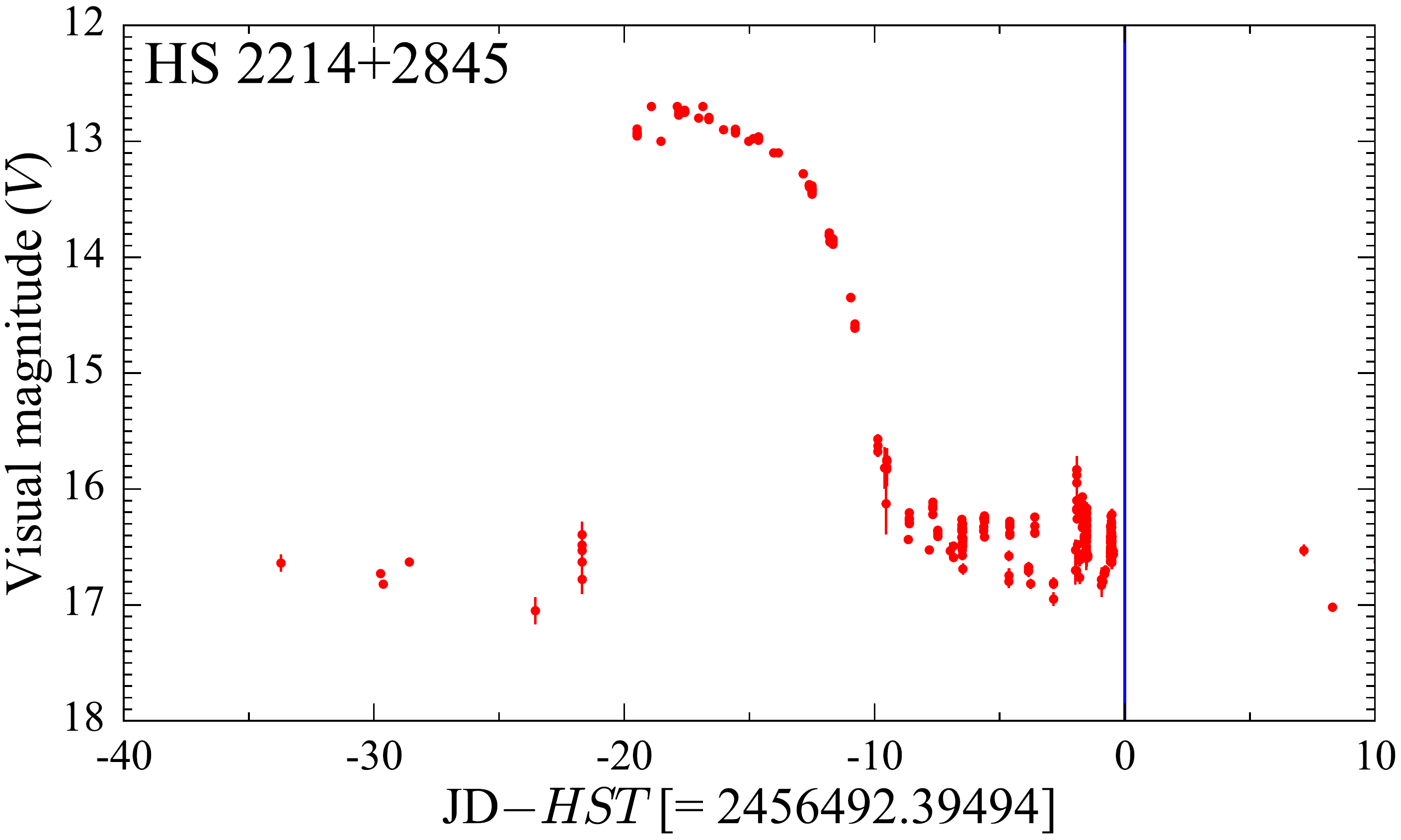} \qquad
  }
 \subfloat{%
  \label{IYUMa}
  \includegraphics[angle=-90,width=0.48\textwidth]{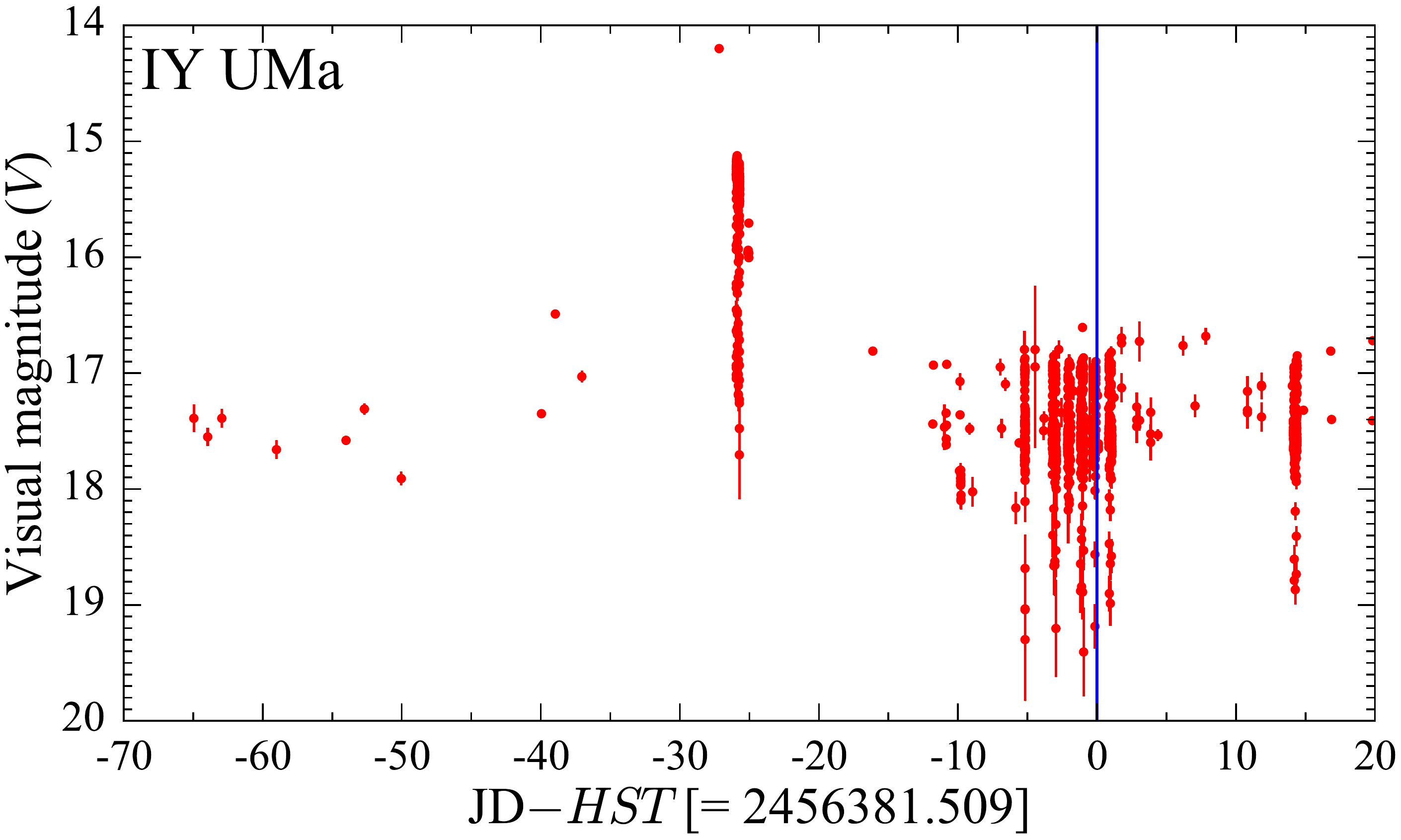}  %
  }\\
 \subfloat{%
  \label{MRUMa}
  \includegraphics[angle=-90,width=0.48\textwidth]{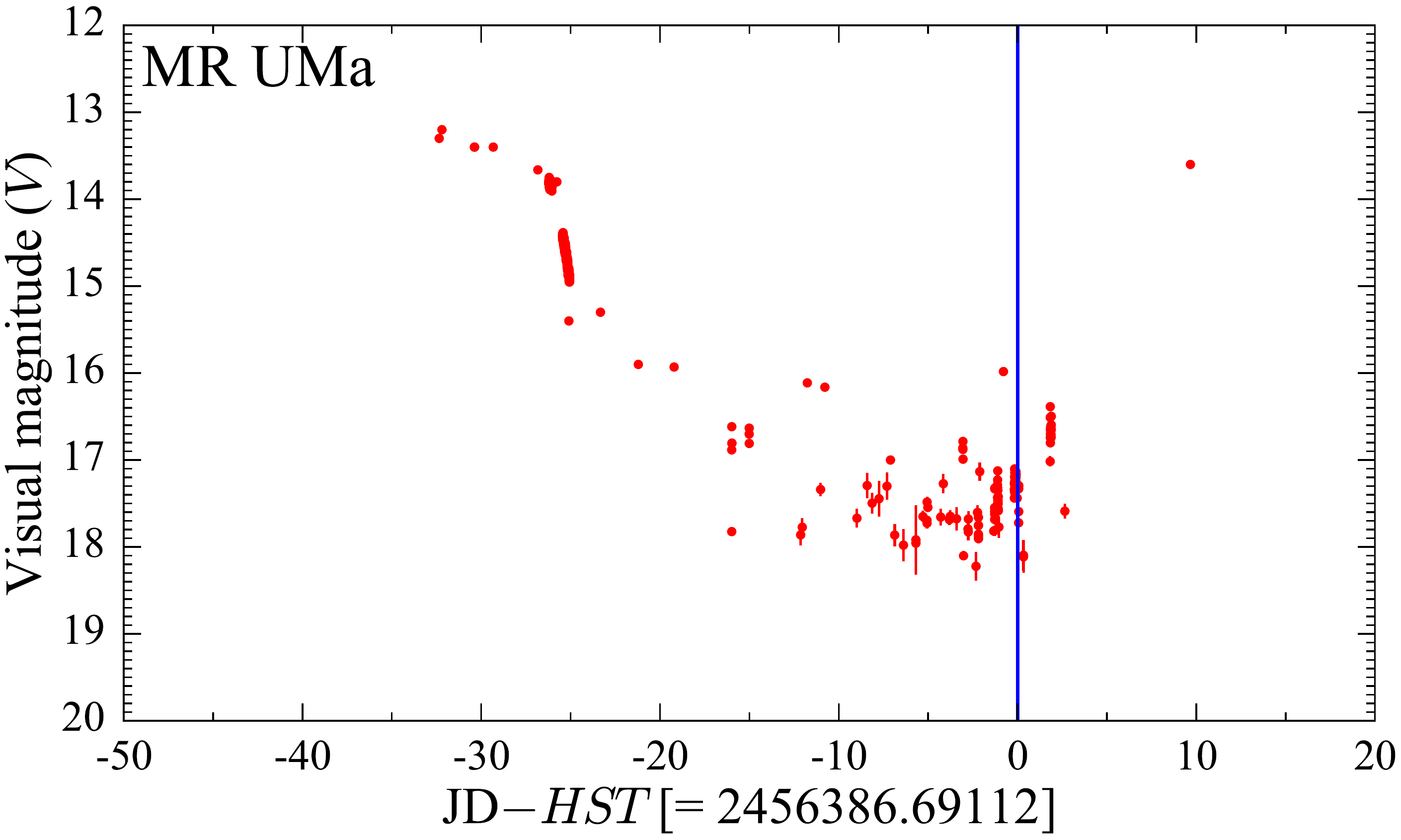} \qquad
  }
 \subfloat{%
  \label{QZSer}
  \includegraphics[angle=-90,width=0.48\textwidth]{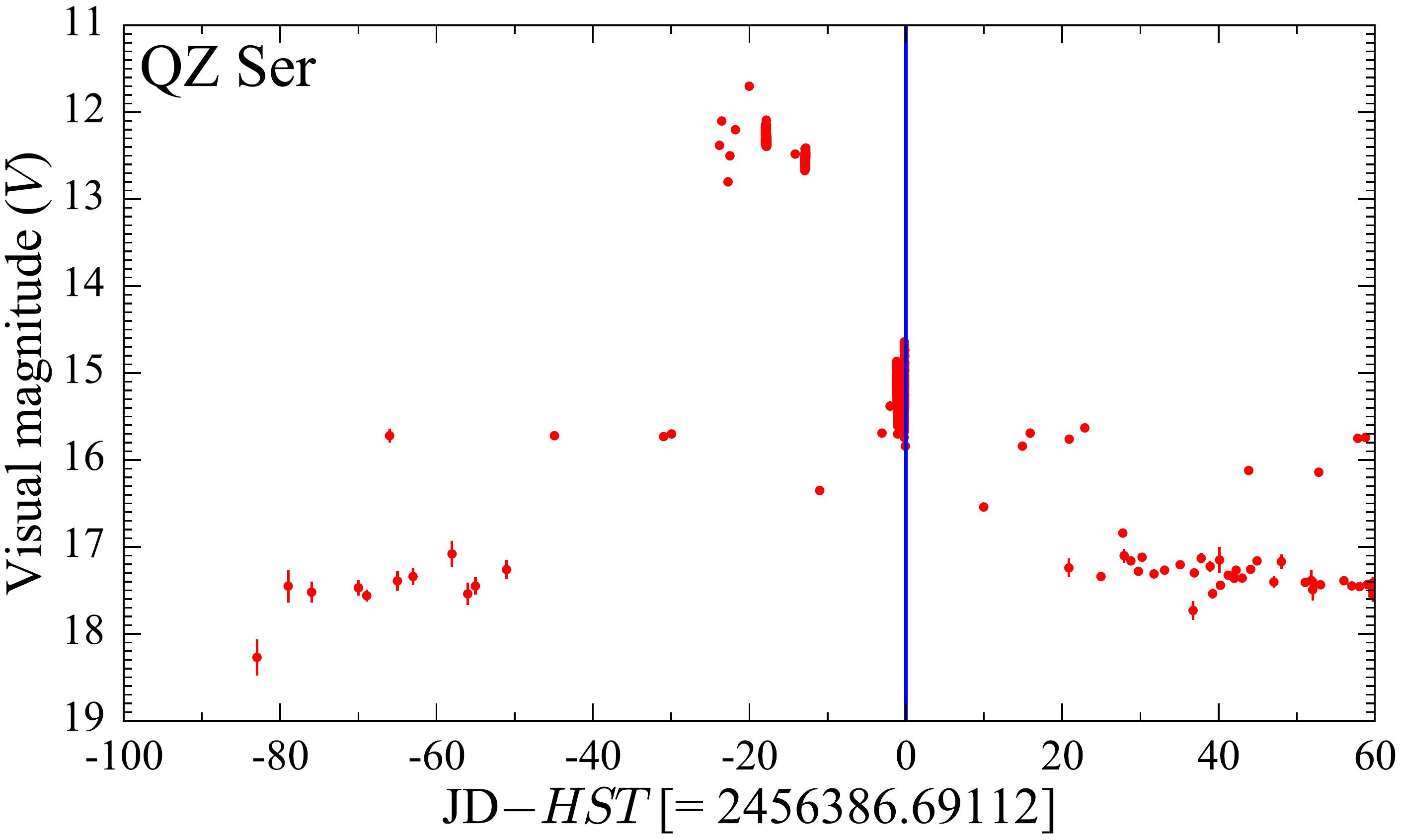} %
  }\\
 \subfloat{%
  \label{V485Cen}
  \includegraphics[angle=-90,width=0.48\textwidth]{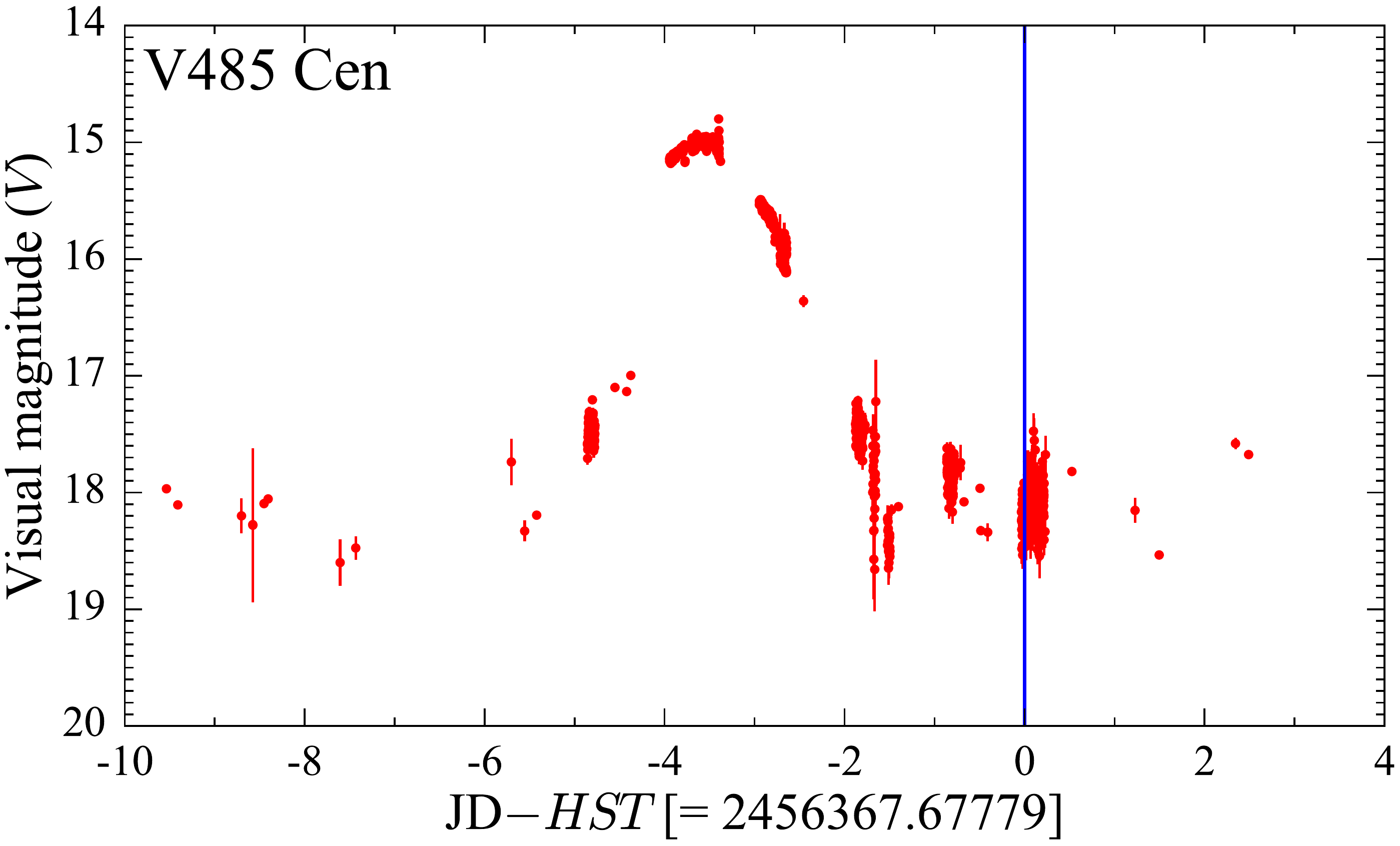}
  }
\caption{Light curves for seven targets in our sample which experienced an outburst within two months before the \textit{HST} observations (indicated with the blue lines). The \textit{HST} observation date has been subtracted from the Julian date. Note the different time range on the x--axis. The data have been retrieved from the AAVSO database. }%
\label{lightcurves1}
\end{figure*}
 
\subsection{Notes on individual objects}\label{subsec:confronto}
\subsubsection{Systems observed close to an outburst}
As explained in Section~\ref{sec:Obs}, if a CV experienced an outburst shortly before the ultraviolet observation, its spectrum could be contaminated by disc emission. Furthermore the white dwarf photosphere is heated by the increased infall of material and different subtypes of CVs have different cooling time scales on which they return to their quiescent temperature (\citeauthor{Sion1995} \citeyear{Sion1995}, \citeauthor{cooling} \citeyear{cooling}). A spectroscopic analysis of these systems can therefore provide only an upper limit on the white dwarf effective temperature.

\begin{figure*}
 \centering
    \subfloat{%
  \includegraphics[angle=-90,width=0.48\textwidth]{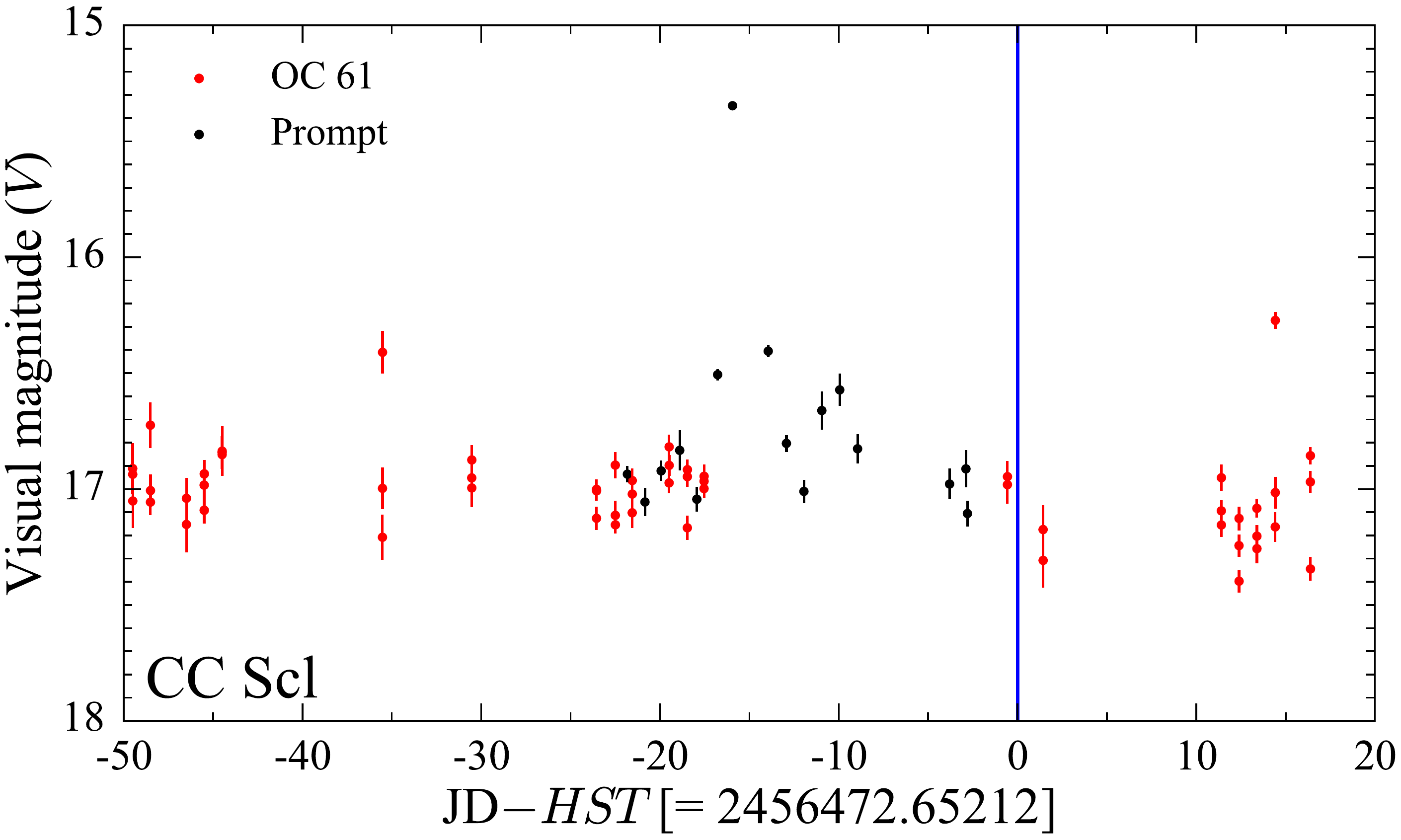} \qquad}
   \subfloat{%
  \includegraphics[angle=-90,width=0.48\textwidth]{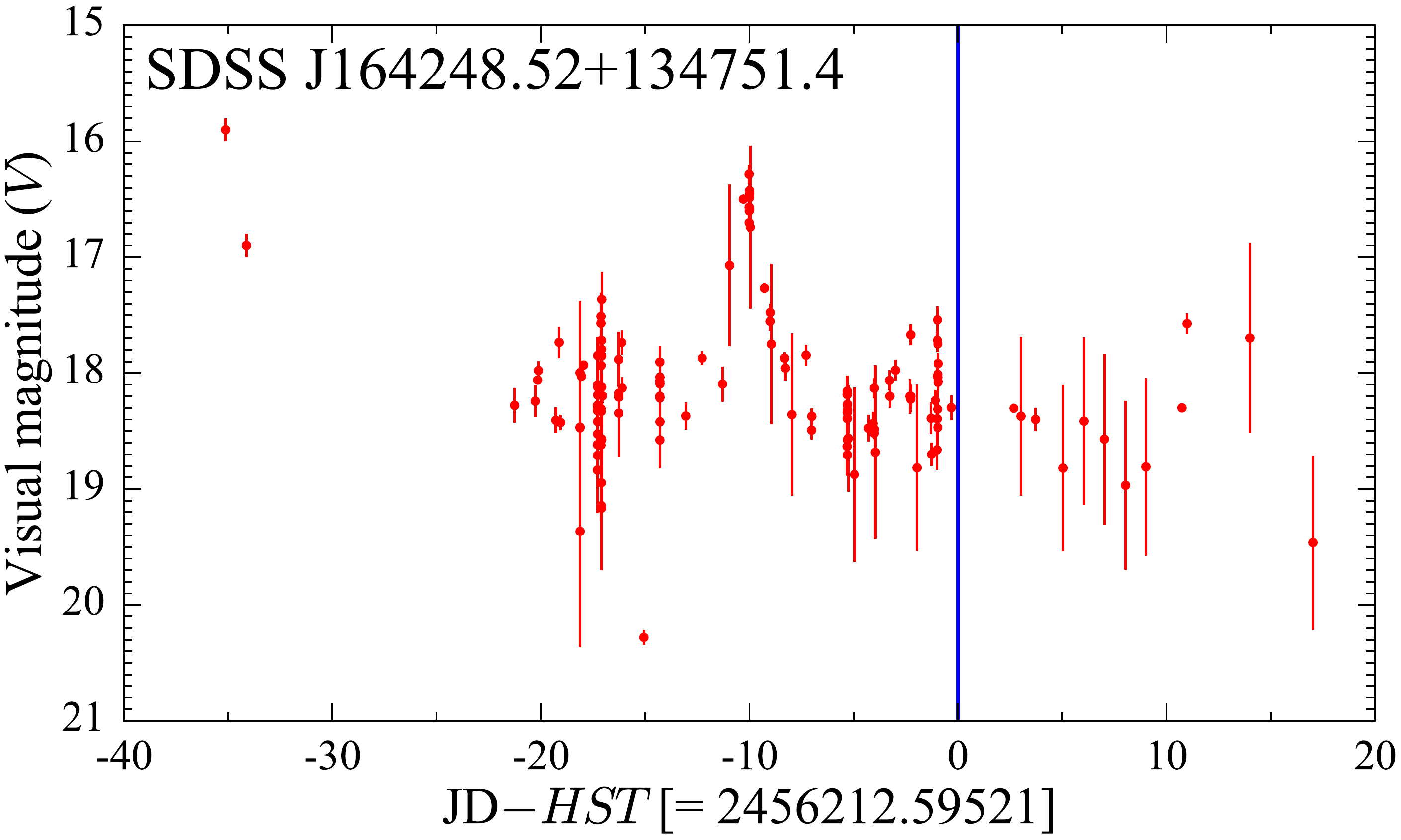} \qquad}

\caption{\emph{Left}: Prompt and OC61 light curve for CC\,Scl. \emph{Right}: AAVSO light curve for SDSS\,J164248.52+134751.4. The blue line represents the date of the \textit{HST} observations which has been subtracted from the Julian date.}%
\label{CCSCl-SDSS1642}
\end{figure*}

For each system, we inspected the light curves retrieved from the AAVSO web site plus a total of additional $\sim\,2000$ images (collected with the Mount John University Observatory OC61 Telescope, the New Mexico State University Observatory TMO61 telescope and the Prompt telescopes located in Chile), covering two months before the \textit{HST}/COS observations. We found that nine systems in which the white dwarf dominates the ultraviolet flux went into outburst within this time interval: AX\,For, BB\,Ari, HS\,2214+2845, IY\,UMa, MR\,UMa, QZ\,Ser, V485\,Cen (Figure~\ref{lightcurves1}), CC\,Scl (Figure~\ref{CCSCl-SDSS1642}, left), and SDSS\,J164248.52+134751.4 (SDSS1642, Figure~\ref{CCSCl-SDSS1642}, right). QZ\,Ser and V485\,Cen have STIS observations (Section~\ref{with_COS_STIS1}) from which we determined the quiescent temperature while, for the remaining systems, the results that we report here only represent an upper limit for their quiescent $T_\mathrm{eff}$.

\begin{figure}
 \centering
  \subfloat{%
  \includegraphics[angle=-90,width=0.48\textwidth]{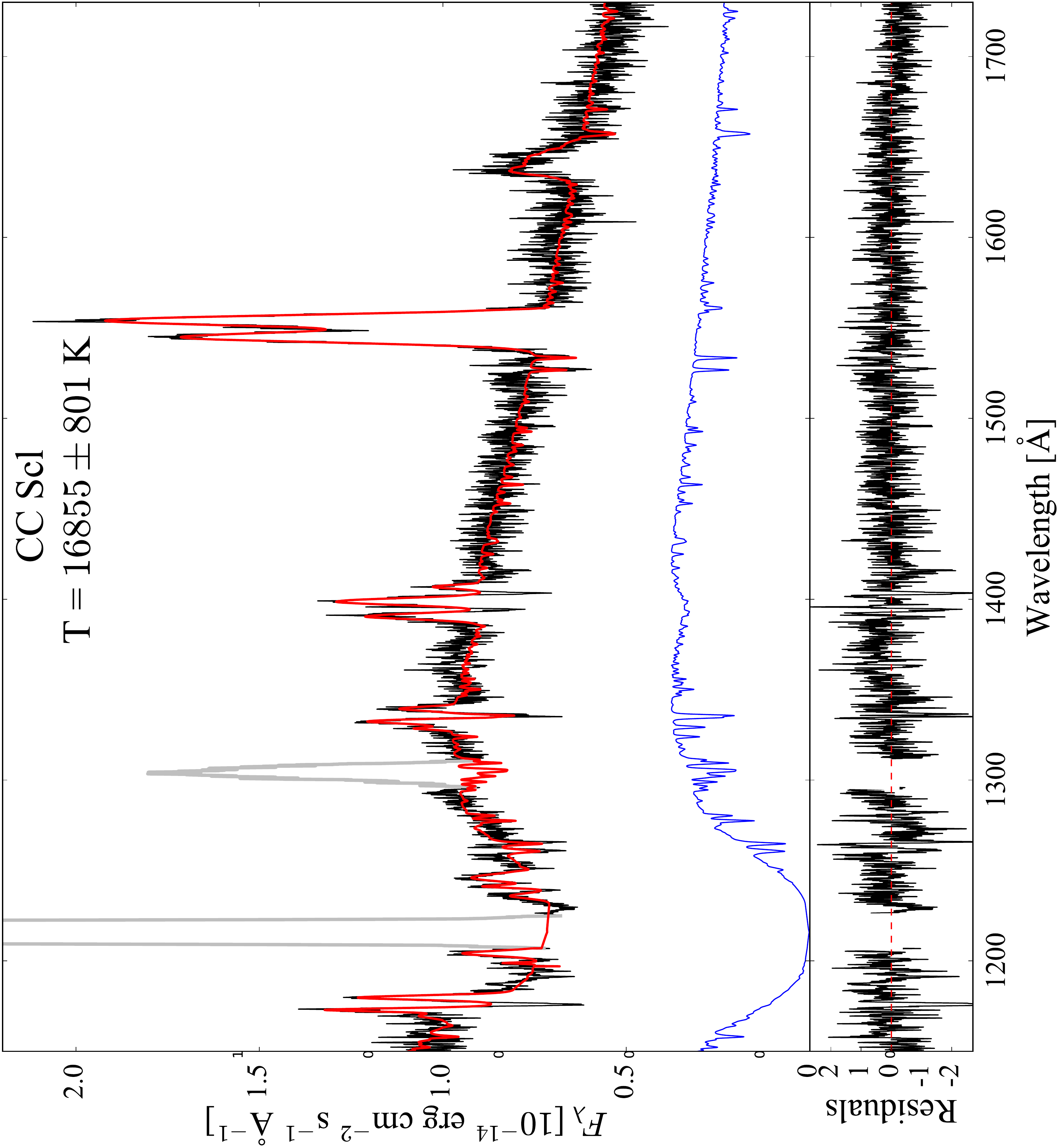}
  }\\
 \subfloat{%
  \includegraphics[angle=-90,width=0.48\textwidth]{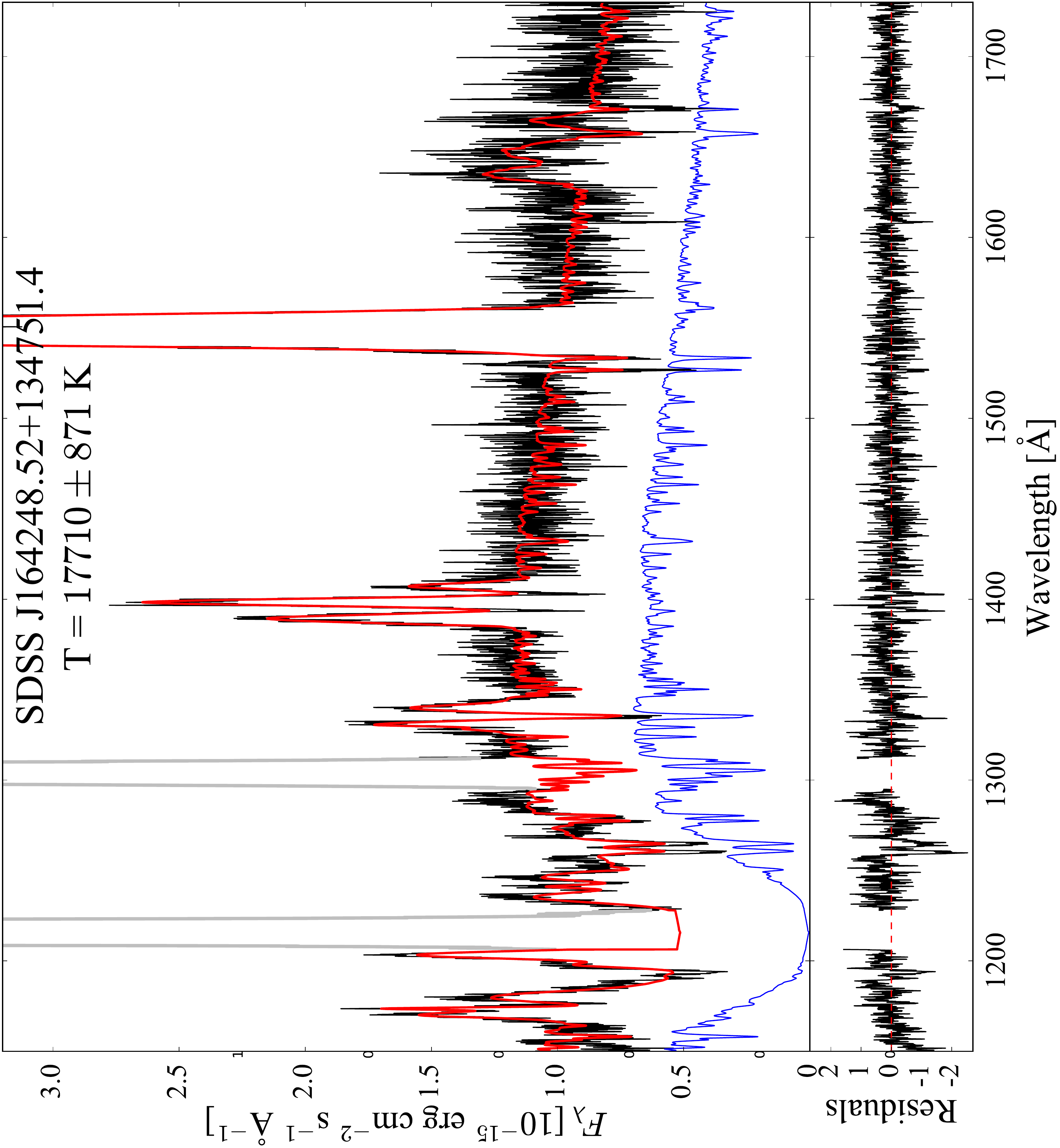}
  }
\caption{\textit{HST}/COS spectra (black) of CC\,Scl (top) and SDSS\,J164248.52+134751.4 (bottom) along with the best--fit model (red) assuming a blackbody second component. The white dwarf emission (blue) only contributes $\simeq 35$ per cent (CC\,Scl) and $\simeq 50$ per cent (SDSS1642) of the total flux. The geocoronal emission lines of Ly$\alpha$ ($1216\,$\AA) and \ion{O}{i} ($1302\,$\AA) are plotted in grey.}%
\label{CCSCl-SDSS1642_spectrum}
\end{figure}

For CC\,Scl and SDSS1642, we found that the white dwarf contributes only $\simeq 35$ per cent (Figure~\ref{CCSCl-SDSS1642_spectrum}, top) and $\simeq 50$ per cent of the total flux (Figure~\ref{CCSCl-SDSS1642_spectrum}, bottom), respectively. Owing to this strong disc contamination, the derived effective temperatures do not fulfil the requirement of a reliable measurement as defined in TG09, and we do not include them in the following analysis. Furthermore CC\,Scl is an Intermediate Polar (IP) and its response to accretion and disc instabilities is more complicated than that of a non--magnetic CV. This system will be analysed in more detail by Szkody et al. (2016, in preparation).

\begin{table}
 \centering
 \caption{Results for BD\,Pav, QZ\,Ser and V485\,Cen derived from the COS and the STIS data.}\label{table_COS_STIS}
  \begin{tabular}{@{}lcc@{}}
  \toprule 
   Object & Instrument & $T_\mathrm{eff}$ (K) \\
\midrule
BD\,Pav               &    STIS             &    $17\,775 \pm 876$ \\
                      &     COS             &    $18\,915 \pm 964$ \\
\midrule
QZ\,Ser               &    STIS             &    $14\,481 \pm 587$ \\
                      &     COS             &    $15\,425 \pm 676$ \\
\midrule
V485\,Cen             &    STIS             &    $15\,200 \pm 655$ \\
                      &     COS             &    $16\,208 \pm 746$ \\
\bottomrule
\end{tabular}
\end{table}

\subsubsection{Systems with COS and STIS observations}\label{with_COS_STIS1}
\begin{figure}
 \centering
 \includegraphics[angle=-90,width=0.48\textwidth]{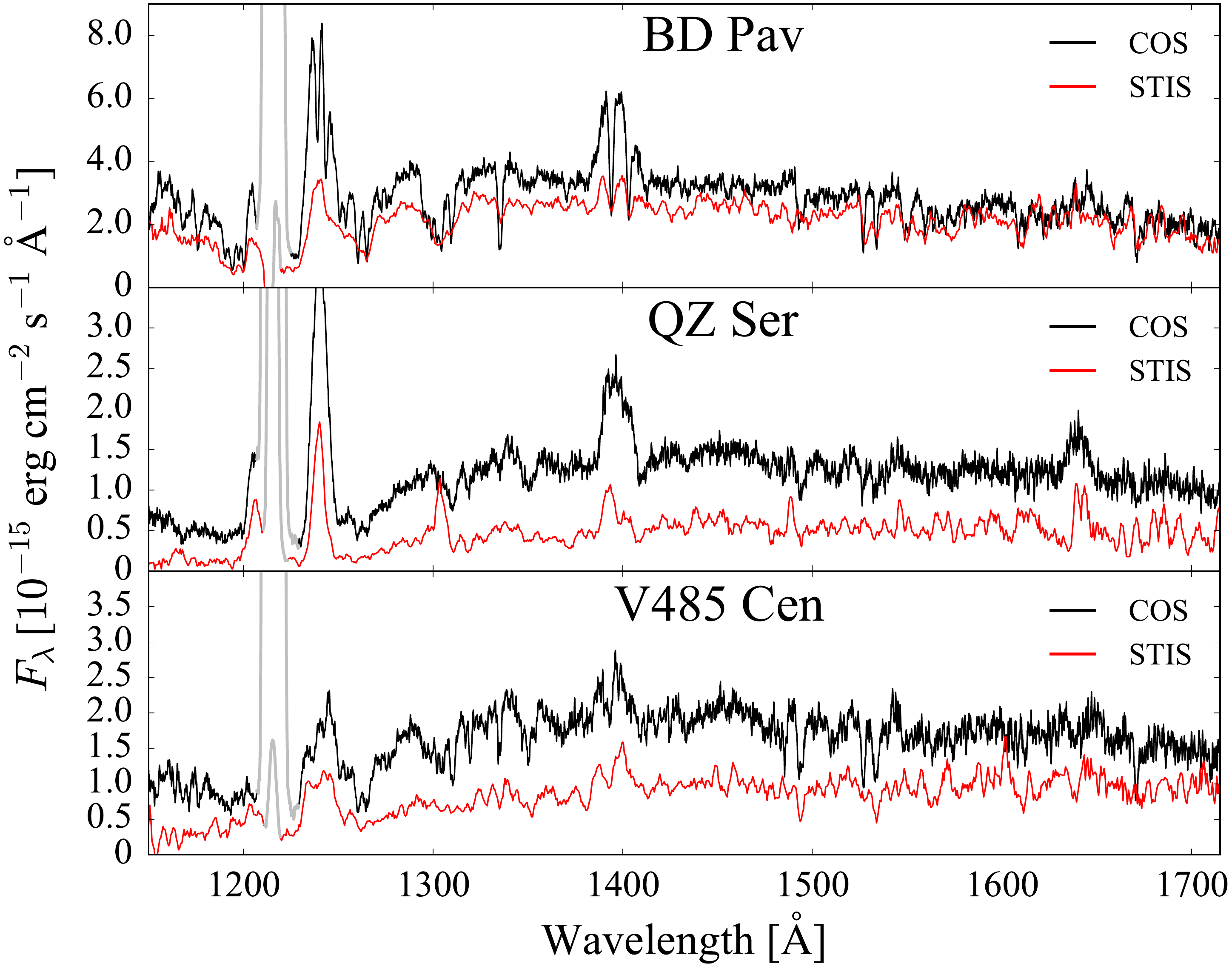}
 \caption{\textit{HST}/COS spectra (black) in comparison with the \textit{HST}/STIS spectra (red) of BD\,Pav, QZ\,Ser and V485\,Cen. The geocoronal Ly$\alpha$ emission ($1216\,$\AA) is plotted in grey.}\label{fig_COS_STIS} 
\end{figure}

QZ\,Ser and V485\,Cen have been observed both with COS and STIS (Table~\ref{table_COS_STIS}). Inspecting their spectra (middle and bottom panel of Figure~\ref{fig_COS_STIS}), the COS observations have a higher flux compared to the STIS observations, which is also reflected in their effective temperatures; the STIS data return a lower value (QZ\,Ser: $T_\mathrm{eff} = 14\,481 \pm 587\,\mathrm{K}$, V485\,Cen: $T_\mathrm{eff} = 15\,200 \pm 655\,\mathrm{K}$) compared to the COS data (QZ\,Ser: $T_\mathrm{eff} = 15\,425 \pm 676\,\mathrm{K}$, V485\,Cen: $T_\mathrm{eff} = 16\,208 \pm 746\,\mathrm{K}$). This difference is not related to difference in the calibration of the two instruments (see Section~\ref{subsec:additional_systematics}) and it is rather expected given that the magnitudes derived from the STIS acquisition images suggest that these systems were observed by STIS during the quiescent phase (see Section~\ref{subsec:STIS_Obs}), whereas the COS spectra were obtained relatively shortly after an outburst (Figure~\ref{lightcurves1}). Since the temperatures derived from the STIS spectra are representative of the quiescent properties of these systems, we use these values in the discussion below.

\begin{figure*}
 \centering
   \subfloat{%
  \includegraphics[angle=-90,width=0.48\textwidth]{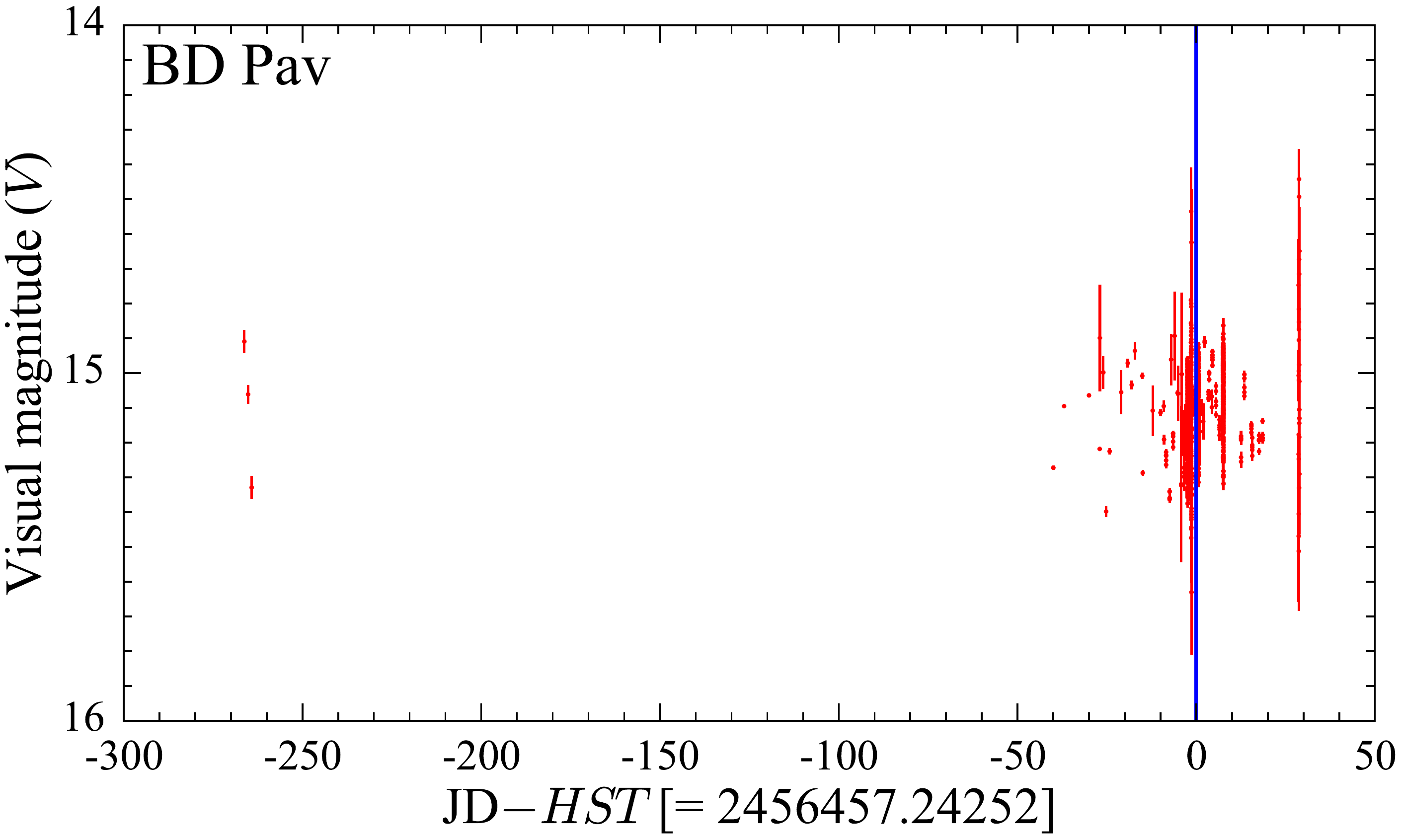} \qquad %
  }
 \subfloat{%
  \includegraphics[angle=-90,width=0.48\textwidth]{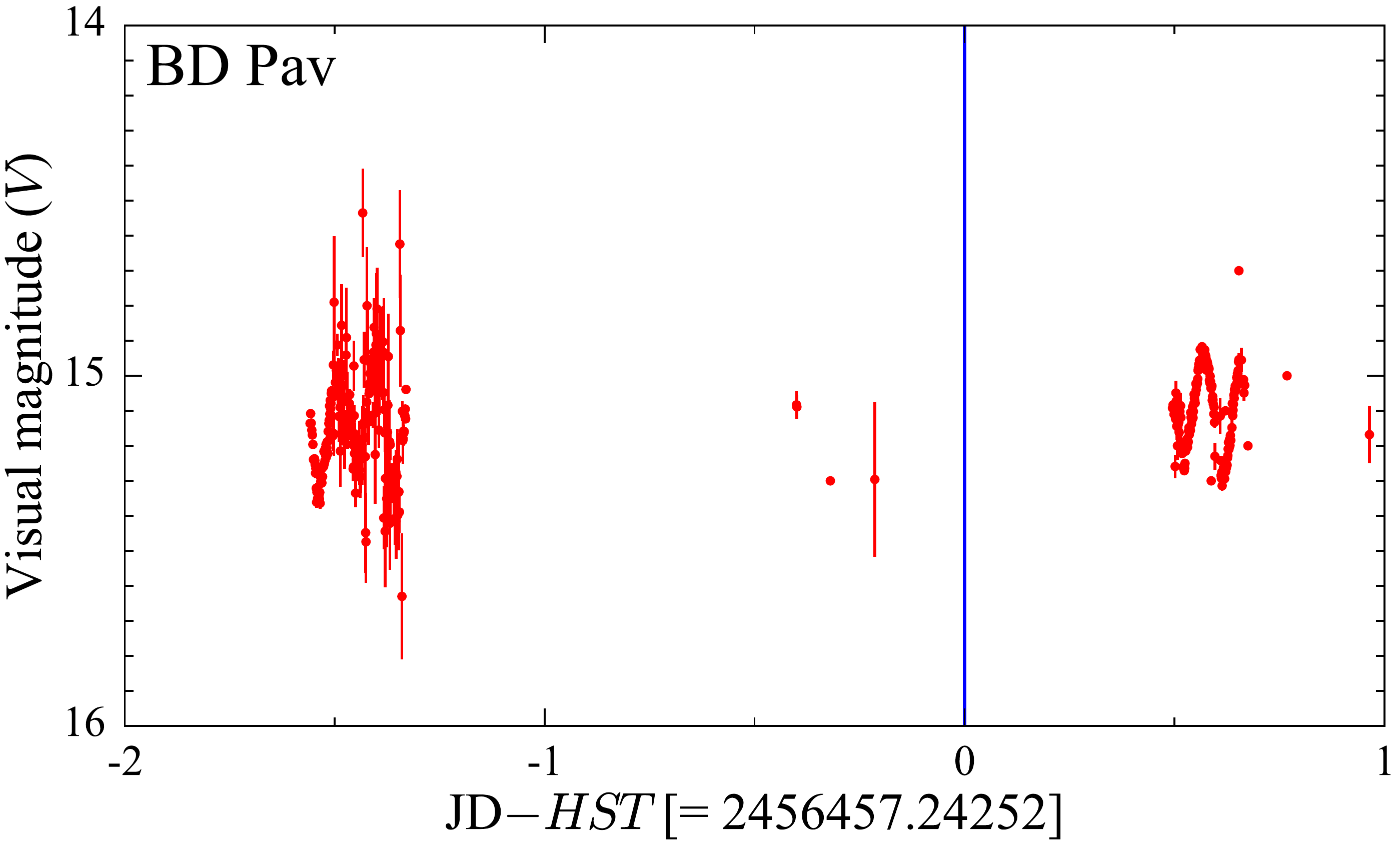}
  }
\caption{\emph{Left}: light curves for BD\,Pav over the period June 2012 to July 2013. \emph{Right}: close up of the days before the \textit{HST} observations (highlighted with the blue line), which has been subtracted from the Julian date. The system only shows ellipsoidal variations. The data have been retrieved from the AAVSO database.}%
\label{BDPav_lightcurve}
\end{figure*}

COS and STIS observations are available for one additional system: BD\,Pav (top panel of Figure~\ref{fig_COS_STIS}). The difference observed among the two data sets (STIS: $T_\mathrm{eff} = 17\,775 \pm 876\,\mathrm{K}$, COS: $T_\mathrm{eff} = 18\,915 \pm 964\,\mathrm{K}$) could be explained by its eclipsing nature if the STIS snapshot observations were obtained (partly) during the eclipse phase, resulting in a lower observed flux. Unfortunately, a reliable ephemeris is not available for this object and we cannot verify this hypothesis. On the other hand, the two spectra are characterised by a different slope and the derived effective temperatures differ by $\simeq 1\,200 \,\mathrm{K}$, which could suggest that an outburst occurred before the COS observations. Since BD\,Pav is a U\,Gem type CV, its outburst frequency and cooling time are expected to be similar to those of U\,Gem itself, $\sim\,120$ and $\sim\,60$ days respectively \citep{UGem_cooling}, i.e. BD\,Pav is expected to experience a few outbursts every year. 
In fact, AAVSO observations of BD\,Pav (available since 1987) show a rough average of one outburst per year, with up to two outbursts observed in years of intense monitoring. No outburst is clearly identifiable in the AAVSO light curve for the period 2012/2013 (Figure~\ref{BDPav_lightcurve}) but, owing to the relatively sparse sampling, we cannot completely rule out this hypothesis and therefore we consider in the discussion below the lower effective temperature derived from the STIS data.

\subsection{Systems with previous $T_\mathrm{eff}$ measurements}
Among the 36 systems in our sample, four CVs have a previous $T_\mathrm{eff}$ determination in the TG09 sample: SDSS\,J103533.02+055158.4 (SDSS1035), CU\,Vel, GW\,Lib and DV\,UMa, which can be used for comparison with our results. Among them, SDSS1035 and DV\,UMa have also a previous mass determination and we analyse them in Section~\ref{subsec:degeneracy}.

\begin{figure}
 \centering
 \includegraphics[angle=-90,width=0.48\textwidth]{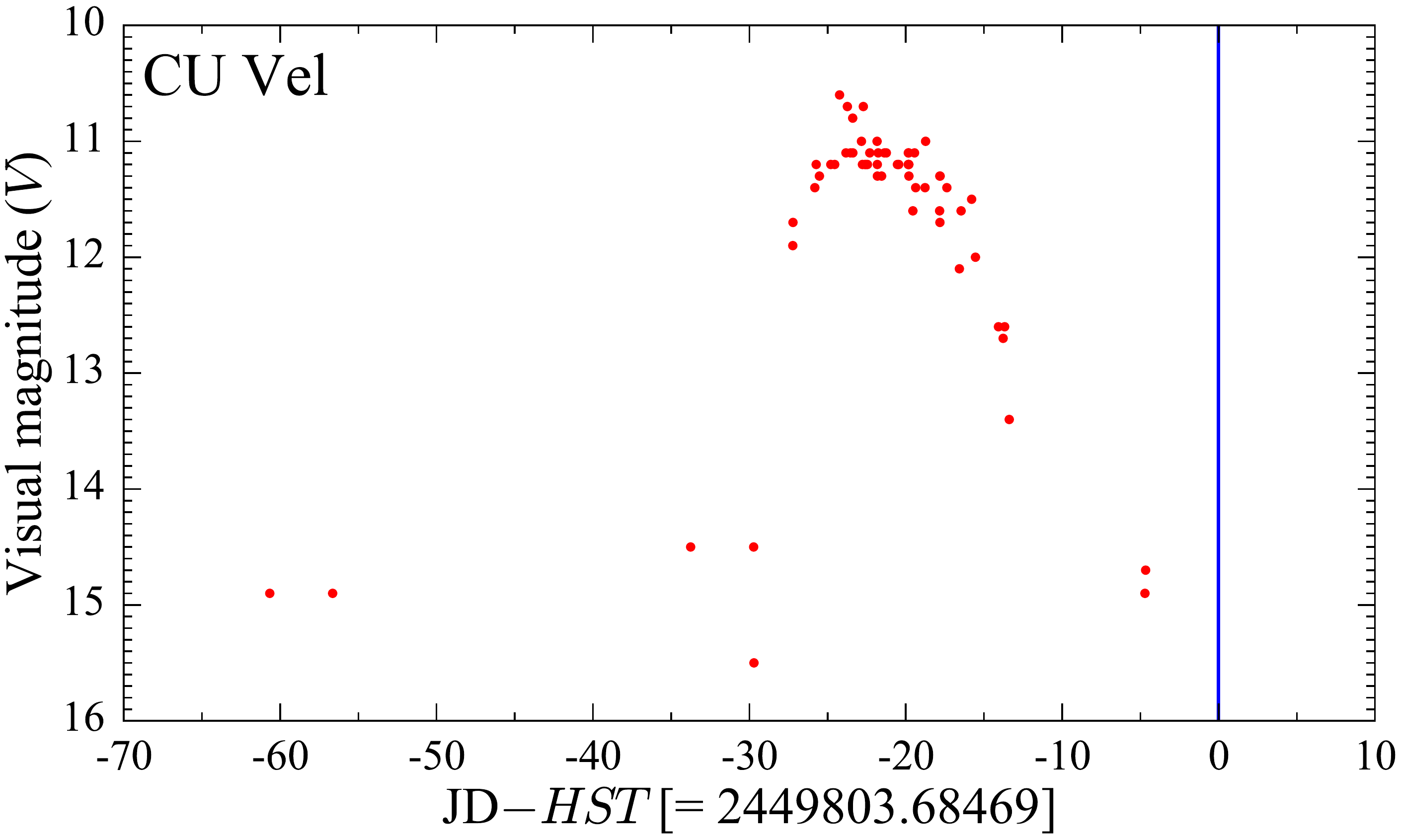}
 \caption{Light curve for CU\,Vel over the period January--April 1995. The blue line represents the date of the \textit{IUE} observations,  which has been subtracted from the Julian date. The data have been retrieved from the AAVSO database.}\label{light_curve_Cuvel} 
\end{figure}

\begin{figure}
 \centering
 \includegraphics[angle=-90,width=0.48\textwidth]{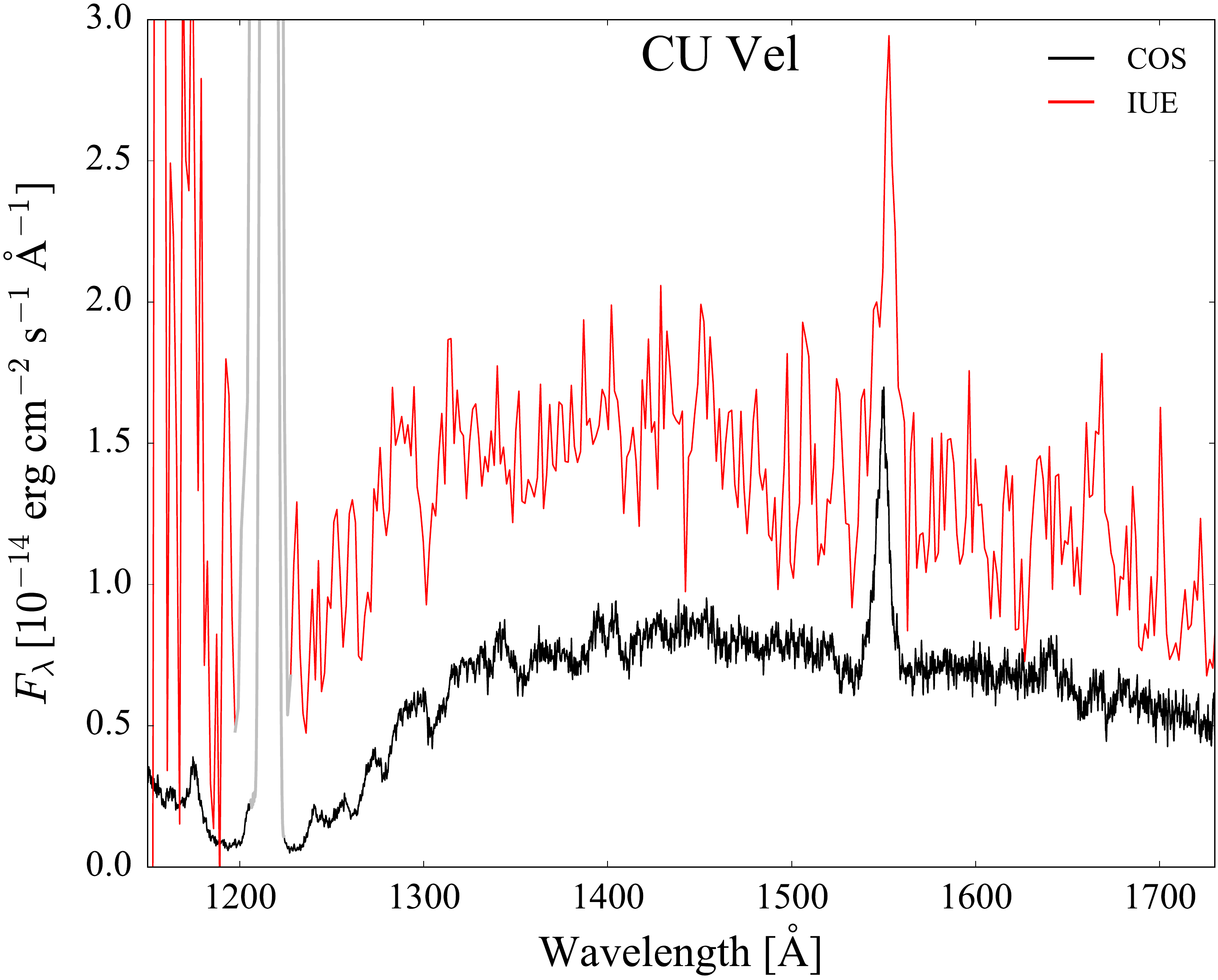}
 \caption{\textit{HST}/COS spectrum (black, $T_\mathrm{eff} = 15\,336 \pm 668\,\mathrm{K}$) in comparison with the \textit{IUE} spectrum (red, $T_\mathrm{eff} = 18\,500 \pm 1\,500\,\mathrm{K}$, \citeauthor{CUVel} \citeyear{CUVel}) of CU\,Vel. The geocoronal Ly$\alpha$ emission ($1216\,$\AA) is plotted in grey.}\label{CUVEL_IUE} 
\end{figure}

\subsubsection{Previous ultraviolet measurements}
\paragraph{CU\,Vel}\label{CUVel_section}\mbox{}\\
CU\,Vel is located in the Vela constellation at a distance of $150 \pm 50\,\mathrm{pc}$ \citep{Mennickent2}. Although Vela lies in the galactic plane, given the low distance to the system, CU\,Vel is not likely severely affected by interstellar reddening.
To verify this, we retrieved from the archive the \textit{IUE} spectra of CU\,Vel, which cover the wavelength range $\simeq 1150 -  3300\,$\AA. Inspecting the data, we did not detect the main signature of interstellar dust absorption, i.e. the bump at $\simeq 2175\,$\AA. To estimate the colour excess, we de--reddened the spectrum assuming $E(B-V)$ in the range $0.01$ to $0.5$. We noticed that for $E(B-V) \gtrsim 0.02$, a positive bump started to be visible in the spectrum, suggesting that we were overestimating the extinction. The low resolution and low signal--to--noise ratio of the \textit{IUE} data did not allow us to accurately measure the colour excess, and we assumed $E(B-V)= 0.02$ as an upper limit, well below the $E(B-V) \simeq 0.1$ threshold defined in Section~\ref{subsec:reddening}, concluding that the results for CU\,Vel are not affected in an appreciable way by interstellar dust absorption.

From the same \textit{IUE} spectrum, \citet{CUVel} derived $T_\mathrm{eff} = 18\,500 \pm 1\,500\,\mathrm{K}$. 
This temperature is significantly higher than $T_\mathrm{eff} = 15\,336 \pm 668\,\mathrm{K}$ derived here from the COS data. To rule out any possible calibration differences among the two instruments, we compared the \textit{IUE} and COS spectra for the flux standard white dwarf WD\,0308--565. The flux levels of the different datasets are in good agreement, with an average difference of $\simeq 3$ per cent, and therefore the difference in the effective temperatures are not related to instrument calibration issues. Inspecting the AAVSO light curve of CU\,Vel, we find that the system experienced an outburst 20 days before the \textit{IUE} observations (Figure~\ref{light_curve_Cuvel}), which explains the higher ultraviolet flux observed with \textit{IUE} compared to that of our new \textit{HST}/COS data (Figure~\ref{CUVEL_IUE}), and the higher temperature determined by \citet{CUVel}.

\paragraph{GW\,Lib}\mbox{}\\
GW\,Lib has been observed with \textit{HST} before (January 2002, \citeauthor{SzkodyGWLib2002} \citeyear{SzkodyGWLib2002}) and after (March 2010 and April 2011, \citeauthor{GWLib} \citeyear{GWLib}) its superoutburst in April 2007. In these works, the authors assume $\log\,g = 8.0$ and $\log\,g = 8.7$, respectively, so we cannot directly compare our and their results.\par
\citet{Odette} present a new analysis of all these \textit{HST} observations. They re--fit the data assuming $\log\,g = 8.35$ and find $T_\mathrm{eff} = 14\,695^{+13}_{-11}\,\mathrm{K}$ for the 2002 dataset, $T_\mathrm{eff} = 17\,980\,\pm 14\,\mathrm{K}$ and $T_\mathrm{eff} = 15\,915\,\pm 9\,\mathrm{K}$ for the 2010 and 2011 observations, respectively.\par
GW\,Lib is a very peculiar system and a comparison between our ($T_\mathrm{eff} = 16\,995 \pm 813\,\mathrm{K}$) and these previous temperature measurements has to be considered with some caution. Indeed, GW\,Lib not only hosts a pulsating white dwarf but it also showed variations in $T_\mathrm{eff}$ of $\simeq\,3000\,\mathrm{K}$ throughout the course of the \textit{HST} observations \citep{Odette}.
Moreover, assuming $\log\,g = 8.35$, \citet{Paula_GWLib2016} derived $T_\mathrm{eff} = 17\,560 \pm 9\,\mathrm{K}$ from 2015 \textit{HST}/COS data, showing that the system has not cooled back to its quiescent temperature yet. Therefore the $T_\mathrm{eff}$ derived here is only an upper limit for the quiescent white dwarf temperature and therefore, in the following analysis, we assume as $T_\mathrm{eff}$ for GW\,Lib the one derived for the 2002 dataset by \citet{Odette} for $\log\,g = 8.35$, $T_\mathrm{eff} = 14\,695\,\mathrm{K}$. 
The mass of the white dwarf in GW\,Lib has been measured by \citeauthor{GWLib} (\citeyear{GWLib}, $M_\mathrm{WD} = 0.79\,\pm\,0.08\,\mathrm{M}_\odot$) and \citeauthor{GWLib_mass} (\citeyear{GWLib_mass}, $M_\mathrm{WD} = 0.84\,\pm0.02\,\mathrm{M}_\odot$). However, these mass estimates depend on using velocity amplitude of the secondary star derived from the narrow Ca II emission lines during outburst (assuming they originate on the irradiated face of the donor star), and do not fulfil the accuracy requirement to be included in the sample from \citet{Zorotovic}, which represents the reference for our analysis. We therefore preferred a more conservative approach and assumed as uncertainty of the effective temperature, the one related to the unknown mass of the white dwarf, $\pm\,812\,$K (Section~\ref{subsec:unknown_mass}).

\subsubsection{Systems with a measured mass}\label{subsec:degeneracy}
As discussed in Section~\ref{sec:DataAnalysis}, it is not possible to break the degeneracy between $T_\mathrm{eff}$ and $\log\,g$ solely from the fit to the \textit{HST} data and therefore we assumed $\log g = 8.35$ ($M_\mathrm{WD} = 0.8\,\text{M}_\odot$). Although several CVs in our sample have published mass estimates (e.g. BD\,Pav, SDSS\,J123813.73--033932.9), only four of them have a mass measurement that is accurate enough to be included in the study of \citeauthor{Zorotovic} (\citeyear{Zorotovic}, see their table\,1): SDSS\,J103533.02+055158.4 (SDSS1035), IY\,UMa, DV\,UMa and SDSS\,J100658.41+233724.4 (SDSS1006). A previous $T_\mathrm{eff}$ that can be considered reliable is only available for SDSS1035 and DV\,UMa, which we discuss in the following.

\citet{1035} carried out a colour analysis of the SDSS1035 light curves during the white dwarf eclipse. From the observed shape of the white dwarf and bright spot eclipses, they measured a white dwarf mass of $M_\mathrm{WD} = 0.835 \pm 0.009\,\mathrm{M}_\odot$ and an effective temperature of $T_\mathrm{eff} = 10\,100 \pm 200\,\mathrm{K}$. 
The effective temperature we derived here, $T_\mathrm{eff} = 11\,620 \pm 44\,\mathrm{K}$, is $\simeq 1500\,\mathrm{K}$ higher than the temperature measured by \citet{1035}. Since our assumption of $\log g = 8.35$ is consistent with the measured white dwarf mass, we can conclude that this difference is related to the different methods used (light curves analysis vs. spectroscopic analysis).

While the masses of SDSS1035 correspond, within the errors, to $\log\,g = 8.35$, the mass of DV\,UMa is substantially higher: $M_\mathrm{WD} = 1.098 \pm 0.024\,\text{M}_\odot$ \citep{SDSS1035_2011}, which corresponds to $\log\,g = 8.78$. To take into account the independently measured surface gravity, we repeated the fit assuming $\log\,g = 8.78$, which returned $T_\mathrm{eff}(\log\,g = 8.78) = 18\,874\,\pm\,182\,\mathrm{K}$. As expected, this value is higher than the temperature we obtained for $\log g = 8.35$, i.e. $T_\mathrm{eff}(\log g = 8.35) = 18\,176\,\pm\,108\,\mathrm{K}$, and demonstrates that, if the masses were known, we could measure effective temperatures with an accuracy of $\simeq\,200\,\mathrm{K}$, appreciably more accurate than the uncertainty estimate recommended here for all systems without measured masses.

\citet{Feline} and \citet{SDSS1035_2011} measured the temperature of DV\,UMa from light curve analyses and report $T_\mathrm{eff} = 20\,000\,\pm\,1\,500\,\mathrm{K}$ and $T_\mathrm{eff} = 15\,500\,\pm\,2\,400\,\mathrm{K}$. Our result is in agreement with the temperature from \citeauthor{Feline} while it is hotter than the measurement from \citet{SDSS1035_2011}. Comparing the DV\,UMa spectrum (Figure~\ref{figure_spectra}, bottom) to the spectra of cooler systems (Figure~\ref{figure_spectra}, top and middle panels), it is easily recognizable that the narrower Ly$\alpha$ profile and the absence of quasi--molecular bands require $T \gtrsim 19\,000\,\mathrm{K}$.
Thus we are confident that our measurement is a realistic measure of the temperature of the white dwarf. DV\,UMa exemplifies how space--based ultraviolet spectroscopic measurements can give $T_\mathrm{eff}$ to about one per cent accuracy compared to the $\simeq 10-15$ per cent accuracy achievable from ground--based optical light curve analysis.

As a final caveat, we have to bear in mind that comparing effective temperatures measured at different epochs is always a delicate matter since a genuine temperature change can occur owing to increases of the system mass--transfer rate above its quiescent value.

\begin{figure}
 \centering
 \includegraphics[angle=-90,width=0.48\textwidth]{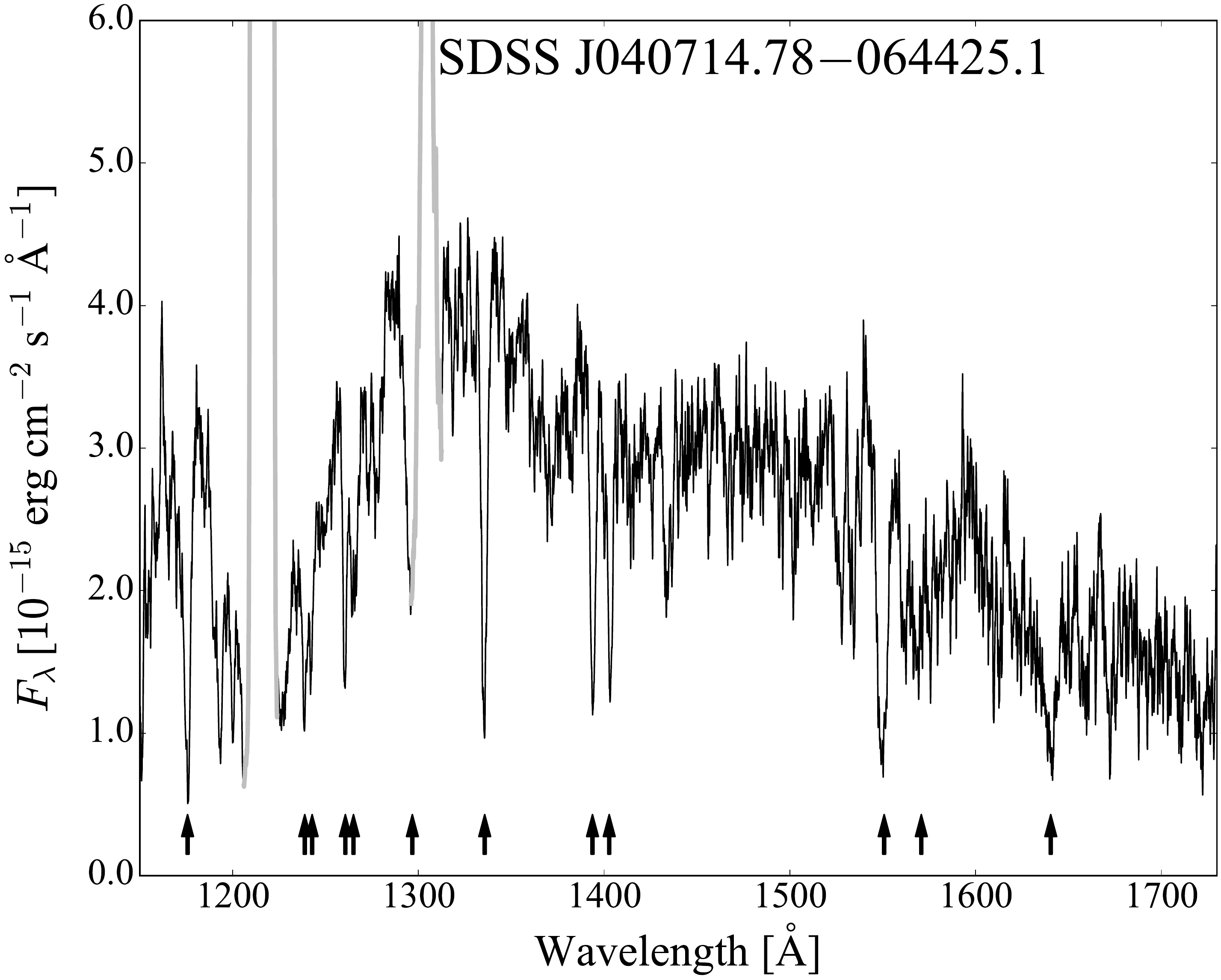}
 \caption{\textit{HST}/COS spectrum of SDSS\,J040714.78--064425.1. The arrows highlight the position of those lines which are contaminated by or arise from absorption within the veiling gas: \ion{C}{iii} ($1175\,$\AA), \ion{N}{v} ($1242\,$\AA), \ion{Si}{ii} ($1260\,$\AA), \ion{Si}{iii} ($1298\,$\AA), \ion{C}{ii} ($1335\,$\AA), \ion{Si}{iv} ($1400\,$\AA), \ion{C}{iv} ($1550\,$\AA) and broad absorption bands from \ion{Fe}{ii} ($1568\,$\AA\, and $1636\,$\AA). The geocoronal emission lines of Ly$\alpha$ ($1216\,$\AA) and \ion{O}{i} ($1302\,$\AA) are plotted in grey.}\label{iron_curtain} 
\end{figure}

\begin{figure*}
 \centering
 \subfloat{%
    \includegraphics[angle=-90,width=0.48\textwidth]{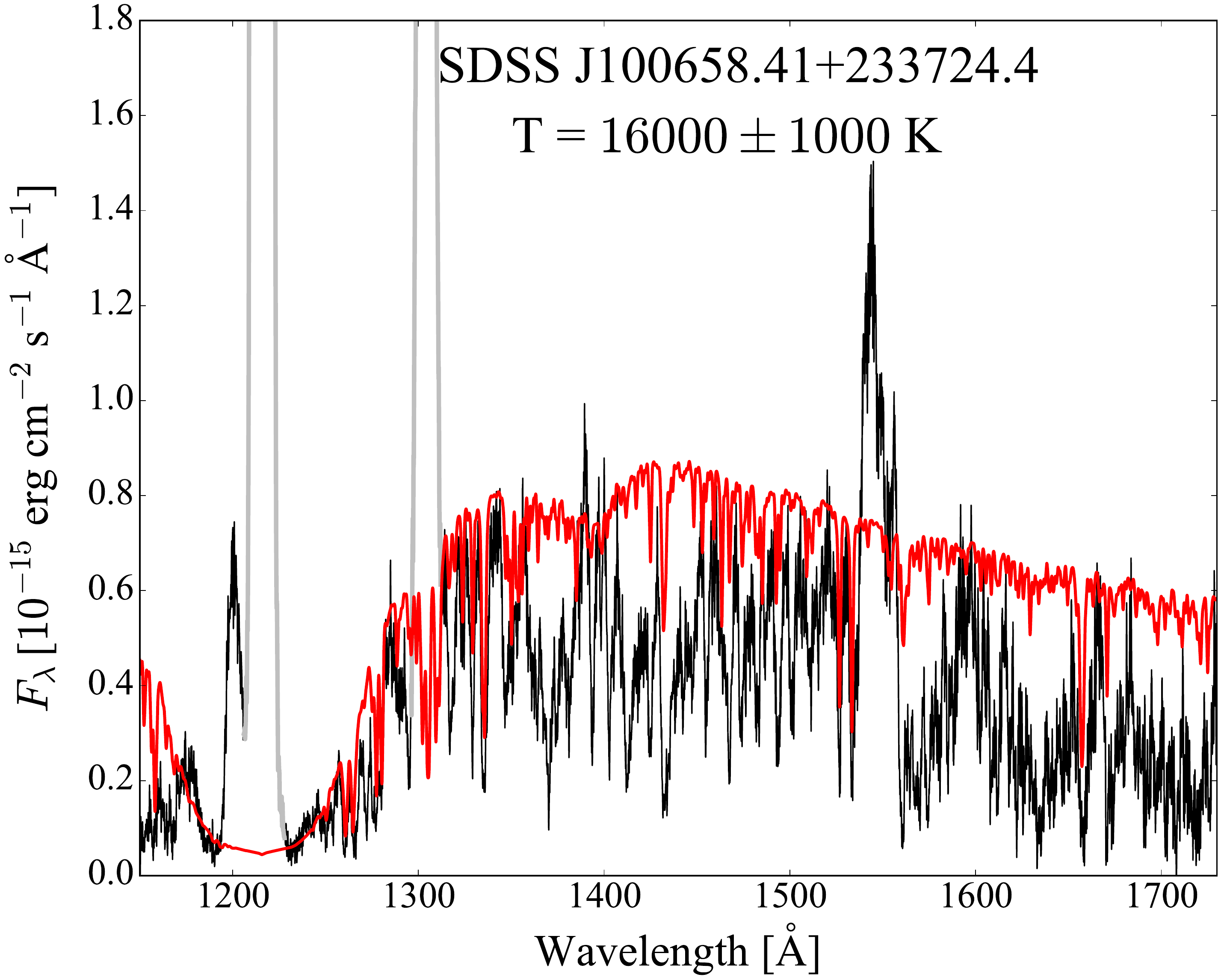}\qquad
    \includegraphics[angle=-90,width=0.48\textwidth]{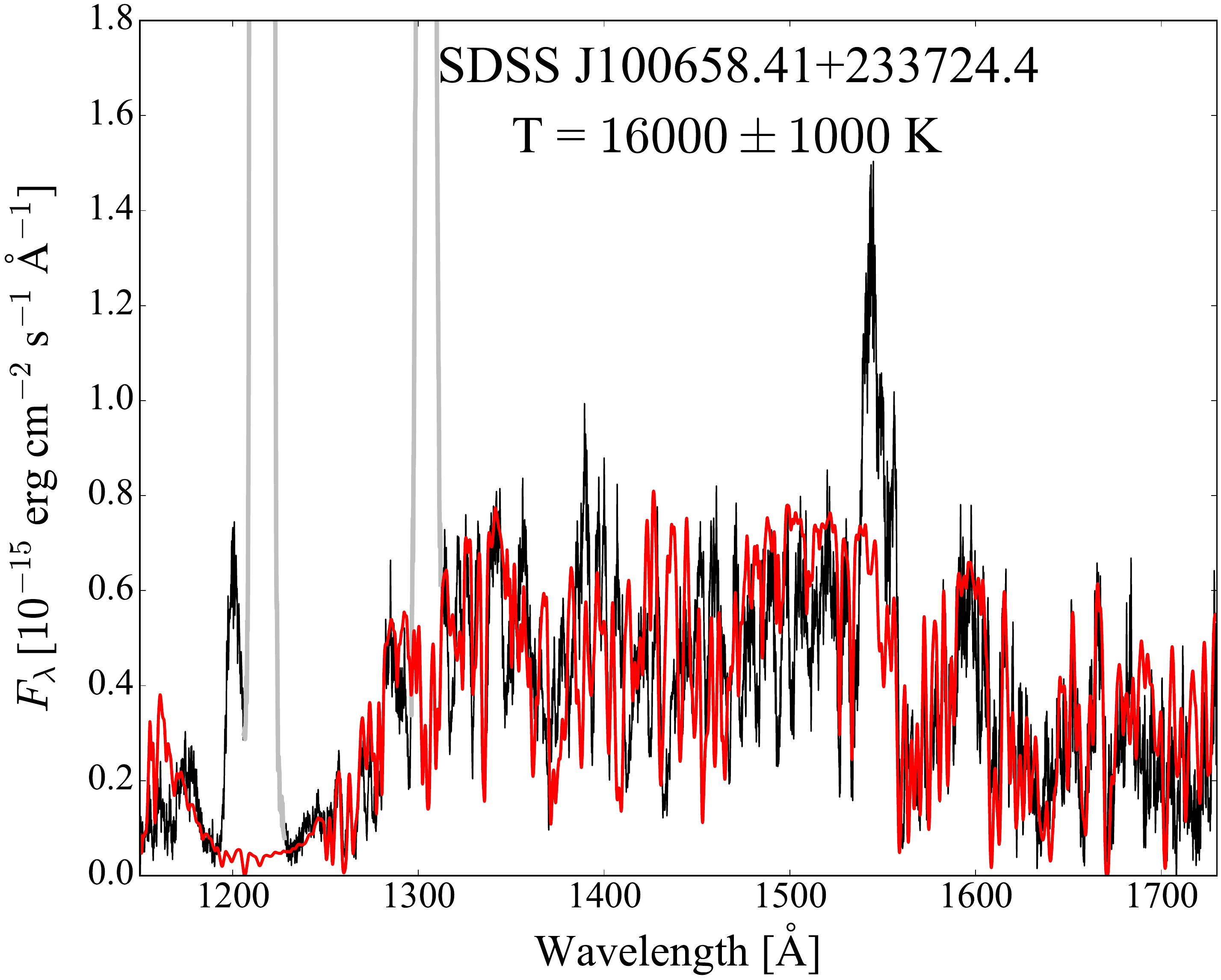}
    }
 \caption{\textit{HST}/COS spectrum (black) of SDSS\,J100658.41+233724.4 along with the best model (red) assuming a constant second component, with (right) and without (left) absorption from two slabs with different temperatures. The geocoronal emission lines of Ly$\alpha$ ($1216\,$\AA) and \ion{O}{i} ($1302\,$\AA) are plotted in grey.}\label{curtain}
\end{figure*}

\begin{figure*}
 \subfloat{%
  \includegraphics[angle=-90,width=0.48\textwidth]{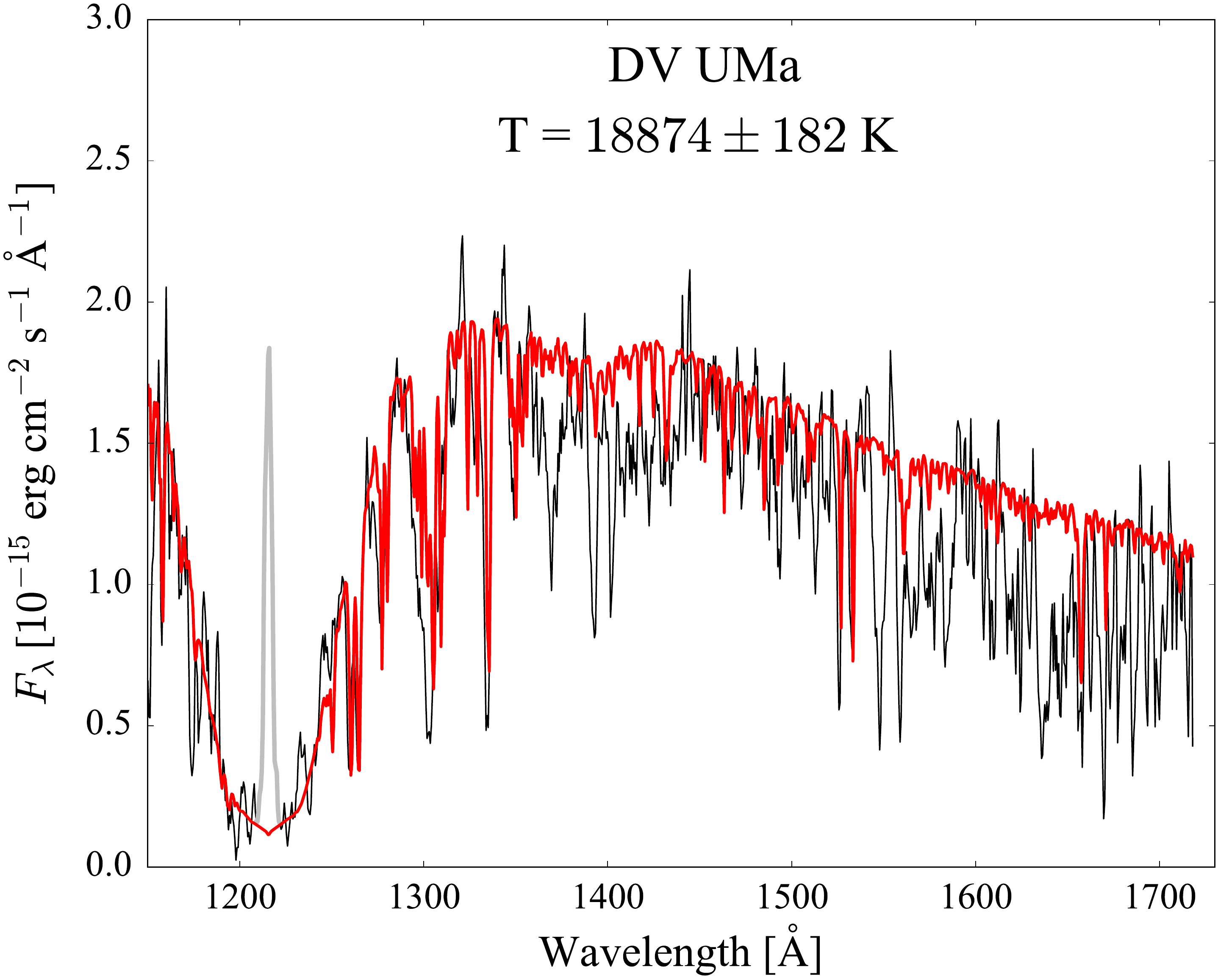}\qquad
  \includegraphics[angle=-90,width=0.48\textwidth]{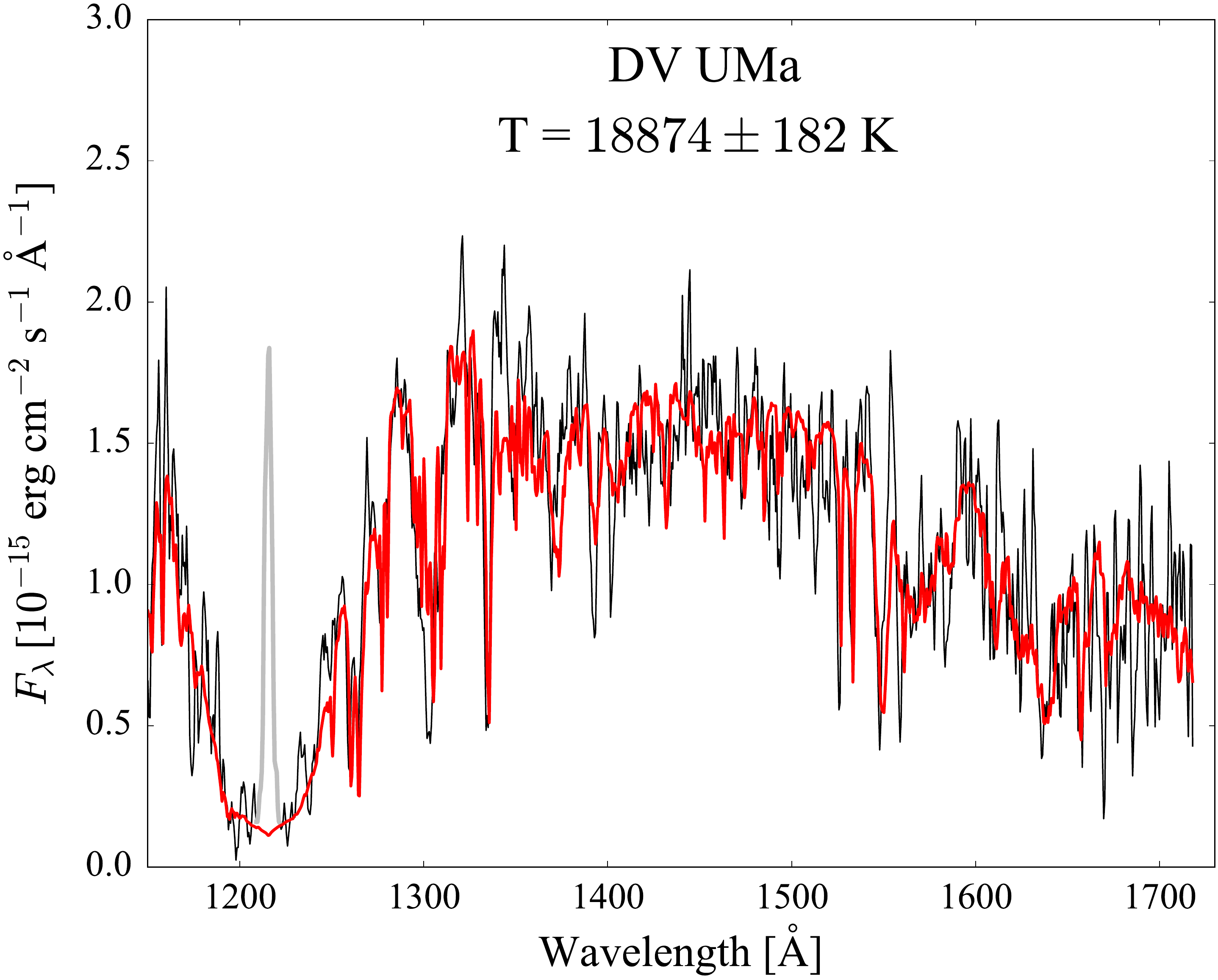}
  }
 \caption{\textit{HST}/STIS spectrum (black) of DV\,UMa along with the best model (red) assuming a constant second component, with (right) and without (left) absorption from two slabs with different temperatures. The geocoronal emission line of Ly$\alpha$ ($1216\,$\AA) is plotted in grey.}\label{curtain_dvuma}
\end{figure*}

\begin{table*}
\setlength{\tabcolsep}{0.10cm}
 \centering
 \caption{Optimal set of parameters for the two homogeneous slabs, one cold and one hot, veiling the white dwarf spectrum in the eclipsing systems.}\label{table_curtain}
  \begin{tabular}{lccccccccc}
  \toprule 
          & \multicolumn{4}{c}{Cold slab parameters} & & \multicolumn{4}{c}{Hot slab parameters}\\\cmidrule{2-5}\cmidrule{7-10}
  System  & $T_\mathrm{eff}\,(\mathrm{K})$ & $\log (n_\mathrm{e} \cdot \mathrm{cm}^3)$ & $\log (N_\mathrm{H} \cdot \mathrm{cm}^2)$ & $V_\mathrm{t}\,(\mathrm{km\,s}^{-1})$ &  & $T_\mathrm{eff}\,(\mathrm{K})$ & $\log (n_\mathrm{e} \cdot \mathrm{cm}^3)$ & $\log (N_\mathrm{H} \cdot \mathrm{cm}^2)$ & $V_\mathrm{t}\,(\mathrm{km\,s}^{-1})$\\ 
\midrule
IY\,UMa                    & 10\,000 & 12.0 & 20.5 & 200 & & 100\,000 & 18.7 & 20.3 & 200 \\
DV\,UMa                    & 10\,000 & 12.0 & 19.8 &  50 & &  80\,000 & 18.0 & 20.7 & 20  \\
SDSS\,J040714.78--064425.1 & 10\,000 & 12.0 & 19.9 & 250 & & 100\,000 & 18.7 & 19.9 & 200 \\
SDSS\,J100658.41+233724.4  & 10\,000 & 12.0 & 21.6 & 200 & & 100\,000 & 18.0 & 21.8 & 200 \\       
\bottomrule
\end{tabular}
\end{table*}

\subsection{Eclipsing systems}\label{eclipsing}
In eclipsing CVs, the line of sight can pass through material extending above the disc, giving rise to strong absorption features. This has been observed for example in OY\,Car, where a forest of blended \ion{Fe}{ii} absorption lines (the so--called ``iron curtain'') veil the white dwarf emission \citep{iron_curtain}. Our sample contains seven eclipsing systems: BD\,Pav, DV\,UMa IR\,Com, IY\,UMa, SDSS\,J040714.78--064425.1 (SDSS0407), SDSS1006 and SDSS1035. While the observations of BD\,Pav, IR\,Com and SDSS1035 are not strongly affected by the veiling gas, in the spectra of the remaining systems the curtain signature is revealed by several absorption features (Figure~\ref{iron_curtain}), which modify the overall slope of the spectrum and contaminate the core of the Ly$\alpha$ line, the main tracers for the white dwarf effective temperature. 

In order to estimate how our measurements and their uncertainties are affected by the presence of the veiling gas we carried out an exploratory study of each eclipsing CV. Inspecting their spectra, we identify strong \ion{C}{iv} ($1550\,$\AA) lines, suggesting the presence of hot gas along the line of sight, but also broad absorption bands from \ion{Fe}{ii} ($\simeq 1568\,$\AA\, and $\simeq 1636\,$\AA), which instead imply cold veiling gas. We therefore assumed that the white dwarf spectrum is attenuated by two homogeneous slabs, one cold ($T \simeq 10\,000\,$K) and one hot ($T \simeq 80\,000 - 100\,000\,$K). We used \textsc{synspec} to generate models for the monochromatic opacity of the slabs which, combined with the column densities, returns the absorption due to the curtain. 

The purpose of this study is not to determine the best--fitting model for the curtains (which will be done in a subsequent paper), but to determine the white dwarf effective temperature and realistic uncertainties. We varied the curtain parameters manually, exploring the space of four free parameters defining each slab (effective temperature $T_\mathrm{eff}$, electron density $n_\mathrm{e}$, turbulence velocity $V_\mathrm{t}$ and column density $N_\mathrm{H}$) and the white dwarf $T_\mathrm{eff}$ and scaling factor. 
We started from the white dwarf effective temperature determined following the prescription from Section~\ref{sec:DataAnalysis}. Since the problem already contains a large number of free parameters, we used a constant as second additional component. We attenuated the white dwarf emission using the best matching model for the curtain we could identify and then changed the white dwarf temperature and scaling factor to better reproduce the observed spectrum. We repeated this procedure iteratively until we found the optimal set of parameters (Table~\ref{table_curtain}), which define the final white dwarf $T_\mathrm{eff}$. This exercise  allowed us to estimate that the systematic uncertainties introduced by the curtain are of the order of $\simeq 1000\,$K for IY\,UMa, SDSS0407 and SDSS1006. Except for SDSS0407, this uncertainty is bigger than the systematic due to the unknown mass of the white dwarf and it is the value we use in the following analysis. 
In Figure~\ref{curtain}, we show the spectrum of SDSS1006 (black), which is dominated by several strong absorption bands. A white dwarf model alone (red, left) cannot adequately reproduce the observed flux level, which instead is well modelled including additional absorption from two slabs with different temperatures (red, right).

Finally, in the case of DV\,UMa (Figure~\ref{curtain_dvuma}), the main source of contamination comes from the iron bands in the red portion of the spectrum. Weaker absorption features are present at shorter wavelengths but they do not contaminate the Ly$\alpha$ region. For this reason we found that the veiling gas has no effect on the white dwarf temperature and therefore our result from Section~\ref{subsec:degeneracy} remains unchanged.

\subsection{Testing models for the present day CV population}\label{subsec:confronto_teorie}
\begin{figure*}
\thisfloatpagestyle{empty}
 \centering
 \subfloat{%
 \includegraphics[width=0.74\textwidth]{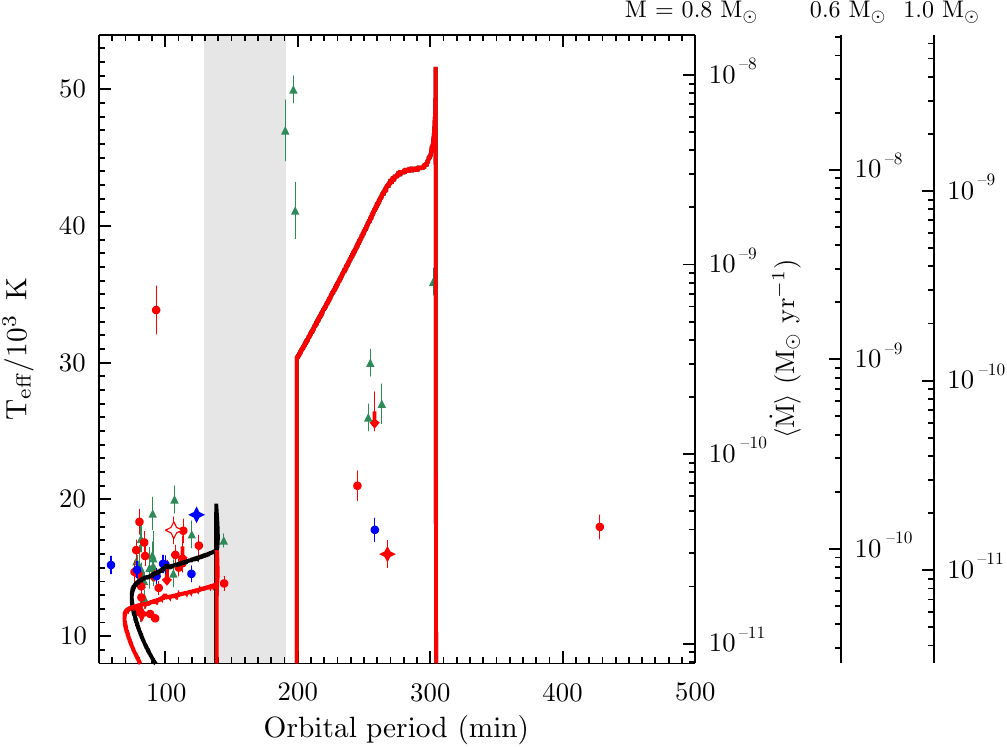}}\\
 \vspace{10pt}
 \subfloat{%
  \includegraphics[width=0.74\textwidth]{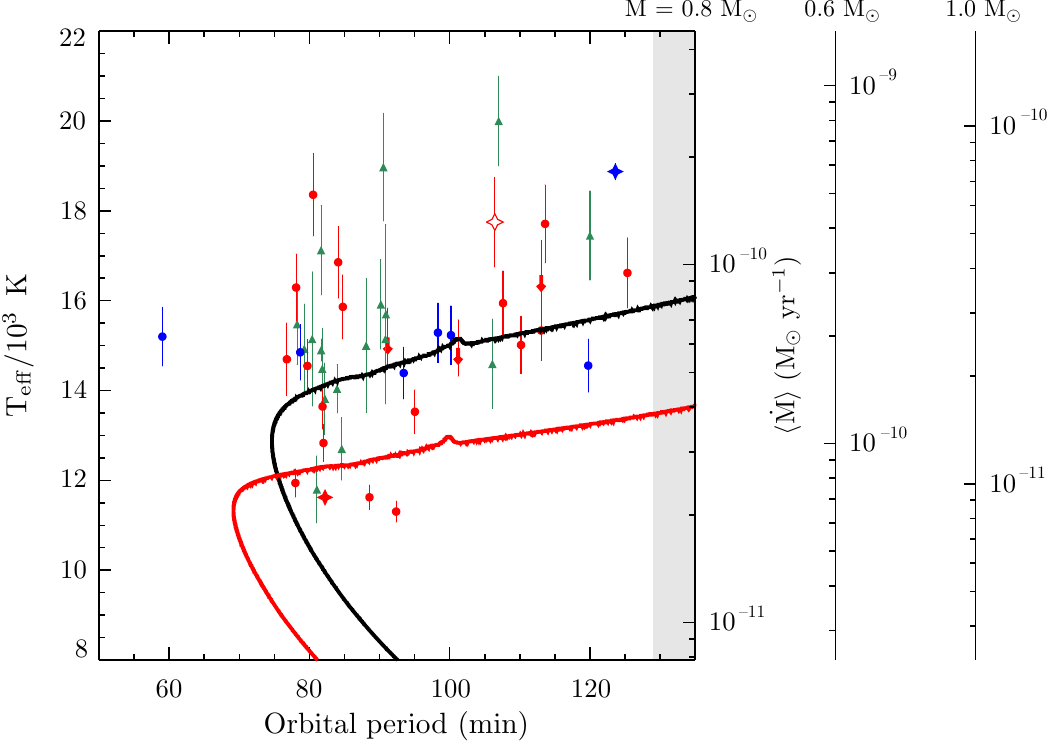}}
 \vspace{10pt}
  \caption{\emph{Top:} effective temperature as function of the orbital period from this work (circles, red for COS data, blue for STIS data) and for the non--magnetic system from \citeauthor{Tow_and_Boris2009} (\citeyear{Tow_and_Boris2009}, green triangles).  
Systems for which only an upper limit has been determined are shown with a downward arrow. The systems for which a mass measurement is available (DV\,UMa, SDSS1035 and SDSS1006) are shown with a starred symbol. IY\,UMa, for which a mass measurement is available but only an upper limit on its effective temperature has been determined, is shown with an empty star. The uncertainties associated to SDSS1035 and DV\,UMa are too small to be seen. Some temperatures in the TG09 sample do not have an associated error. In these cases, we estimated the corresponding uncertainty using our Equation~\ref{equation2} as described in Section~\ref{subsec:correzione}. The solid lines represent the evolutionary tracks for a system with a $0.8\,\mathrm{M}_\odot$ white dwarf accreting from a donor with an initial mass $0.65\,\mathrm{M}_\odot$, and an initial orbital period of 12 hours. The red track has been generated with the classical recipe for CV evolution (gravitational wave radiation only below the period gap) while the black track includes a ``residual'' magnetic braking equal to the gravitational wave radiation AML when the donor has no radiative core. The two tracks overlap above the period gap. The small bump at $P_\mathrm{orb} \simeq 100\,$ min is a computational artefact arising in the change of equation of state table describing the secondary internal structure. The grey band highlights the period gap while, on the right, a mapping to $\langle \dot{M} \rangle$ calculated through equation 1 from TG09 is shown for $M = 0.8$, $0.6$ and $1.0 \,\mathrm{M}_\odot$, respectively. \emph{Bottom:} closeup of the systems below the period gap.}\label{T-P}
\end{figure*}

In Figure~\ref{T-P} we show our sample (circles, red for COS data and blue for STIS data) and the corrected effective temperatures for the non--magnetic systems from TG09 (green triangles). While TG09 also included magnetic CVs observed in low state, we only observed non--magnetic systems (except for CC\,Scl, which is an Intermediate Polar). We exclude the magnetic systems from the following discussion, as their evolution may differ from that of non--magnetic CVs due to (potential) suppression of magnetic braking (\citeauthor{mag1} \citeyear{mag1}, \citeauthor{STIS_filter} \citeyear{STIS_filter}, \citeauthor{magnetic_wds} \citeyear{magnetic_wds}).

The uncertainties in the two samples are dominated by the unknown white dwarf mass. It is important to notice that these uncertainties are of the order of $\simeq\,300 - 1800\,\mathrm{K}$ (see Section~\ref{subsec:unknown_mass}), therefore the overall shape of the distribution is unlikely to change once the white dwarf masses are accurately determined.

The white dwarf $T_\mathrm{eff}$ is set by the compressional heating of the accreted material \citep{Dean_Lars} and provides a measurement of the secular mean accretion rate $\langle \dot{M} \rangle$ as defined by \citet{TowBild}. By comparing the distribution shown in Figure~\ref{T-P} with results from population synthesis calculations (see e.g. the $\dot{M}-P_\mathrm{orb}$ plane in figure\,2b from \citeauthor{Goliasch-Nelson} \citeyear{Goliasch-Nelson}), we can test the models for the present day CV population. Our results qualitatively agree with theoretical predictions, with the evolution of long period systems driven by magnetic braking and typical mass transfer rates of $\dot{M} \sim 10^{-10}$ to $\sim 10^{-9}\,\mathrm{M}_\odot \mathrm{yr}^{-1}$. In contrast, systems below the period gap, driven only by gravitational wave radiation\footnote{When the secondary star becomes fully convective ($P_\mathrm{orb} \simeq 3 \mathrm{h}$), it loses its radiative core, and the magnetic stellar wind, which sustains magnetic braking, is thought to be greatly reduced. In fact, there is good observation evidence for a decrease in the efficiency of magnetic braking in CVs \citep{Matthias_MB} as well as in single stars \citep{MB_ApJ}. At this point, in the standard model of CV evolution, the evolution of the system is thought to be driven only by gravitational wave radiation. However, there are arguments for an angular momentum loss that is larger than produced by pure gravitational wave radiation below the period gap, suggesting the presence of a `residual' magnetic braking at short periods (\citeauthor{Patterson1998} \citeyear{Patterson1998}, \citeauthor{Knigge2011} \citeyear{Knigge2011}).}, are characterised by a lower average $\dot{M} \sim 5 \times 10^{-11}\,\mathrm{M}_\odot \mathrm{yr}^{-1}$. 

However, there are some clear discrepancies between the standard model and our results, as illustrated by an evolutionary track\footnote{Evolutionary track computed using \textsc{MESAbinary} revision 8845 and inlists provided by \citet{mesa} for the cataclysmic variable case (their figure~5).} (red line) for a system with a $0.8\,\mathrm{M}_\odot$ white dwarf accreting from a donor with initially $M_2 = 0.65\,\mathrm{M}_\odot$ and an initial orbital period of 12 hours. Above the period gap, several systems with $T_\mathrm{eff} \simeq 40\,000 - 50\,000\,\mathrm{K}$ are present in the range $P_\mathrm{orb} \simeq 180-240\,\mathrm{min}$. These objects are known as VY\,Scl systems and are usually characterised by high temperatures, which indicate very high accretion rates onto the white dwarf, with $\langle \dot{M} \rangle$ about ten times above the value predicted by the standard evolution theory (TG09, \citeauthor{Howell2001} \citeyear{Howell2001}). \citet{Goliasch-Nelson} suggested that these objects could arise naturally from systems close to the regime of unstable mass transfer -- where the mass of the donor star is similar to the white dwarf mass -- but this hypothesis needs to be investigated in more detail.

The effective temperatures at $P_\mathrm{orb} \gtrsim 3\,\mathrm{h}$ indicate that there is more diversity above the period gap than expected from the results of TG09. Considering only the green points (non--magnetic CV from TG09), the systems above the period gap present clearly higher temperatures than the systems below. In contrast, the new systems from this work (red and blue circles) are characterised by surprisingly low temperatures, and the difference between CVs above and below the gap is not as pronounced as before. 
The interpretation of these results is not straightforward since we are limited by the small number of available effective temperatures in this period range, which is mainly populated by the so--called novalike systems. Novalikes are non--magnetic CVs and are further sub--divided into VY\,Scl systems, which have been detected at least once during a state of low accretion activity, and UX\,UMa systems, which have no recorded low states \citep{novalike}. This dichotomy is naturally subject to observational biases. Since the white dwarfs in novalike CVs are always/often hidden by the disc emission, it is difficult to measure their effective temperature and more observations are needed to increase the number of systems with an accurate $T_\mathrm{eff}$ above the period gap.
Additional constraints will be provided in the next few years by \textit{Gaia}: from its accurate parallaxes, we will determine an upper limit on the white dwarf effective temperatures of all disc--dominated CVs for which ultraviolet spectroscopy is available, improving our understanding of CV evolution above the period gap.

At short orbital periods, we note the presence of an extreme outlier, SDSS\,J153817.35+512338.0, with an effective temperature of $T_\mathrm{eff} = 33\,855 \pm 1\,785\,\mathrm{K}$. This is \textit{much hotter} than expected given its short orbital period ($P_\mathrm{orb} = 93.11\,\mathrm{min}$), i.e. for CVs with $P_\mathrm{orb} \lesssim 2 \mathrm{h}$, $T_\mathrm{eff} \simeq 15\,000\,\mathrm{K}$. This temperature is incompatible with mere accretion heating, but suggests that the system is either a young CV which just formed at this orbital period, that it had a recent period of sustained high mass transfer, or that it recently experienced a nova explosion and the white dwarf has not cooled down yet. We are obtaining follow--up observation for a detailed study of this system and the results will be presented in a future paper.

The bottom panel of Figure~\ref{T-P} shows a closeup of the systems below the period gap. We note two unexpected features: (i) there is a large scatter near $P_\mathrm{orb} \simeq 80\,\mathrm{min}$, in contrast to theoretical predictions that suggest a well defined $T_\mathrm{eff}$ at the period minimum and, (ii) the effective temperatures show a steeper decrease towards short periods, and a mean accretion rate about twice above that expected from pure GR AML. For comparison, we overplot a second evolutionary track (black) computed including a ``residual'' magnetic braking equal to the gravitational wave radiation AML when the donor has no radiative core. The new track agrees better with the observed $T_\mathrm{eff}$ decrease and the observed orbital period minimum than the standard recipe for CV evolution. Although the new track does not accurately reproduce the observed slope towards the period minimum, its better match to the measured effective temperatures strongly supports the ideas of additional AML mechanisms than GR below the period gap, which have been suggested as ``residual magnetic braking'' by \citet{Patterson1998} and \citet{Knigge2011} and, more recently, as consequential angular momentum loss (CAML) by \citet{Matthias2016} and \citet{Nelemans}.

Finally, reaching the minimum period, CVs are expected to evolve back towards longer periods with mean accretion rates of $\dot{M} \lesssim 2 \times 10^{-11}\,\mathrm{M}_\odot\,\mathrm{yr}^{-1}$, corresponding to $T_\mathrm{eff} \lesssim 11\,500\,\mathrm{K}$. These systems are the so--called `period bouncers' and the standard model of CV evolution predicts that about 70 per cent of the present day CVs should be found in this phase. 
However, we find only two period bouncer candidates in our sample: 1RXS\,J105010.8--140431 ($P_\mathrm{orb} = 88.56\,\mathrm{min}$, $T_\mathrm{eff} = 11\,622 \pm 277\,\mathrm{K}$) and QZ\,Lib ($P_\mathrm{orb} = 92.36\,\mathrm{min}$, $T_\mathrm{eff} = 11\,303 \pm 238\,\mathrm{K}$) which are characterised by effective temperatures $\simeq 3\,000 - 4\,000\,\mathrm{K}$ lower than those of the other systems at similar orbital periods, and both have also been identified as possible period bouncers by \citet{Patterson}. However these two systems have still higher temperatures than predicted by the theory and do not fall onto either of the lower two branches of the evolutionary tracks in Figure~\ref{T-P}. This could suggest the presence of enhanced AML in the post--bounce regime. This would consequently imply faster evolution than predicted by the standard theory, which could be a possible explanation for the dearth of this component of the CV population in the observed samples. 

The lack of unambiguous period bouncers in our sample cannot be explained by selection biases: we selected our targets among CVs that experience rare outbursts and for which the white dwarf is visible in their optical spectra. However, given their extremely low mass--transfer rate, period bouncers are expected to be intrinsically faint. 
In fact, assuming the average distance of the CVs in our sample $\langle d \rangle \simeq 320\,$pc, the expected apparent $V$--band magnitude for a period bouncer with $T_\mathrm{eff} \simeq 9\,000\,$K and $\log g = 8.35$ would be $V \simeq 20.5\,$mag (calculated using synthetic colours and evolutionary sequences of Hydrogen atmosphere white dwarfs\footnote{\url{http://www.astro.umontreal.ca/~bergeron/CoolingModels}}, \citeauthor{holberg_bergeron} \citeyear{holberg_bergeron}, \citeauthor{kowalski} \citeyear{kowalski}, \citeauthor{tremblay} \citeyear{tremblay}, \citeauthor{bergeron2011} \citeyear{bergeron2011}) and therefore it is possible that we are limited by the present observational technologies which are not sensitive enough to identify these elusive systems.

\subsection{CVs as Supernova Ia progenitors}\label{subsec:SNeIa}
The evolution and the final fate of CV white dwarfs are, as for all types of accreting white dwarfs, of key interest in the framework of Type Ia Supernova (SNe Ia) progenitors. Although early work showed the average mass of CV white dwarfs to lie in the range $\langle M_\mathrm{WD} \rangle \simeq 0.8-1.2\,\mathrm{M}_\odot$ (\citeauthor{Warner1973} \citeyear{Warner1973}, \citeauthor{Ritter1987} \citeyear{Ritter1987}), higher than that of isolated white dwarfs $\langle M_\mathrm{WD} \rangle \simeq 0.6\,\mathrm{M}_\odot$ (\citeauthor{Koester1979} \citeyear{Koester1979}, \citeauthor{Liebert2005} \citeyear{Liebert2005}, \citeauthor{Kepler2007} \citeyear{Kepler2007}), historically CVs have not been considered as a major component of the SNe Ia progenitor population. This is because most models for classical novae -- i.e. the thermonuclear ignition of the accreted hydrogen -- predict that not only most (if not all) of the accreted matter \citep{Nova_erosion} but also some of the underlying core \citep{Epelstain} would be ejected during the eruption (as supported by the observed abundances of nova ejecta, \citeauthor{nova_ejecta} \citeyear{nova_ejecta}), preventing mass growth. Nevertheless, \cite{Zorotovic} showed that CV white dwarfs masses are genuinely higher than both those of their detached progenitors and those of single white dwarfs, suggesting that CV white dwarfs may grow in mass. 

\citet{Wijnen2015A} investigated the effect on CV evolution of white dwarf mass growth either during nova cycles or through thermal time-scale mass transfer during the pre--CV phase, finding that this cannot explain the discrepancy between the mean masses of isolated and CV white dwarfs. Recent alternative explanations are consequential angular momentum loss and asymmetric mass ejection during nova explosions, which could lead to dynamically unstable mass transfer in CVs hosting low--mass white dwarfs, potentially merging these semi--detached binaries into single objects (\citeauthor{Matthias2016} \citeyear{Matthias2016}, \citeauthor{Nelemans} \citeyear{Nelemans}). On the other hand, the recent work by \citet{Hillman_mass_growth} shows that quasi--steady helium burning after several nova cycles can lead a CV white dwarf to reach the Chandrasekhar limit. At present, neither of these hypotheses have definitive arguments to rule out the other one and the debate on whether mass growth can occur or not is still wide open. It is therefore fundamental to improve our understanding of the long--term evolution of CVs in order to investigate their potential role as SNe\,Ia progenitors.

Against this background, it appears urgent to resolve the major discrepancy between observations and theory highlighted by our results. 
Particularly, below the period gap, the additional AML mechanisms required in order to better reproduce the measured effective temperatures could play a crucial role in the SNe\,Ia debate and a detailed theoretical analysis is needed to understand their origin. 
Finally, an observational effort is needed in order to enlarge the sample of CV white dwarfs with an accurate mass determination, presently limited to 32 \citep{Zorotovic}. Next year, the ESA \textit{Gaia} mission will provide accurate parallaxes in the second data release DR2 for all the targets in our sample. From the knowledge of the temperature and the distance, the white dwarf mass can then be estimated assuming a mass--radius relationship, thus doubling also the number of systems with an accurate mass determination.

\section{Conclusions}\label{sec:Conclusions}
From a large \textit{HST} survey of 45 cataclysmic variables we find 36 systems in which the white dwarf dominates the ultraviolet flux, as revealed by broad Ly$\alpha$ absorption. We determine the white dwarf effective temperatures by fitting the \textit{HST} data with atmosphere models.

Combining our results with the 43 CVs from \citet{Tow_and_Boris2009}, we almost double the number of objects with an accurate temperature measurement. Comparing the $T_\mathrm{eff}$--$P_\mathrm{orb}$ distribution with the mean accretion rate $\dot{M}$--$P_\mathrm{orb}$ distribution from population models, we test the model for the present day CV population. The systems above the period gap have, on average, mean accretion rates about one order of magnitude higher than those of the systems below the period gap. This is in accordance with the angular momentum loss mechanisms which are thought to dominate the different phases of CV evolution (magnetic braking above and gravitational wave radiation below the period gap).

However, besides this qualitative agreement between observations and theoretical predictions, some clear discrepancies are present:
\begin{itemize}
\item VY\,Scl systems dominate the orbital period range $P_\mathrm{orb} \simeq 180-240\,\mathrm{min}$, with $T_\mathrm{eff} \simeq 40\,000 - 50\,000\,\mathrm{K}$. These temperatures correspond to a $\langle \dot{M} \rangle$ about ten times above the value predicted by the standard evolution theory. 
\item The results at $P_\mathrm{orb} \gtrsim 3\,\mathrm{h}$ present a much more variegated scenario than in the smaller sample analysed by \citet{Tow_and_Boris2009}, with the new systems showing surprisingly low temperatures. To improve our understanding of CV evolution  above the period gap, we need to increase the number of well--characterised systems in this period range. 
\item At short orbital periods, the temperatures imply higher mass transfer rates than predicted by pure GR.
Moreover, near the period minimum the data show a large scatter that is not accounted for by the models. 
\item The standard model of CV evolution predicts that period bouncers should make up for about 70 per cent of the present day CV population. We identify only two period bouncer candidates in our sample and, since the lack of these systems cannot be explained by selection effects, we conclude that the present methods used to identify CVs are not sensitive enough to spot these faint systems.
\end{itemize}

We note that our results are dominated by a systematic uncertainty of $\sim\,300 - 1800\,\mathrm{K}$ due to the unknown white dwarf mass. Once additional constraints on the masses -- dynamical and/or distances -- become available, the white dwarf effective temperatures will be known with an accuracy of about $\simeq 200\,$K. While a significant improvement, the overall shape of the distribution will not drastically change and our conclusions will remain unchanged.

Finally, we identify an exceptional outlier (SDSS\,J153817.35+512338.0) below the period gap, whose extremely high temperature ($T_\mathrm{eff} = 33\,855\,\pm\,1\,785\,\mathrm{K}$) for its short orbital period ($P_\mathrm{orb} = 93.11\,\mathrm{min}$) could be explained by a young CV just formed at this $P_\mathrm{orb}$.

We will continue with a detailed analysis of the individual systems in the sample, measuring rotation rates and abundances. This enlarged sample of well--characterised CV white dwarfs will provide robust observational constraints on the response of white dwarfs to the accretion of mass, angular momentum and energy.

\section*{Acknowledgments}
Based on observations made with the NASA/ESA Hubble Space Telescope, obtained at the Space Telescope Science Institute, which is operated by the Association of Universities for Research in Astronomy, Inc., under NASA contract NAS 5--26555. These observations are associated with program GO--9357, GO--9724, GO--12870 and GO--13807.

We gratefully acknowledge the variable star observations from the AAVSO International Database contributed by observers worldwide and used in this research.
We are indebted to the global amateur community for their outstanding support, which made the \textit{HST} survey possible.

We are also extremely grateful for the tireless efforts of the operations and COS teams at STScI, including Amber Armstrong, Elena Mason, Charles Proffitt, Nolan Walborn and Alan Welty, in implementing this challenging program.

The research leading to these results has received funding from the European Research Council under the European Union's Seventh Framework Programme (FP/2007--2013) / ERC Grant Agreement n. 320964 (WDTracer). 

DdM acknowledges support from ASI--INAF I/037/12/0. 
MRS thanks for support from Fondecyt (1141269). 
TRM acknowledges STFC ST/L000733.
PG is pleased to thank William (Bill) Blair for his kind hospitality at the Rowland Department of Physics and Astronomy at the Johns Hopkins University, Baltimore, MD.

This research has made use of the APASS database, located at the AAVSO web site. Funding for APASS has been provided by the Robert Martin Ayers Sciences Fund. 

This research has made use of the NASA/IPAC Extragalactic Database (NED), which is operated by the Jet Propulsion Laboratory, California Institute of Technology, under contract with the National Aeronautics and Space Administration.




\bibliographystyle{mnras}
\bibliography{mnras_template} 








\bsp	
\label{lastpage}
\end{document}